%% file: yuma2013_ms_apj_accepted.tex
\documentclass[iop, apj, natbib]{emulateapj}
\usepackage{graphicx}
\usepackage{amsmath}
\usepackage{amssymb}
\usepackage{color}
\usepackage{multirow}
\usepackage{rotating}
\usepackage{booktabs}
\usepackage{times}
\usepackage{natbib}

\newcommand{\Msun}{\ensuremath{M_{\odot}}}
\newcommand{\Zsun}{\ensuremath{Z_{\odot}}}

\newcommand{\ar}{\arcsec}
\newcommand{\Llya}{$L {\rm (Ly}\alpha{\rm )}$}
\newcommand{\lya}{Ly$\alpha$}
\newcommand{\ha}{H$\alpha$}
\newcommand{\oii}{[O\,{\sc ii}]}	
\newcommand{\oiib}{O\,{\sc ii}B}	
\newcommand{\oiii}{[O\,{\sc iii}]}
\newcommand{\nev}{[Ne\,{\sc v}]}
\newcommand{\mgii}{Mg\,{\sc ii}}
\newcommand{\feii}{Fe\,{\sc ii}}
\newcommand{\feiis}{Fe\,{\sc ii}$^{\ast}$}
\newcommand{\nai}{Na\,{\sc i}}

\newcommand{\hi}{H\,{\sc i}}
\newcommand{\hii}{H\,{\sc ii}}
\newcommand{\nii}{[N\,{\sc ii}]}
\newcommand{\sii}{[S\,{\sc ii}]}
                                   
\newcommand{\nb}{\textit{NB816}}
\newcommand{\nbcorr}{\textit{NB816}$_{\rm corr}$}
\newcommand{\rz}{\textit{Rz}}

\newcommand{\kms}{km\,s$^{-1}$}
\newcommand{\ergs}{erg\,s$^{-1}$}

\newcommand{\ergscm}{erg\,s$^{-1}$\,cm$^{-2}$}
\newcommand{\ergscmA}{erg\,s$^{-1}$\,cm$^{-2}$\,\AA$^{-1}$}

\slugcomment{Accepted for Publication in ApJ: October 9, 2013}

\shorttitle{Systematic \oii\ Blob Search}
\shortauthors{Yuma et al.}

\begin{document}

\title{First Systematic Search for Oxygen-Line Blobs at High Redshift: Uncovering\\
 AGN Feedback and Star-Formation Quenching}
 
\author{Suraphong Yuma\altaffilmark{1}, 
	Masami Ouchi\altaffilmark{1},
	Alyssa B. Drake\altaffilmark{2}, 
	Chris Simpson\altaffilmark{2}, 
	Kazuhiro Shimasaku\altaffilmark{3}, 
	Kimihiko Nakajima\altaffilmark{3}, \\
	Yoshiaki Ono\altaffilmark{1}, 
	Rieko Momose\altaffilmark{1}, 
	Masayuki Akiyama\altaffilmark{4}, 
	Masao Mori\altaffilmark{5}, and
	Masayuki Umemura\altaffilmark{5}
        }

\altaffiltext{1}{Institute for Cosmic Ray Research, The University of Tokyo, Kashiwa-no-ha, Kashiwa 277-8582, Japan}
\altaffiltext{2}{Astrophysics Research Institute, Liverpool John Moores University, IC2, Liverpool Science Park, 146 Brownlow Hill, Liverpool L3 5RF, UK}
\altaffiltext{3}{Department of Astronomy, Graduate School of Science, The University of Tokyo, Hongo, Bunkyo-ku, Tokyo, 113-0033, Japan}
\altaffiltext{4}{Astronomical Institute, Tohoku University, Aoba-ku, Sendai, 980-8578, Japan}
\altaffiltext{5}{Center for Computational Science, University of Tsukuba, Tsukuba, Ibaraki 305-8577, Japan}

\email{yuma@icrr.u-tokyo.ac.jp}

\begin{abstract}

We present the first {\it systematic} search for extended metal-line
\oii$\lambda\lambda$3726,3729 nebulae, or \oii\ blobs (\oiib s), at $z=1.2$ 
using deep narrowband imaging with 
a survey volume of $1.9\times 10^5$ Mpc$^3$ 
on the 0.62 deg$^2$ sky of Subaru$-${\it XMM} Deep Survey (SXDS) field.
We discover a giant \oiib, dubbed ``\oiib\ 1," 
with a spatial extent over $\sim75$ kpc
at a spectroscopic redshift of $z=1.18$, and also identify a total of
12 \oiib s with a size of $>30$ kpc. 
Our optical spectrum of \oiib\ 1 
presents \nev$\lambda$3426 line at the 
$6\sigma$ level, indicating that this object harbors an obscured type-2 
active galactic nucleus (AGN). 
The presence of gas outflows in this object is suggested by two marginal detections of 
\feii$\lambda2587$ absorption and \feiis$\lambda2613$ emission lines 
both of which are blueshifted at as large as $500-600$ km s$^{-1}$, 
indicating that the heating source of \oiib\ 1 is AGN or associated shock 
excitation rather than supernovae produced by starbursts. 
The number density of \oiib\ 1-type giant blobs is estimated to
be $\sim5\times10^{-6}$ Mpc$^{-3}$ at $z\sim1.2$, 
which is comparable with that of AGNs driving outflow at a similar redshift, 
suggesting that giant \oiib s are produced only by AGN activity. 
On the other hand, the number density of small \oiib s, $6\times10^{-5}$ Mpc$^{-3}$, 
compared to that of $z\sim1$ galaxies in the blue cloud in the same $M_B$ range, 
may imply that 3\% of star-forming galaxies at $z\sim1$ 
are quenching star formation through outflows involving extended \oii\ emission. 

\end{abstract}


\section{Introduction}\label{intro}

Elliptical galaxies are one of the most common galaxy populations in the local universe. 
They are featureless and dominated by old stars with very low or zero star formation rates (SFRs). 
Their formation and evolution processes have been studied over decades both theoretically and observationally. 
Many simulations suggest that major mergers of two galaxies with comparable masses 
can produce giant elliptical galaxies \citep{toomre77, barnes92, naab06}. 
Basically, the progenitors of elliptical galaxies have to stop forming new stars and 
begin evolving passively at some epoch in the past. 
That epoch should be as early as $z\gtrsim1$, because the luminosity$-$size and 
stellar mass$-$size relations of early-type galaxies 
from $z\sim1$ evolve consistently with completely passive evolution 
of their stellar populations \citep{trujillo04, mcintosh05}. 
It is not yet understood what makes galaxies stop forming stars and start to evolve passively. 
One popular proposed mechanism for quenching star formation is gas outflows by 
a starburst and/or an active galactic nucleus \citep[AGN; e.g.,][]{dekel09}. 
Therefore, studying outflow processes at $z\gtrsim1$ (especially in massive galaxies) 
is of importance in understanding the evolution of quiescent elliptical galaxies. 

Besides the quenching process of star formation activity, the gas outflow is also a crucial solution to 
various observational puzzles seen, e.g., in the mass$-$metallicity relation \citep[e.g.,][]{tremonti04} 
and the chemical enrichment of the interstellar medium (ISM) and the intergalactic medium 
\citep[IGM; e.g., ][]{martin05, rupke05a, rupke05b, weiner09, coil11}. 
Galactic outflows are theoretically the primary mechanism for regulating the baryon 
and metallicity abundances of galaxies. 
Without strong feedback, cosmological models of galaxy evolution produce too many galaxies 
with high SFRs and large stellar masses compared to observations 
\citep[e.g.,][]{cole00, springel03, keres09}. Some process is thus needed to prevent 
gas from cooling into the central galaxies of dark matter halos and forming stars. 
At the low-mass end of the galaxy mass function, feedback driven by supernovae (SNe) 
is often referred to as an energy/momentum input for outflowing gas from galaxies 
\citep{dekel86, springel03, murray05}. 
Meanwhile, models with feedback from AGN possibly account for the presence of 
massive elliptical galaxies \citep[e.g.,][]{somerville08}. 

Observationally, galactic-scale outflows have been widely found 
in local ultra luminous infrared galaxies \citep[ULIRGs; e.g.,][]{heckman90, martin05, soto12}, 
in post-starburst galaxies at $0.2<z < 0.8$ \citep[e.g.,][]{coil11}, in radio galaxies at $z\sim0.5$ 
\citep[e.g.,][]{nesvadba08, liu13}, in normal star-forming galaxies \citep[e.g,][]{steidel10, martin12, kornei12, erb12, bradshaw13}, 
and in sub-millimeter galaxies at $z\sim2$ \citep[e.g.,][]{alexander10}. 
Outflowing gas is commonly traced via interstellar absorption lines blueshifted relative to stars and \hii\ regions. 
It is found that a certain critical SFR surface density ($\sim0.1$ \Msun\ yr$^{-1}$ kpc$^{-2}$) is 
necessary to drive an outflow \citep{murray11, kornei12} and that outflows are stronger 
for more massive and/or higher-SFR galaxies \citep{weiner09, kornei12, martin12}. 

In order to study galactic outflows in {\it systematic} surveys at high redshifts ($z\gtrsim1$), 
it is necessary to systematically select galaxies at targeted redshifts. 
The narrowband technique is one of the efficient methods used for this purpose, 
providing a systematic sample of galaxies with strong emission lines. 
The idea of this method is to use a narrowband filter to detect a specific nebular emission 
line redshifted into the narrow wavelength range of the filter. 
This method has provided vast numbers of star-forming galaxies ranging from low redshifts  
to redshifts as high as $z\sim7.0$. 
At $z\leq 2.0$, star-forming galaxies with strong \oii, \oiii, and \ha\ emission lines 
are selected as \oii, \oiii, and \ha\ emitters, respectively \citep[e.g., ][]{bunker95, ly12, drake13}. 
Another important type of star-forming galaxy selected via this method is Lyman alpha emitters (LAEs), 
which show a strong \lya\ emission line. Thousands of LAEs have been selected 
at various redshifts over $z\sim0.3$-7.0 \citep[e.g.,][]{cowie10, ouchi08, ouchi10}. 

Some LAEs show very luminous and extended ($20-300$ kpc) Ly$\alpha$ emission. 
They are called \lya\ blobs \citep[LABs; e.g.,][]{matsuda04} and 
are good candidates of galaxies with outflow at high redshifts. 
LABs are largely found at $z\sim2-3$ with \lya\ luminosities of \Llya\ $\gtrsim10^{43}$ \ergs
and spatial extents over 50 kpc \citep{matsuda04, yang09}. 
The most remarkable LABs known at present are \citeauthor{steidel00}'s (2000) 
LAB1 and LAB2 in SSA22 with 
\Llya\ $\simeq 10^{44}$ \ergs\ and sizes of $\sim100$ kpc. 
The most distant LAB is the one discovered by \cite{ouchi09} at $z\sim7$; it shows \Llya\ of  
$\sim4\times10^{43}$ \ergs\ and physically extends more than $17$ kpc. 
Gas outflows by intense starbursts, black hole accretion
\citep{mori04, taniguchi00, wilman05}, photoionization from
massive stars, and/or AGNs \citep{matsuda04, geach09}
are plausible origins of LABs. 
However, we should note that inflows of cooling gas
may also be able to produce LABs \citep{fardal01, smith07}.

Recently, \cite{brammer13} reported an \oiii\ blob at $z=1.61$ spatially extended over 1\ar\ or 8.5 kpc. 
The rest-frame equivalent width (EW$_0$) is $\sim3500$\AA\ for 
the blended H$\beta$ and \oiii$\lambda\lambda4959,5007$ emission lines. 
This discovery was made by {\it Hubble Space Telescope} grism observations; 
therefore, they were able to resolve 1\ar\ extent beyond the galaxy, which 
would not be possible from the ground. 
Unlike resonantly scattered hydrogen-line emission such as \lya\ emission 
that can be plausibly explained by both outflows and inflows, 
extended profiles of metal emission (i.e., oxygen-line emission) are 
more compatible with the outflow scenario rather than the inflow of 
pristine gas from the IGM whose metallicity is only $10^{-3}$\Zsun 
\citep[e.g.,][]{aguirre08, fumagalli11}. 

To date, outflows from objects with oxygen-line emission have been 
studied only after the objects are identified as AGNs. 
In this paper, we use the narrowband technique
to conduct the first systematic search for galaxies with outflow features
at $z\simeq1$ including both AGN host and star-forming galaxies.
We look for \oii\ emission instead of \oiii\ emission,
because we aim to study galaxies as distant as possible
and \oii\ is the shortest-wavelength strong metal emission line. 
This systematic search 
enables us to provide a meaningful constraint 
on the frequency of outflow occurrence in star-forming galaxies at 
$z\gtrsim1$, a crucial epoch that the star formation activity of 
massive galaxies is being quenched. 

This paper is organized as follows. 
Section \ref{sample} describes the data and the selection method 
used to identify \oii\ blobs (\oiib s). We also measure photometry and 
derive stellar properties of the selected \oiib s. 
Section \ref{blob1} explains the photometric, stellar, and spectroscopic properties
of a giant \oiib\ we discover. 
Section \ref{other_blobs} presents the spectroscopic data of other \oiib s. 
Section \ref{discuss} discusses the AGN activity, the signature of 
an outflow, and its possible origins for this giant blob.
We also estimate the number density of \oiib s.
Finally, we summarize all results in Section \ref{summary}.

Throughout this paper, magnitudes are described in the AB system. 
We assume a standard $\Lambda$CDM cosmology with parameters of 
$(\Omega_m, \Omega_{\Lambda}, H_0) = 
(0.3, 0.7, 70\ {\rm km}\ {\rm s}^{-1}\ {\rm Mpc}^{-1})$. 

\section{\oiib\ Candidates at $\lowercase{z}\sim1.2$}\label{sample}

\subsection{Sample Selection}\label{select}

To search for \oiib s, we first select \oii\ emitters whose 
\oii\ emission falls into the narrowband filter \nb\ 
with $\lambda_c = 8150$ \AA\ and FWHM $=120$ \AA. 
We use the catalog of \oii\ emitters at $z\sim1.2$ provided by \cite{drake13}. 
Full details in the selection of \oii\ emitters are described in \cite{drake13}; 
a brief description is as follows. 
An \nb\ image was taken 
with Subaru/Suprime-Cam for an area of $\approx 1$ deg$^2$
in the Subaru$-${\it XMM} Deep Survey (SXDS) field by \cite{ouchi08}.
Its $5\sigma$ limiting magnitude is 26.0 mag on a 2\farcs0 diameter aperture.
Deep broadband ($B$, $V$, $R$, $i$, and $z$) images were also obtained 
by \cite{furusawa08} whose $5\sigma$ limiting magnitudes on a 2\farcs0 aperture 
are 27.39, 27.10, 26.85, 26.66, and 25.95 mag, respectively. 
To obtain good-quality photometric redshifts, they also use {\it u}-band 
data from Megacam on the Canada France Hawaii Telescope (CFHT), {\it J,H,K} 
data from DR8 version of the UKIDSS Ultra Deep Survey \citep[UDS;][]{lawrence07} 
taken with WFCAM on the UK Infrared Telescope (UKIRT), and Infrared Array Camera (IRAC) 
channels 1 (3.6\micron) and 2 (4.5\micron) from {\it Spitzer} UDS  survey (SpUDS; PI: J. Dunlop). 
\oii\ emitters are selected in a 0.62 deg$^2$ region (after masking) 
where the SXDS/UDS data overlap. 

Object detection was made on the \nb\ image. 
Then, for all objects above the $5\sigma$ limiting magnitude, 
2\farcs0 diameter aperture magnitudes are measured for the other bandpasses 
at exactly the same positions as in the \nb\ image by using the dual mode of 
SExtractor \citep{bertin96}. 
Emitter candidates are required to exhibit a significant excess 
in \nb\ with respect to an offband defined by a combination of 
the $R$ and $z$ bands \citep[Figure 2 in ][]{drake13}. 
Possible interlopers (i.e., late-type, red stars) are removed from the emitter candidates 
by applying the $B-z$ and $z-K$ color diagram \citep{daddi04} and $B-z>2.5$ criterion. 
To summarize, \cite{drake13} have 3597 \nb-excess objects classified as galaxies. 
1013 of them are classified as \oii\ emitters at $0.80< z < 1.50$ by using photometric redshifts, 
while the others are either \ha\ emitters at $z\sim0.25$ or \oiii\ emitters at $z\sim0.63$. 

\begin{figure}
\centering
\includegraphics[width=0.45\textwidth]{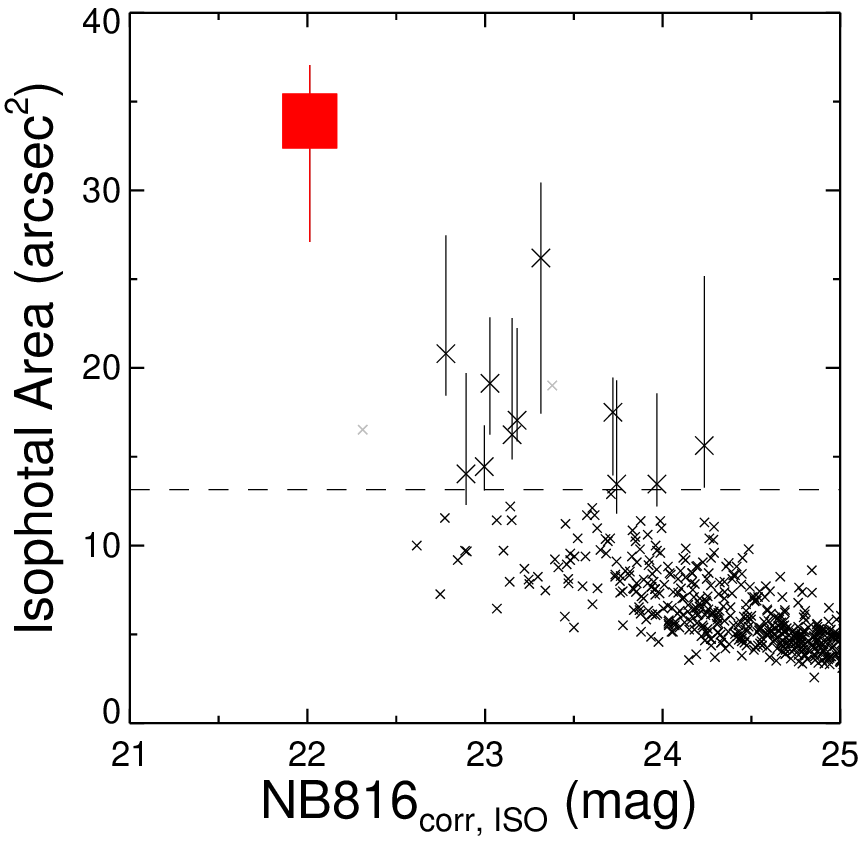}
\caption{Distribution of isophotal area and magnitude on the \nbcorr\ image 
for extended \oii\ emitters at $z\sim1.2$ (black crosses). 
Red square shows a giant \oiib, \oiib\ 1,
and the large black crosses denote the small \oiib s.
The gray crosses indicate two \oii\ emitters that are not
\oiib s, but objects with a large isophotal area made by source blending.
The isophotal$-$area criterion ($>13$ arcsec$^2$) 
identical to that adopted by \cite{matsuda04} is shown with a horizontal dashed line. 
Fourteen emitters are above this criterion including \oiib\ 1 and two objects with false detection. 
The error bars are defined by the 68\% distributions of 
recovered isophotal areas given by our Monte-Carlo simulations
(see text). 
} 
\label{isophotal}
\end{figure}
\input{tab_o2blob_all}

To select \oiib s, 
we adopt a similar methodology to what was used to search for LABs 
at $z=3.1$ by \cite{matsuda04}. 
We search for \oiib s in the emission-line \nbcorr\ image, which 
is obtained by subtracting a continuum-emission image from the \nb\ image; 
the continuum-emission image is constructed from the $R$ and $z$ images
at $\rz\equiv(R+2z)/3$. 
Then objects are detected in the \nbcorr\ smoothed with a Gaussian kernel 
with a FWHM of 1\farcs0. 
We regard an object as being extended if it consists of $\ge 20$ 
continuous pixels above the $2\sigma$ background surface brightness 
fluctuation (28 mag arcsec$^{-2}$ or $1.2\times10^{-18}$ \ergscm\ arcsec$^{-2}$) 
of the original \nbcorr\ image. Extended objects are then crossmatched with 
the catalog of \oii\ emitters described above. 
Figure \ref{isophotal} plots the isophotal area of extended \oii\ emitters 
as a function of \nbcorr\ magnitude. 
\citeauthor{matsuda04} define their LABs as objects with an isophotal area above 16 arcsec$^2$, 
corresponding to a spatial extent of 30 kpc at $z=3.1$. 
By adopting the same criterion for our extended \oii\ emitters, 
we find fourteen objects to satisfy a criterion of 
isophotal area above 13 arcsec$^2$ (corresponding to a spatial extent of 30 kpc at $z\sim1.2$). 
We classify 12 of them as \oiib s after visual inspection. 
The remaining two objects are found to be blended with nearby objects in the smoothed \nbcorr\ image, 
giving a falsely large isophotal area (the gray crosses in Figure \ref{isophotal}). 
%
We estimate the uncertainties of isophotal area measurements by Monte-Carlo simulations. 
We cut out images of the \oiib s, and place them on $\sim 1000$ empty sky regions of 
the NB816corr image to produce artificial \oiib images. We measure 
their isophotal areas in the same manner as we performed for the real \oiib s, 
and define a $1 \sigma$ error of isophotal area
by the 68 percentile of the $\sim 1000$ artificial object distribution.
The estimated uncertainties are presented in Figure \ref{isophotal}.

\begin{figure*}
\begin{center}
\begin{tabular}{cccccc}
\includegraphics[width=0.15\textwidth]{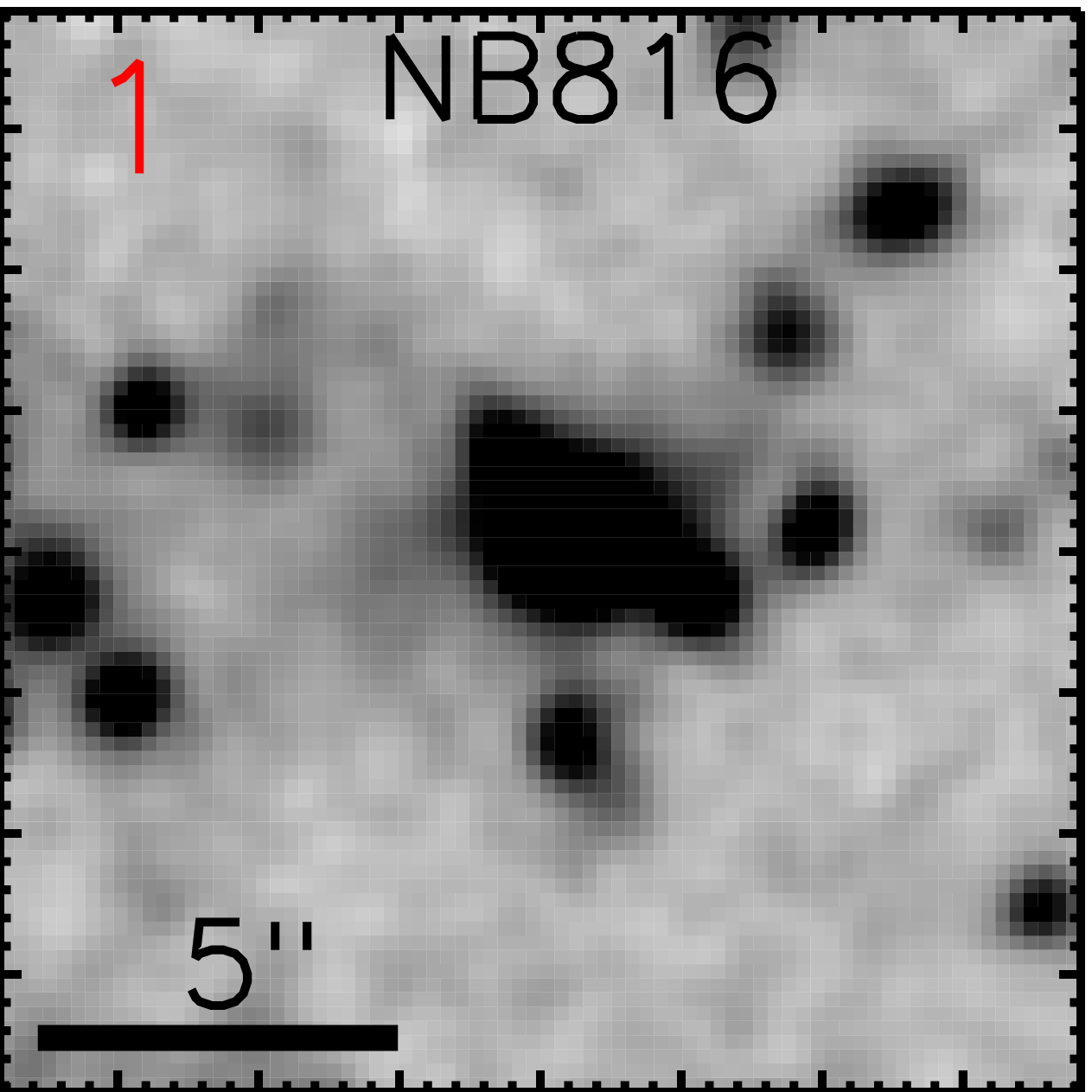} &
\includegraphics[width=0.15\textwidth]{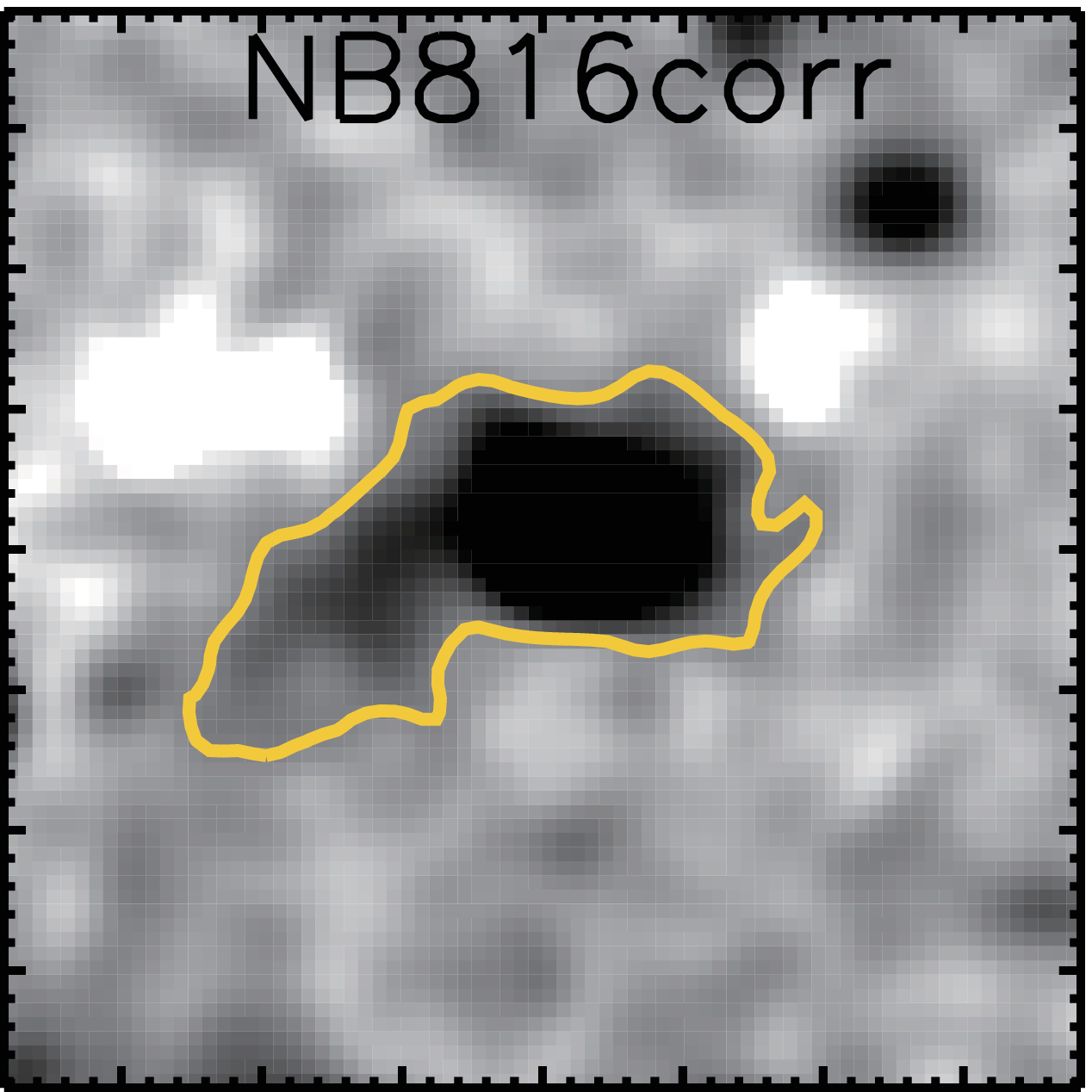} & 
\includegraphics[width=0.15\textwidth]{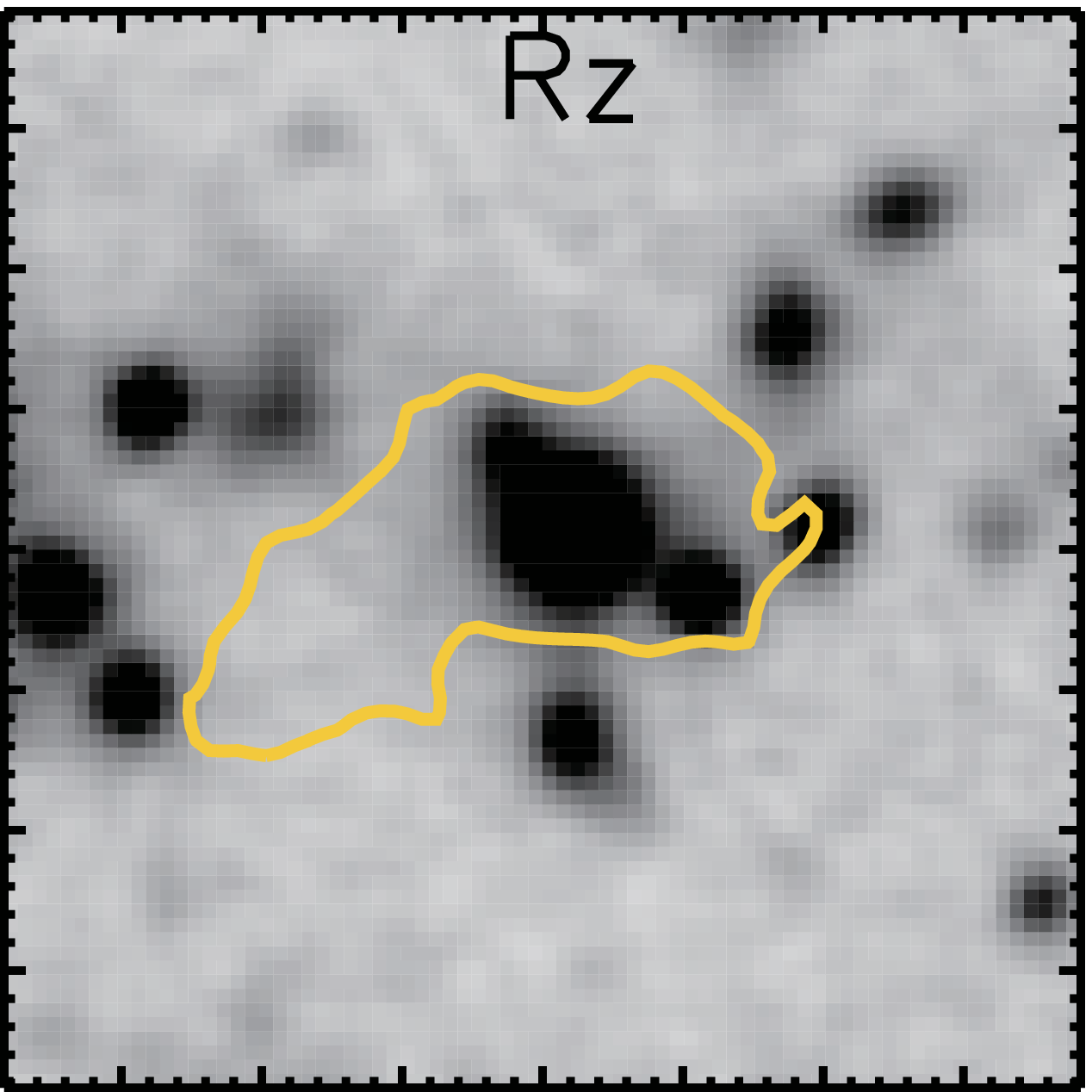} & 
\includegraphics[width=0.15\textwidth]{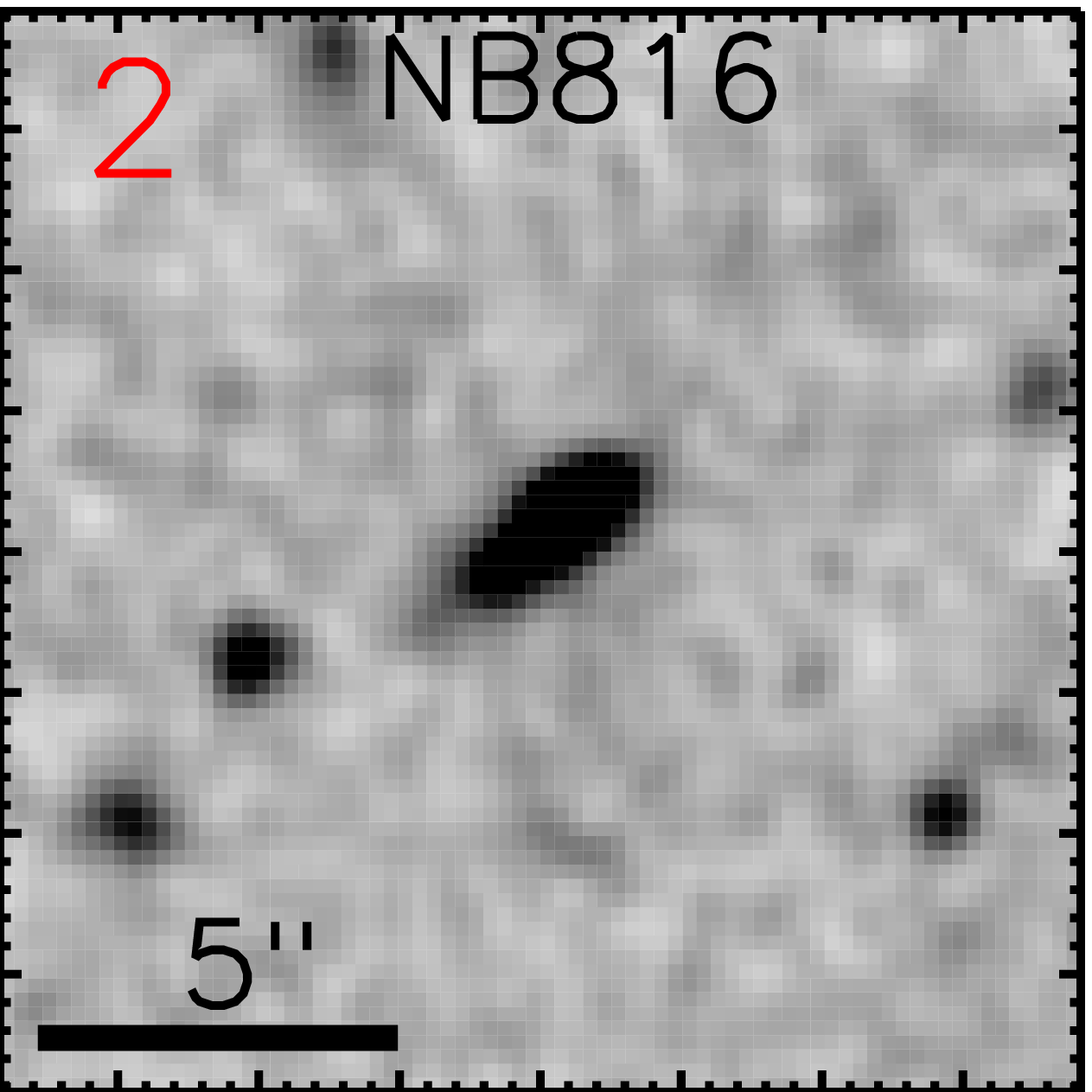} & 
\includegraphics[width=0.15\textwidth]{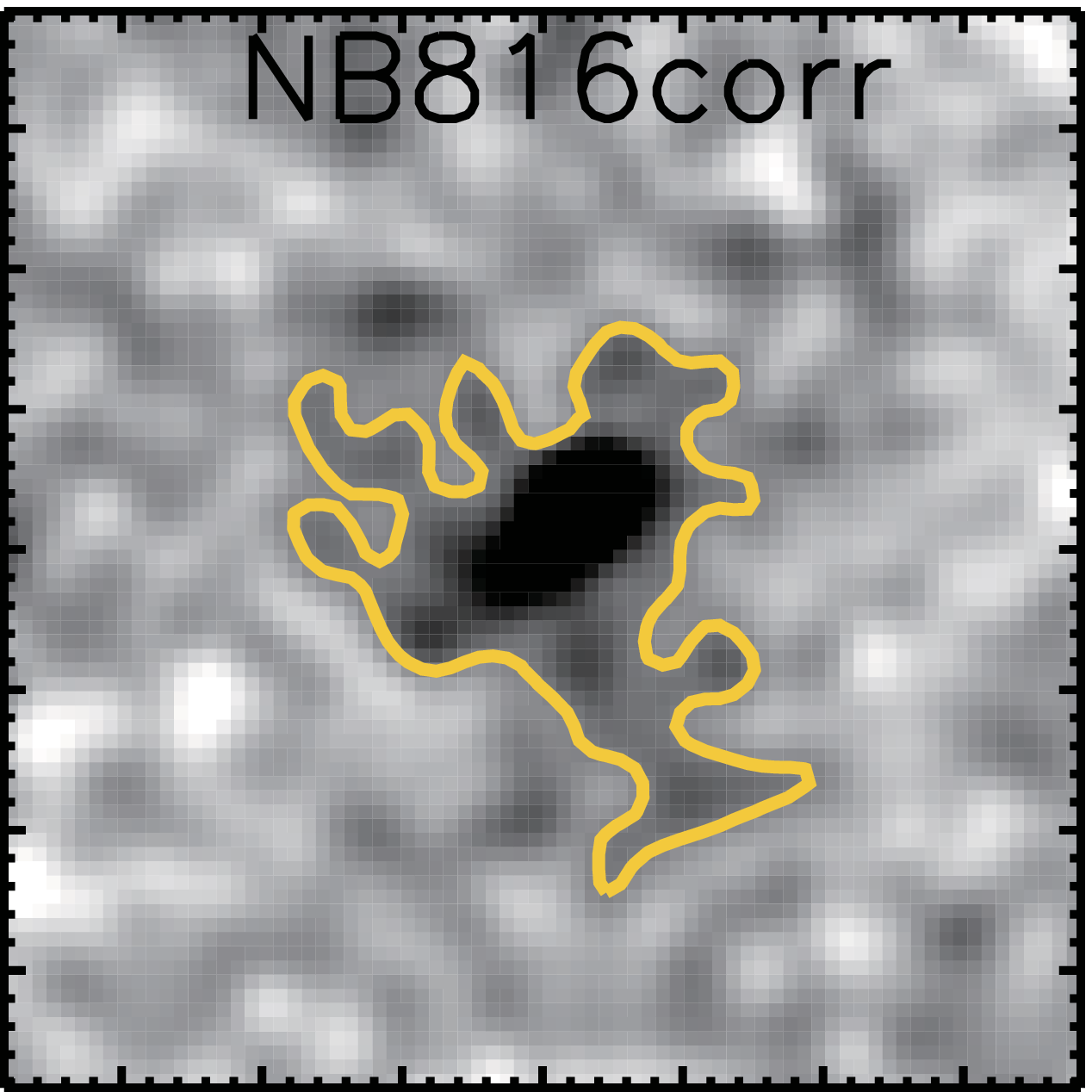} &
\includegraphics[width=0.15\textwidth]{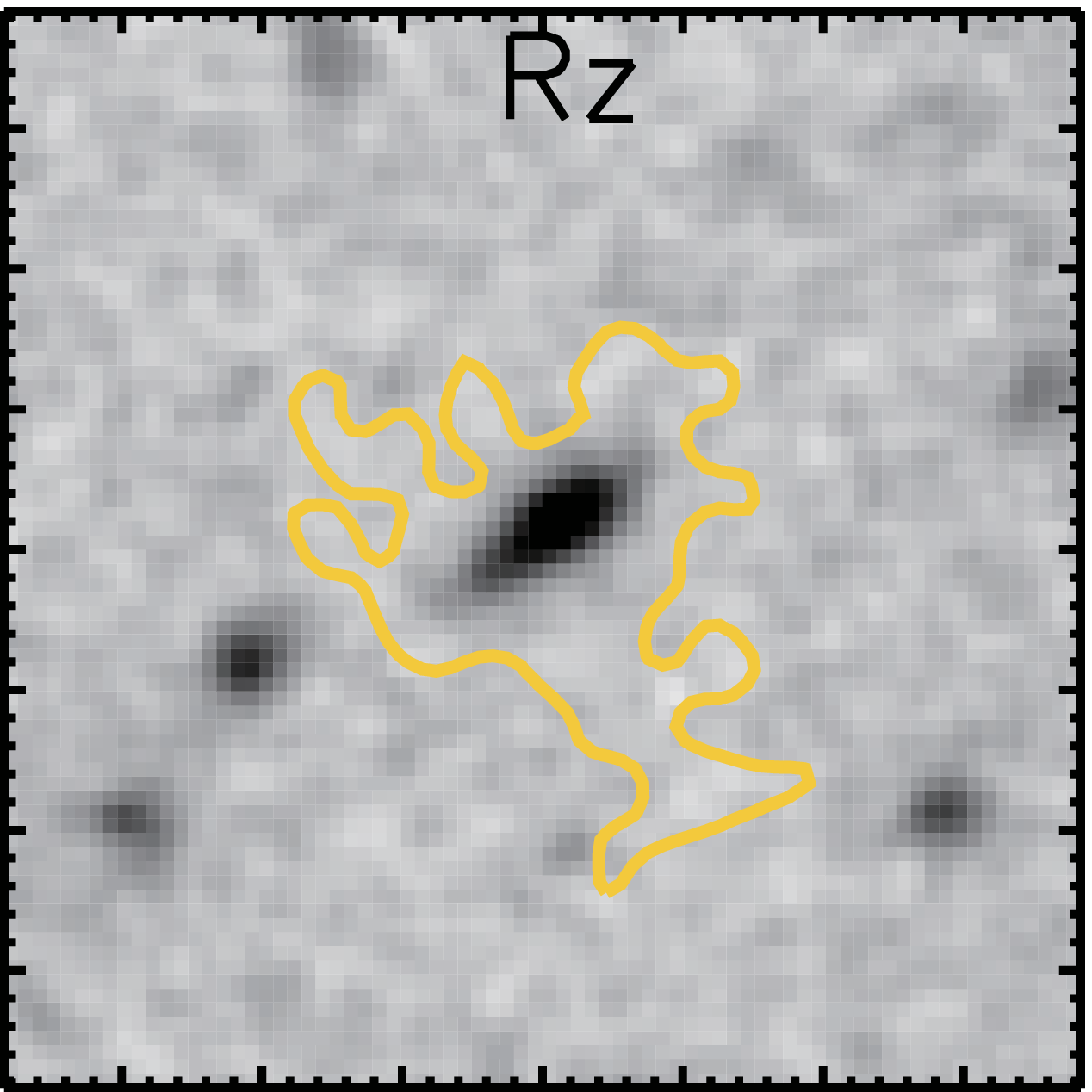} \\

\includegraphics[width=0.15\textwidth]{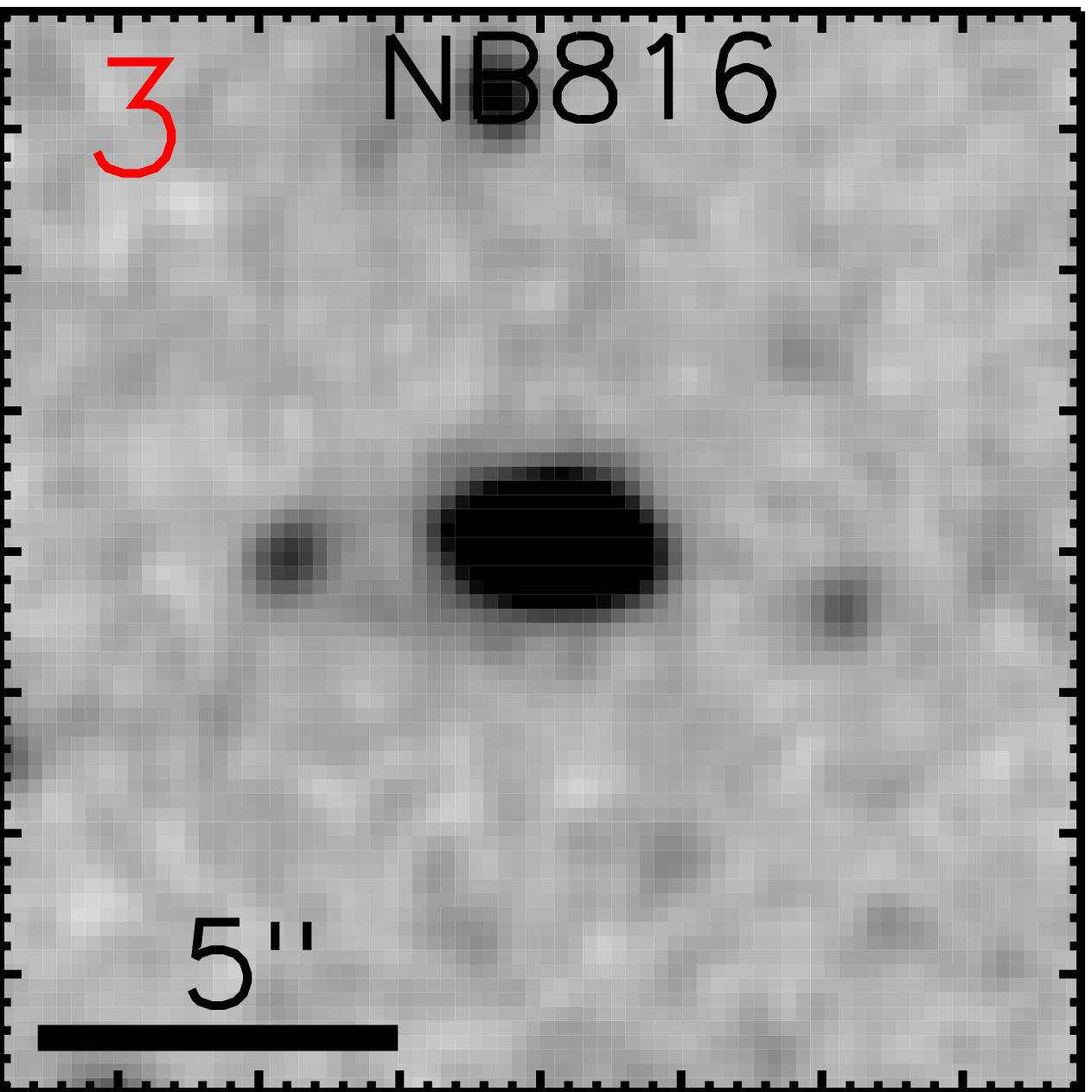} &
\includegraphics[width=0.15\textwidth]{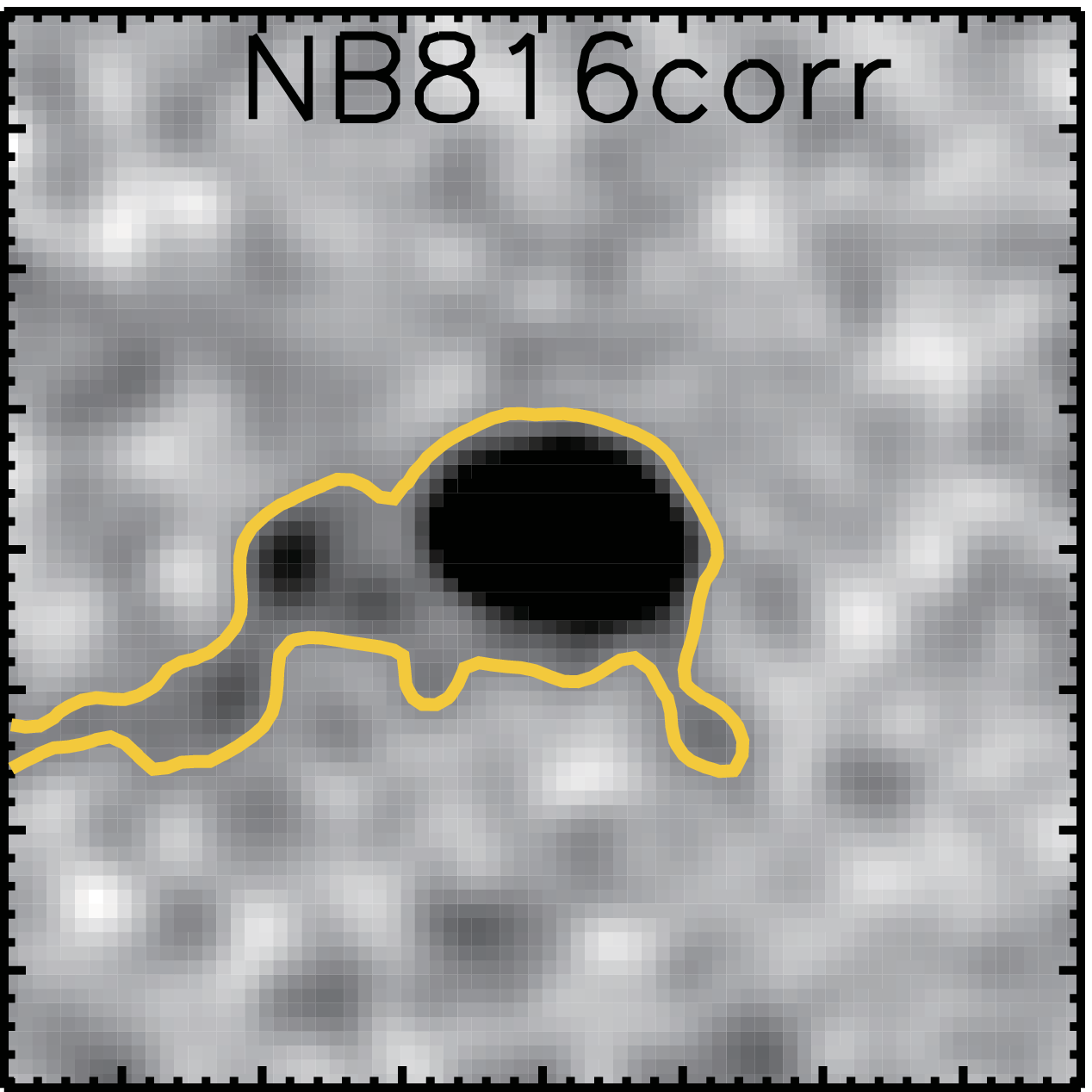} & 
\includegraphics[width=0.15\textwidth]{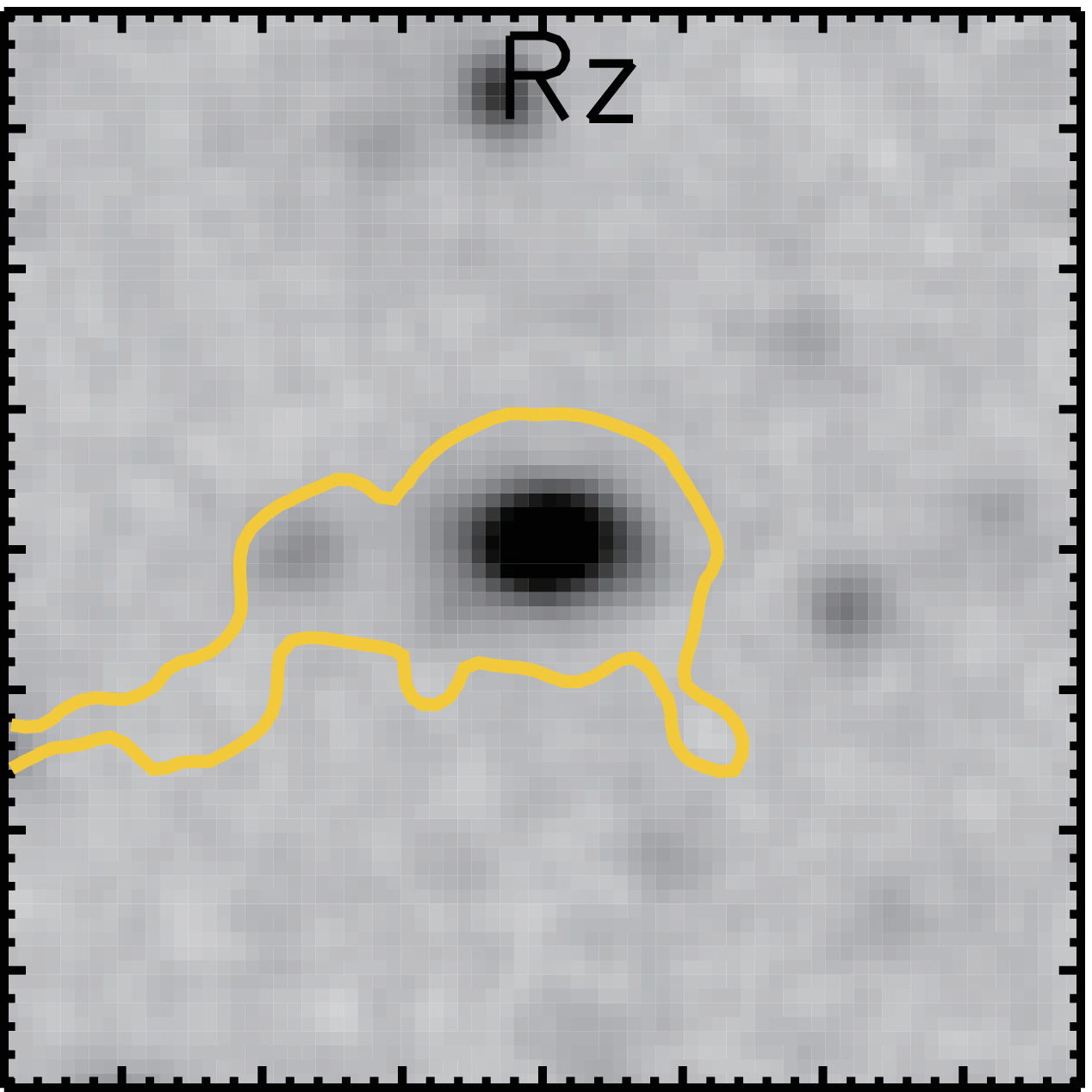} & 
\includegraphics[width=0.15\textwidth]{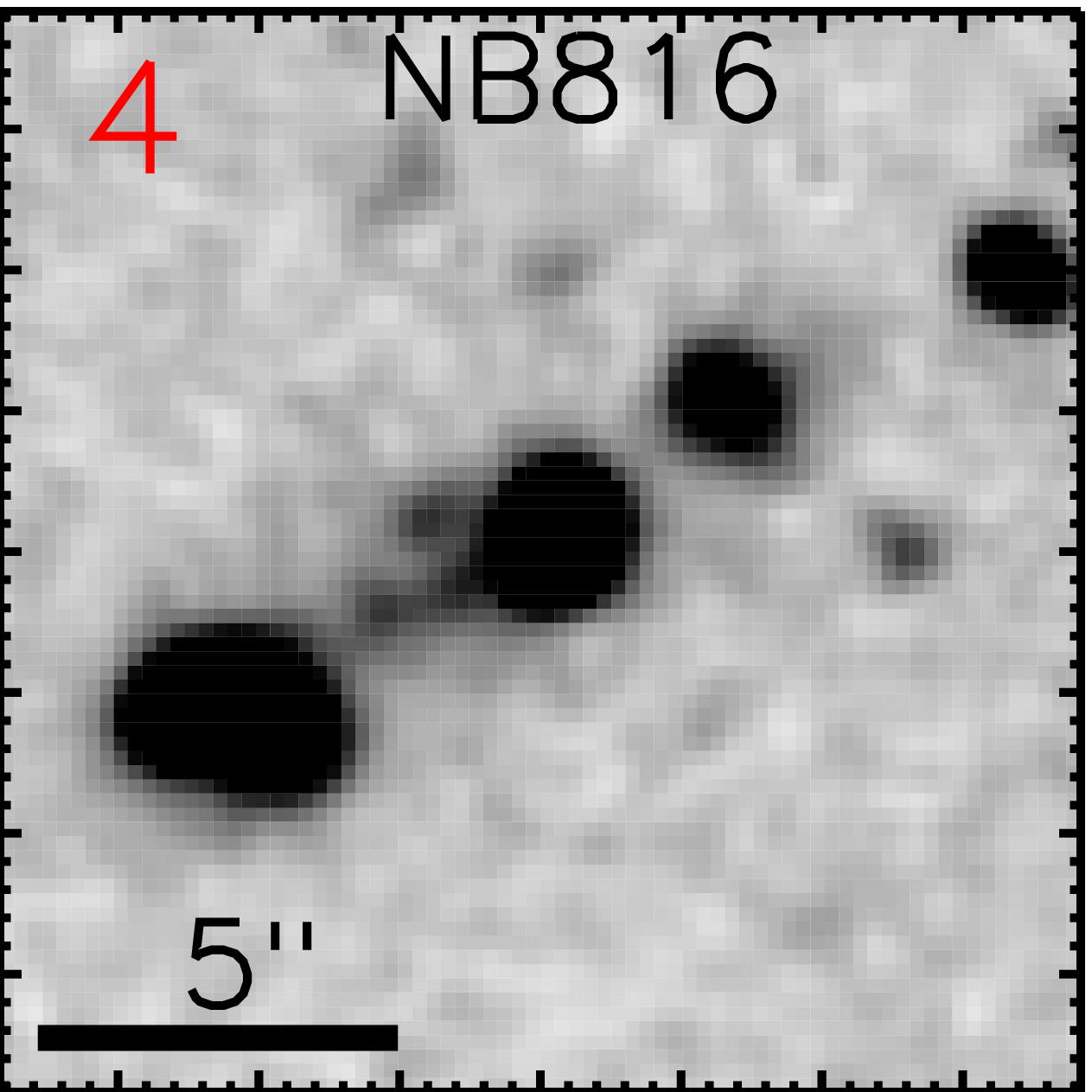} & 
\includegraphics[width=0.15\textwidth]{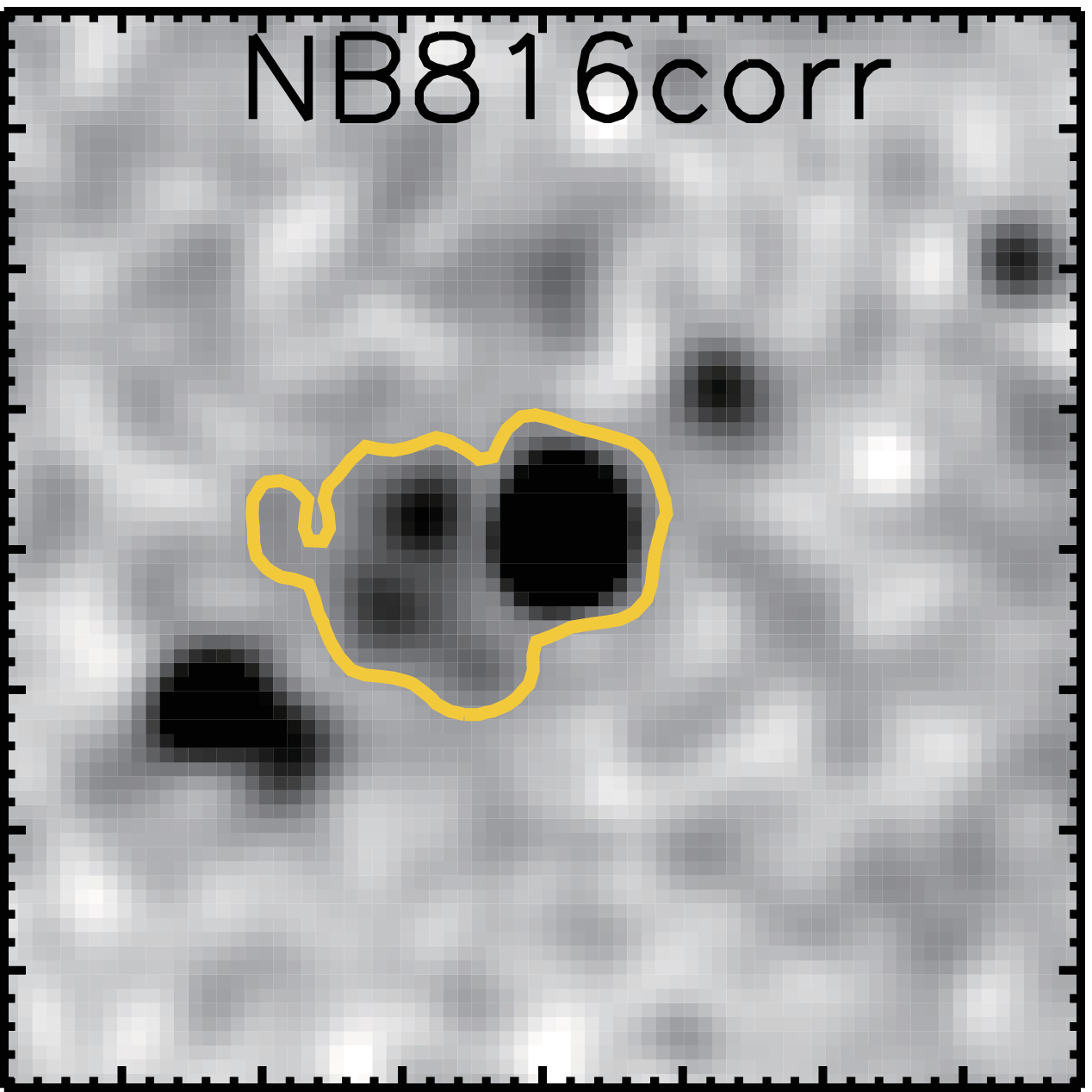} &
\includegraphics[width=0.15\textwidth]{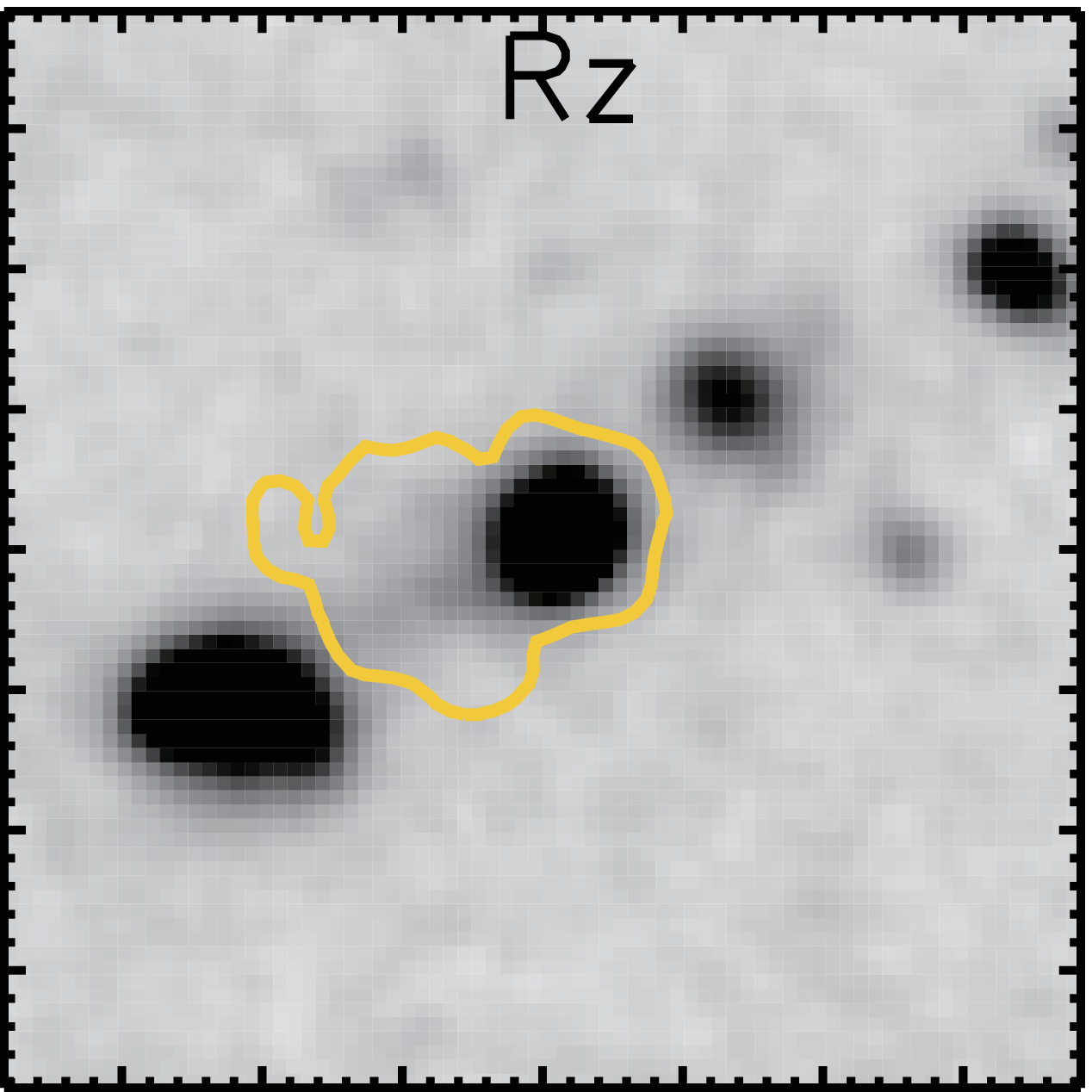} \\

\includegraphics[width=0.15\textwidth]{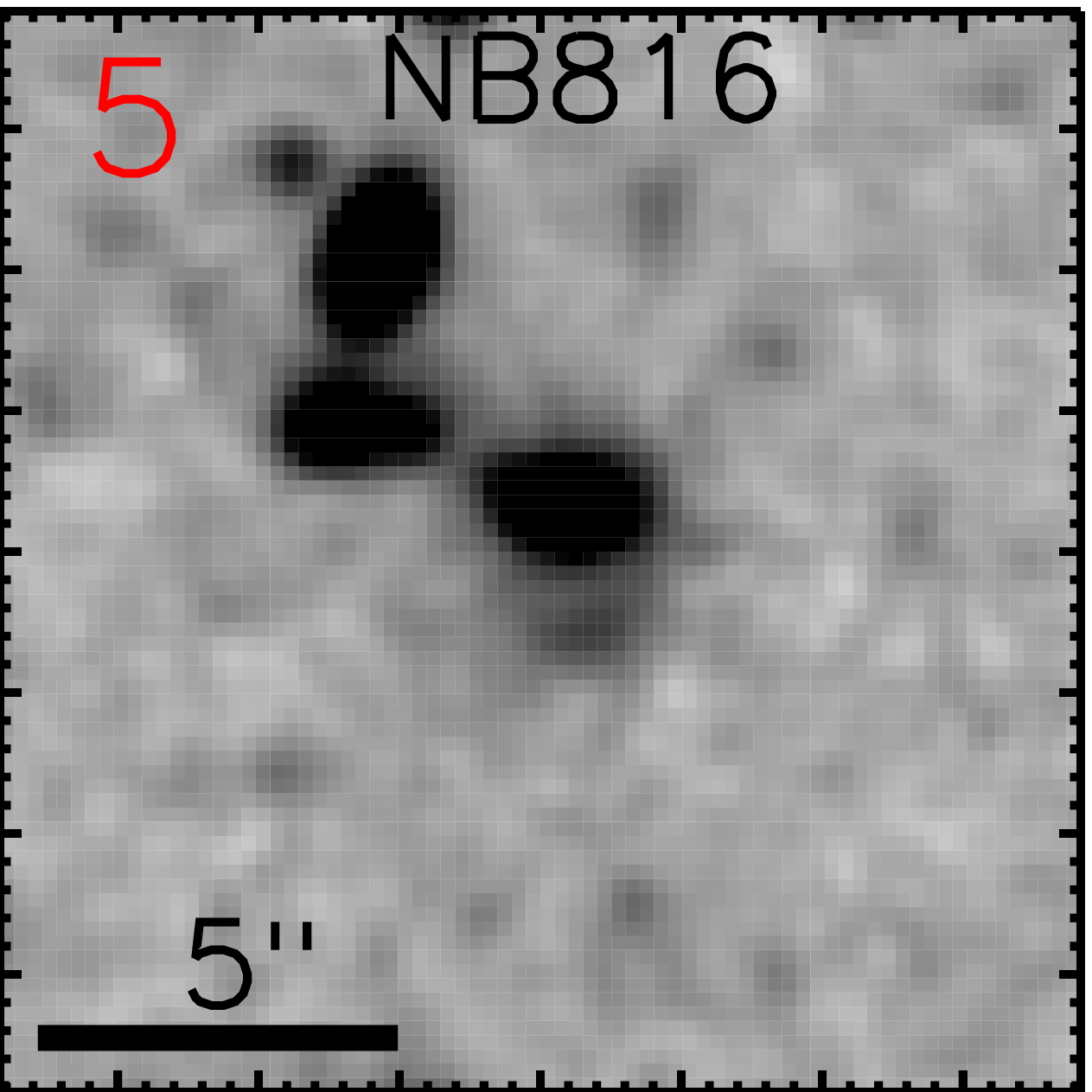} &
\includegraphics[width=0.15\textwidth]{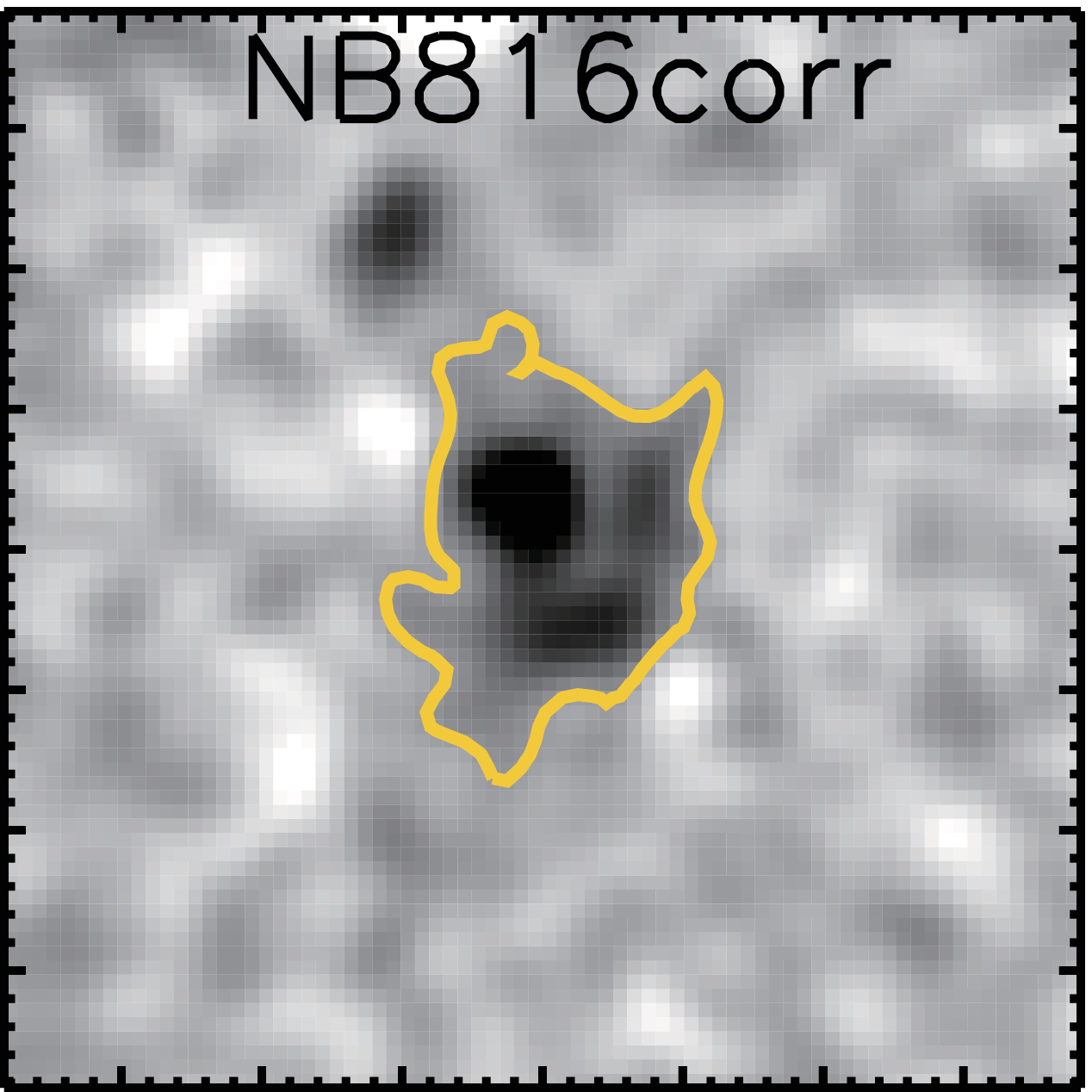} & 
\includegraphics[width=0.15\textwidth]{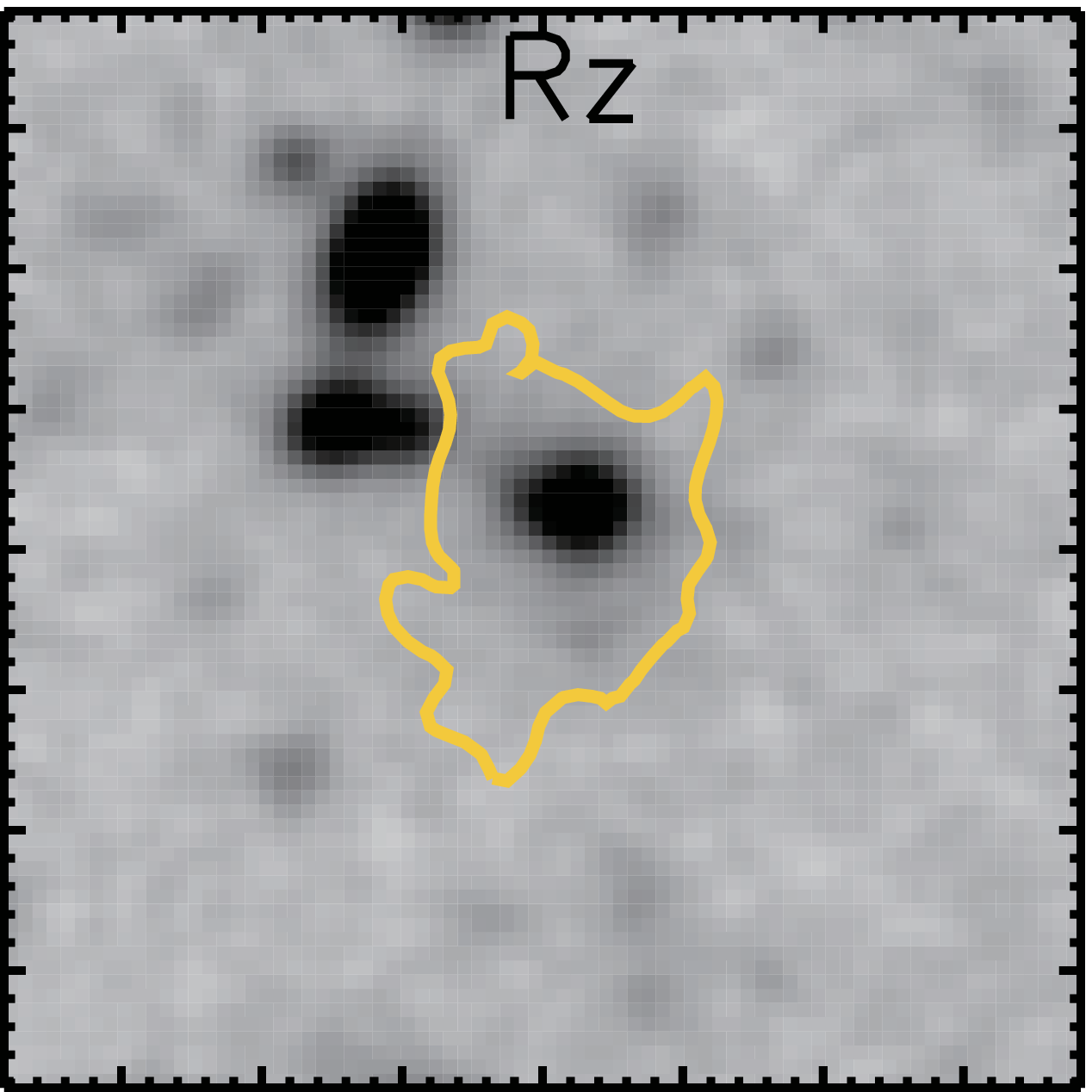} & 
\includegraphics[width=0.15\textwidth]{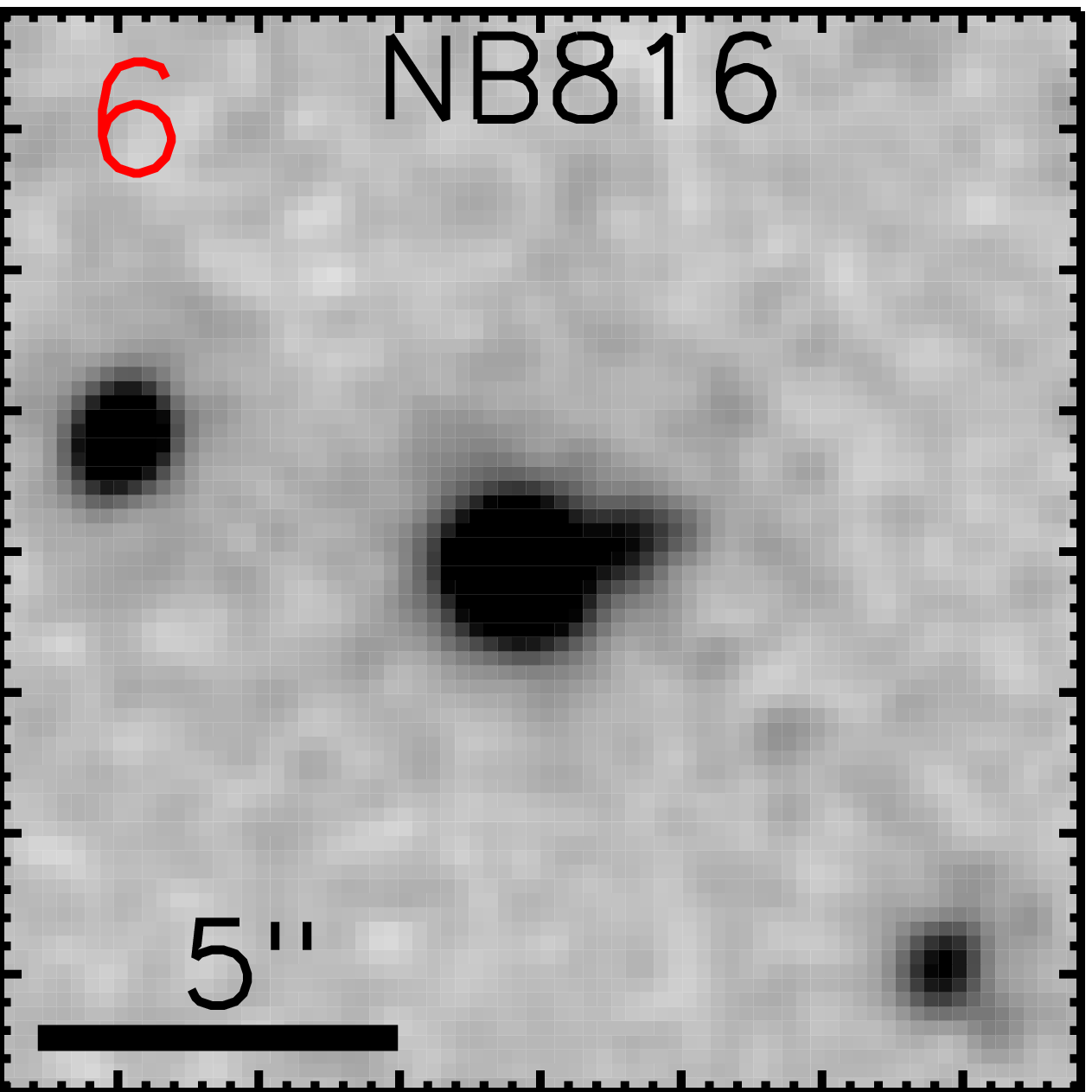} & 
\includegraphics[width=0.15\textwidth]{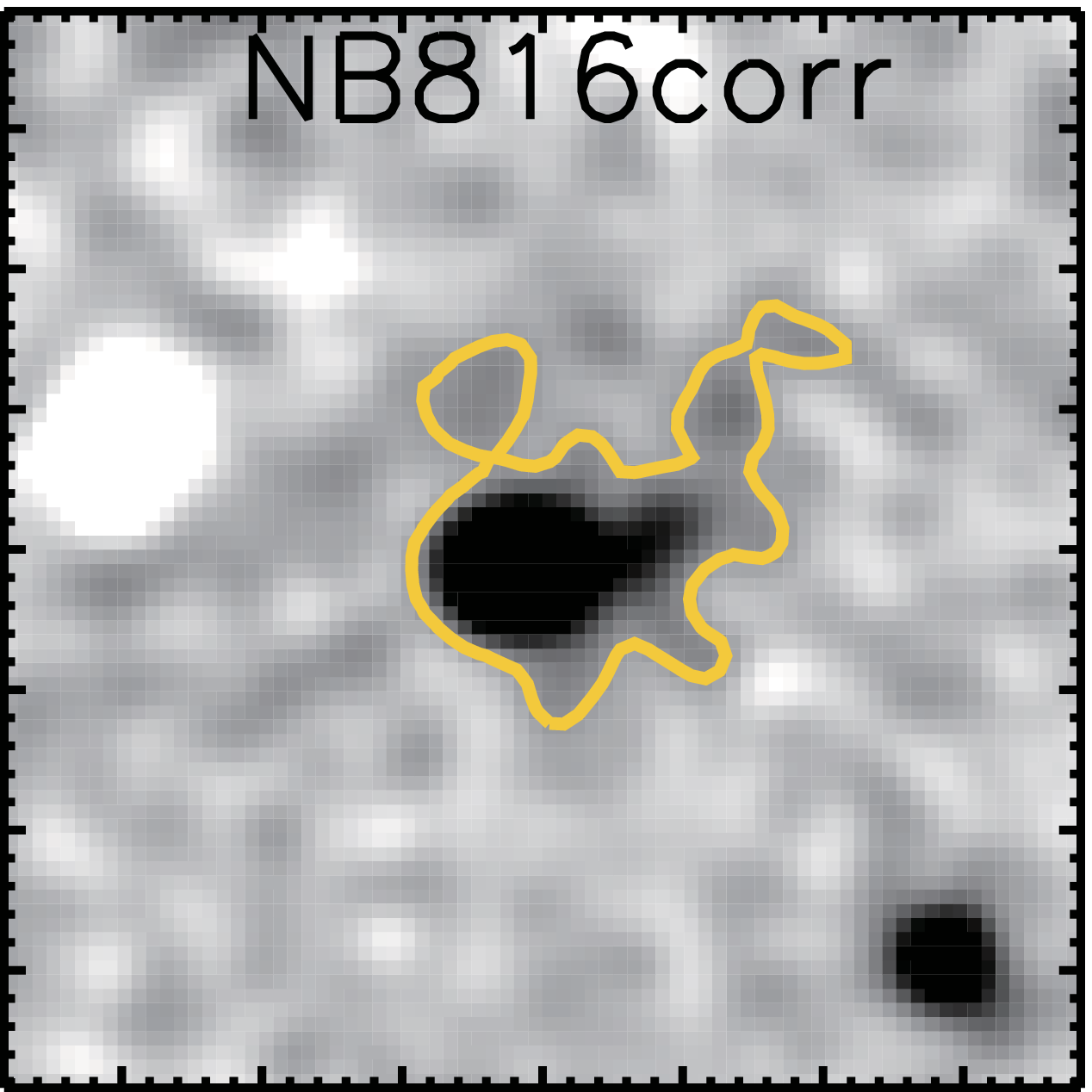} &
\includegraphics[width=0.15\textwidth]{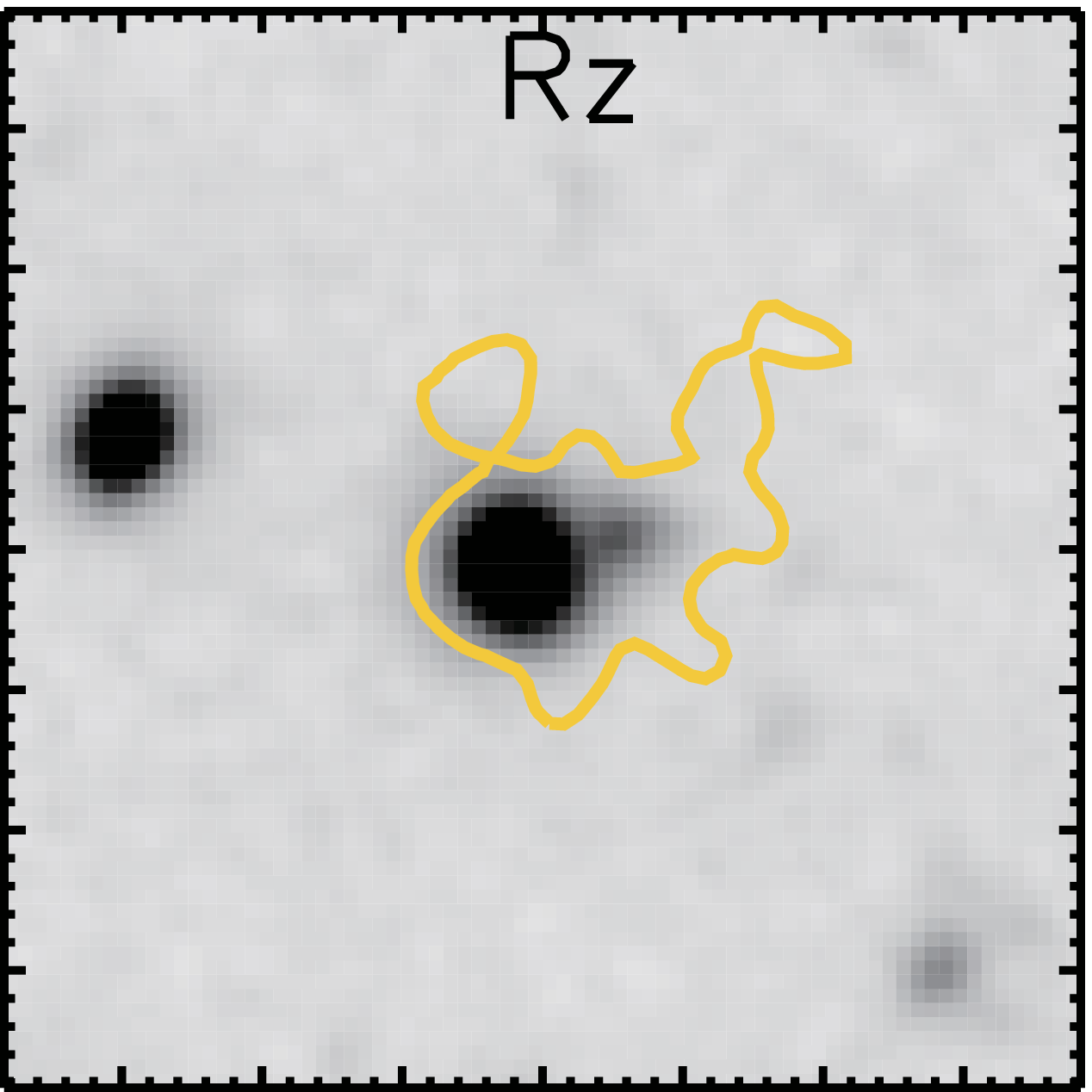} \\

\includegraphics[width=0.15\textwidth]{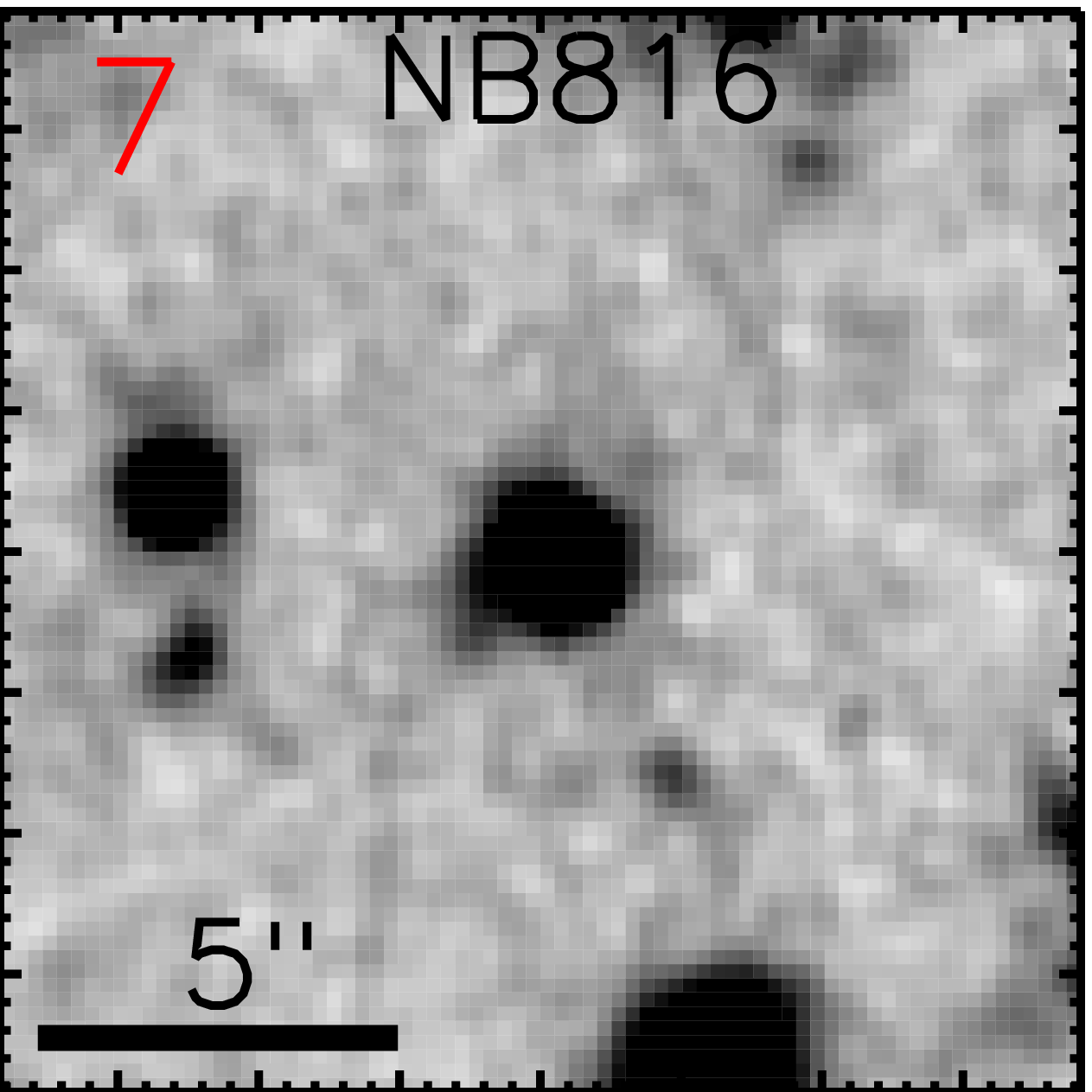} &
\includegraphics[width=0.15\textwidth]{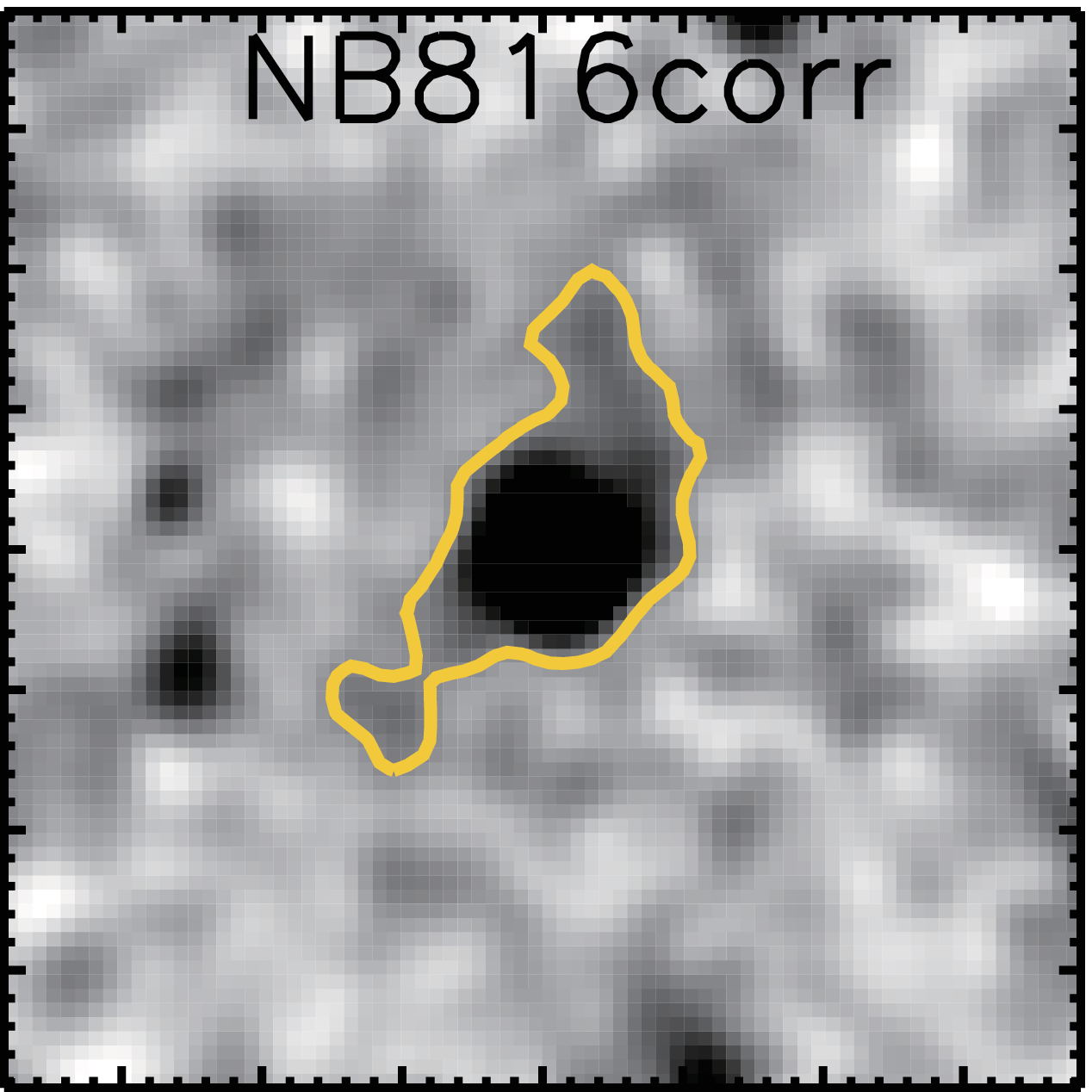} & 
\includegraphics[width=0.15\textwidth]{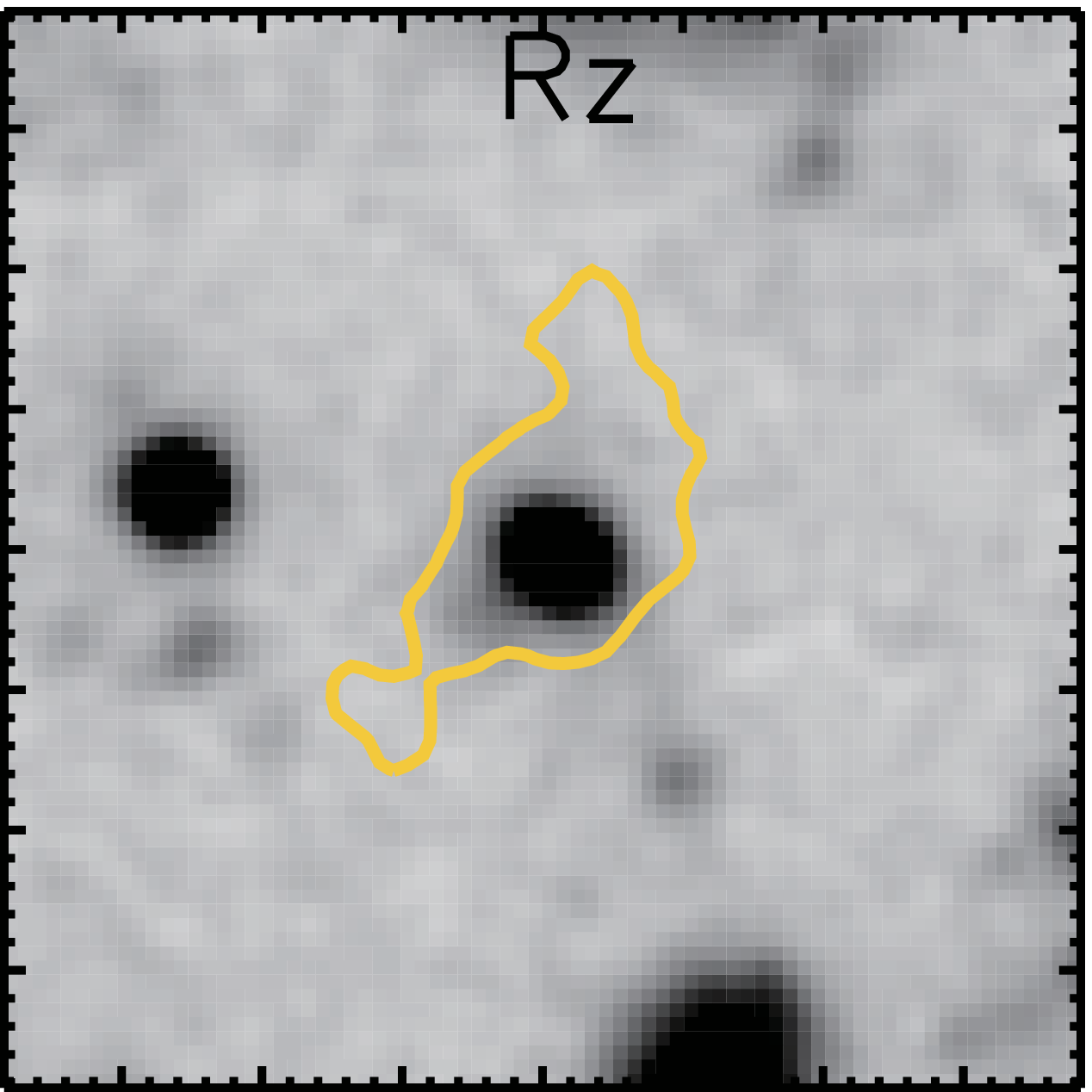} & 
\includegraphics[width=0.15\textwidth]{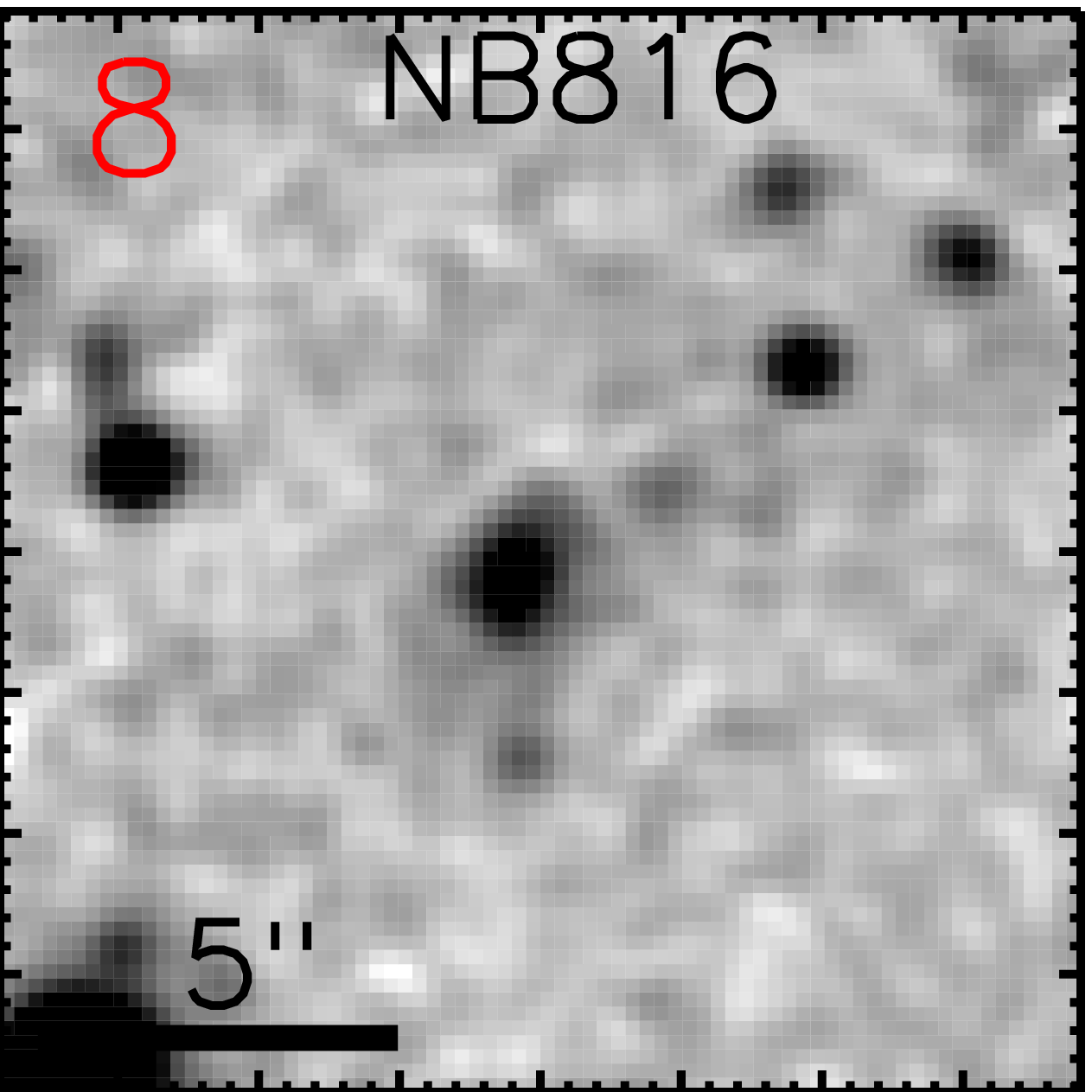} & 
\includegraphics[width=0.15\textwidth]{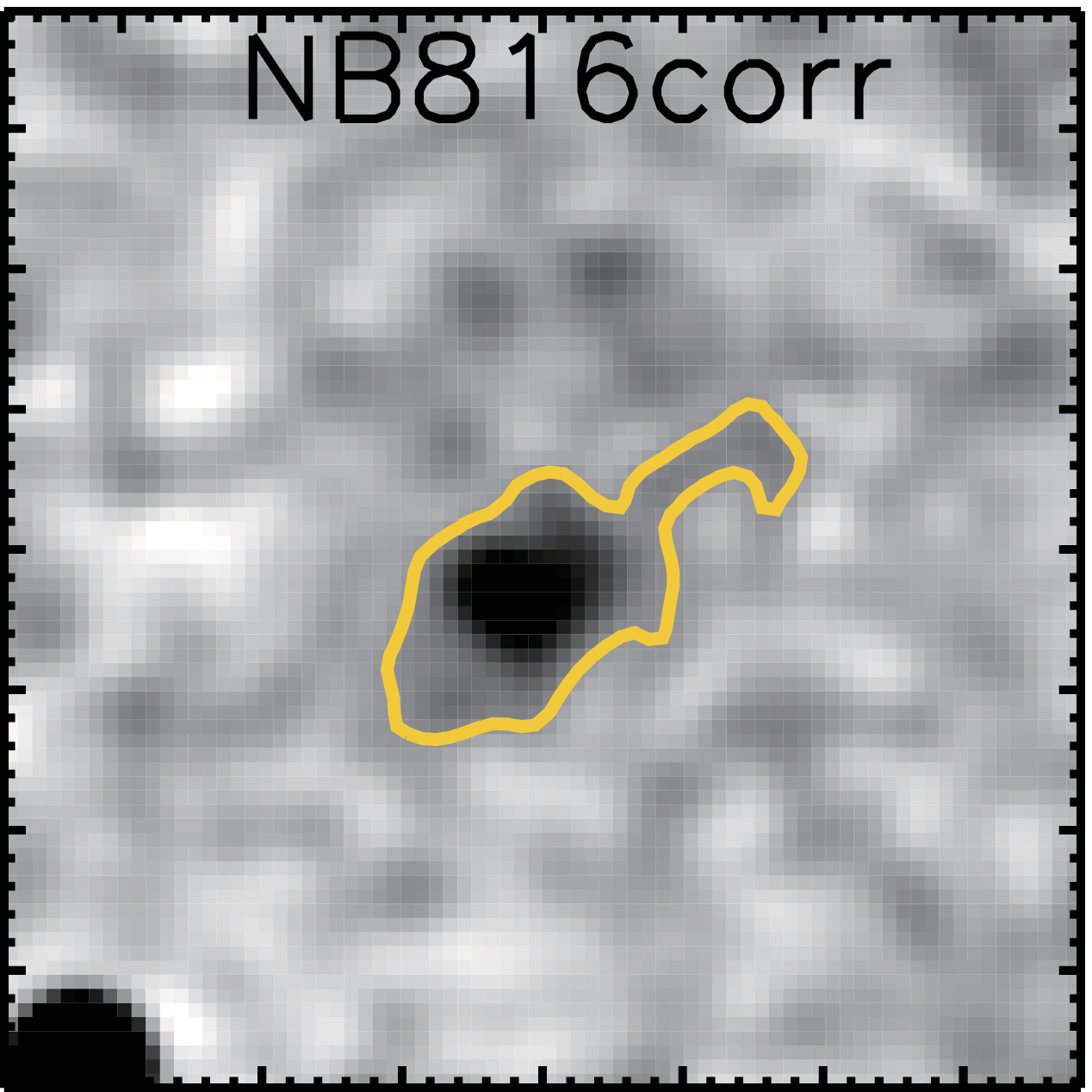} &
\includegraphics[width=0.15\textwidth]{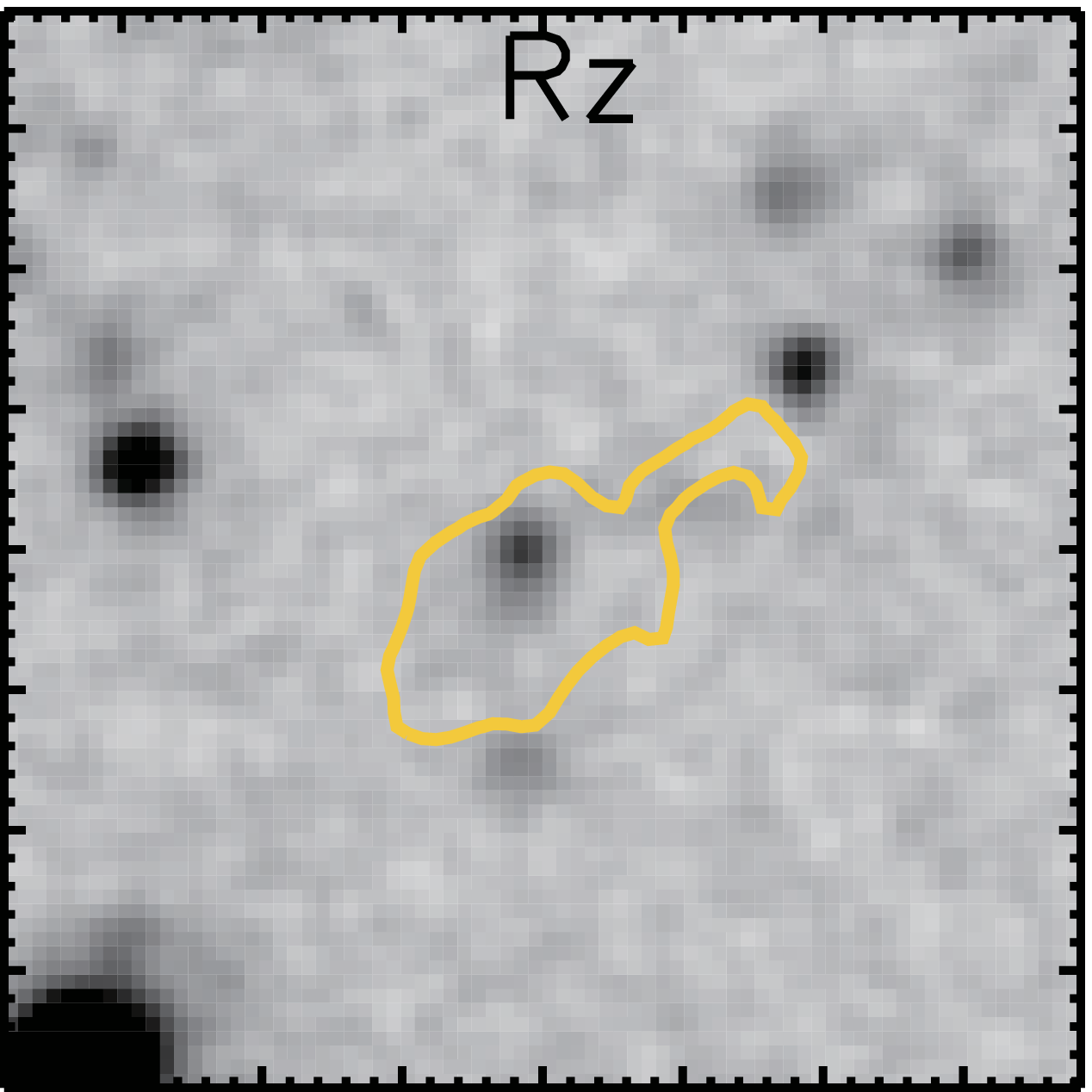} \\

\includegraphics[width=0.15\textwidth]{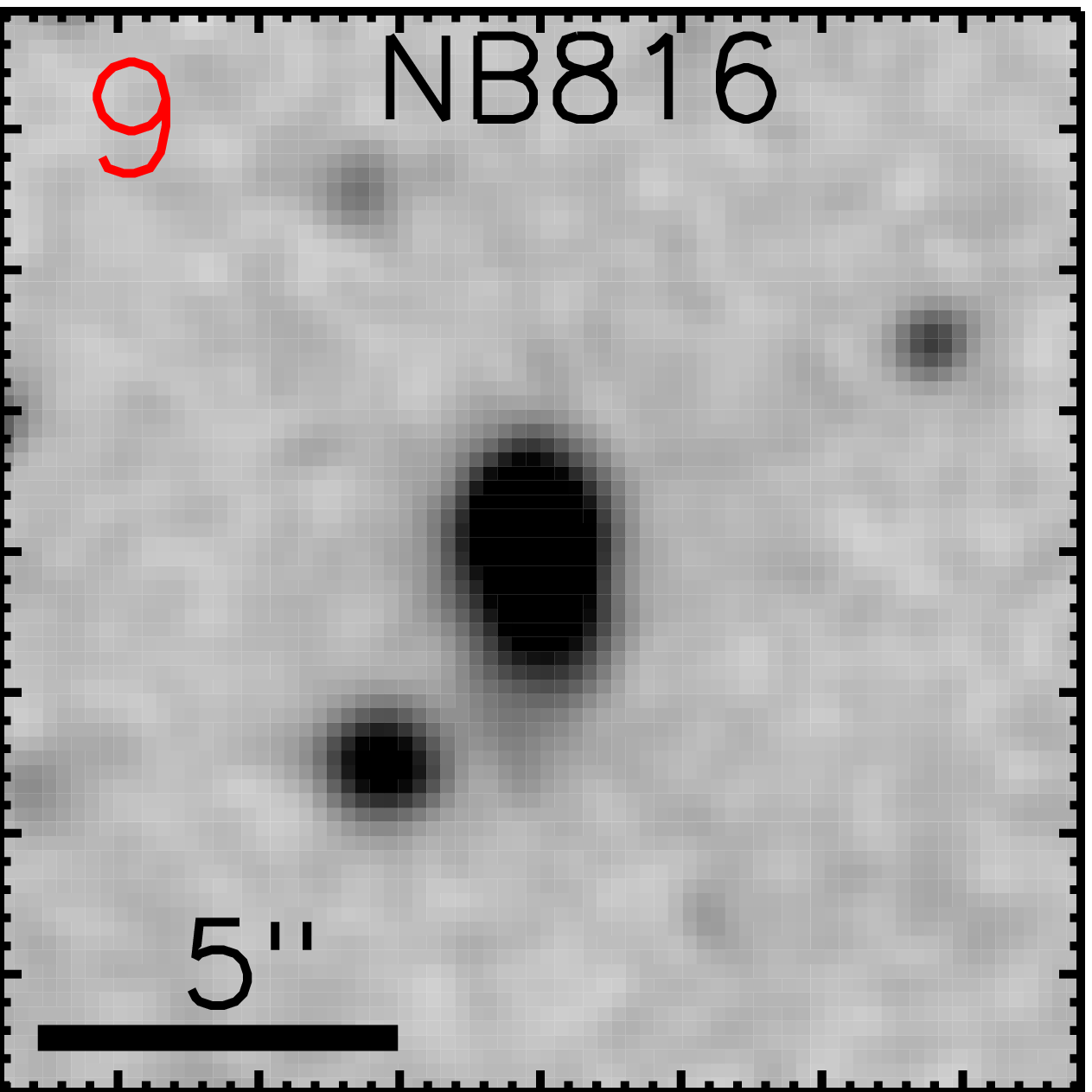} &
\includegraphics[width=0.15\textwidth]{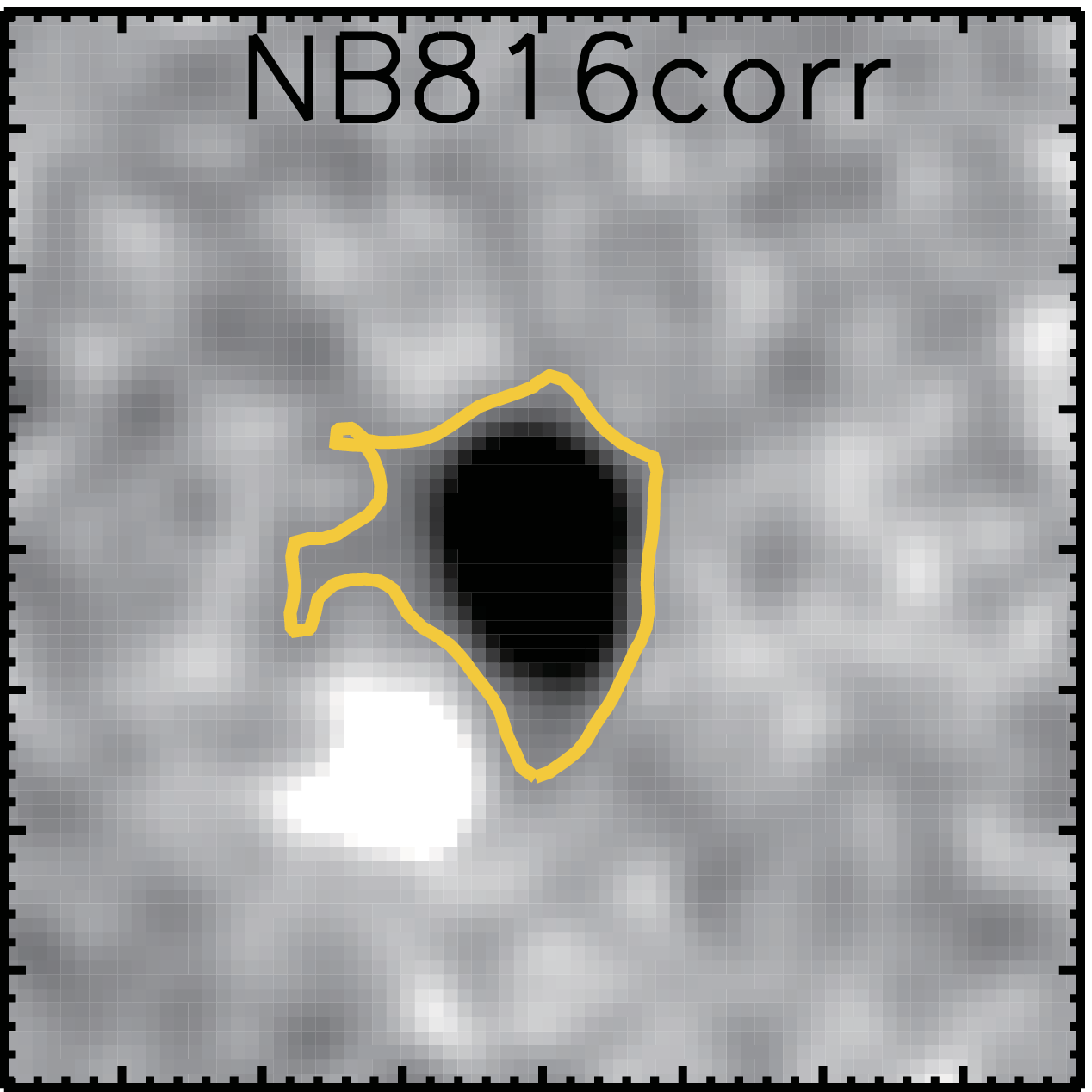} & 
\includegraphics[width=0.15\textwidth]{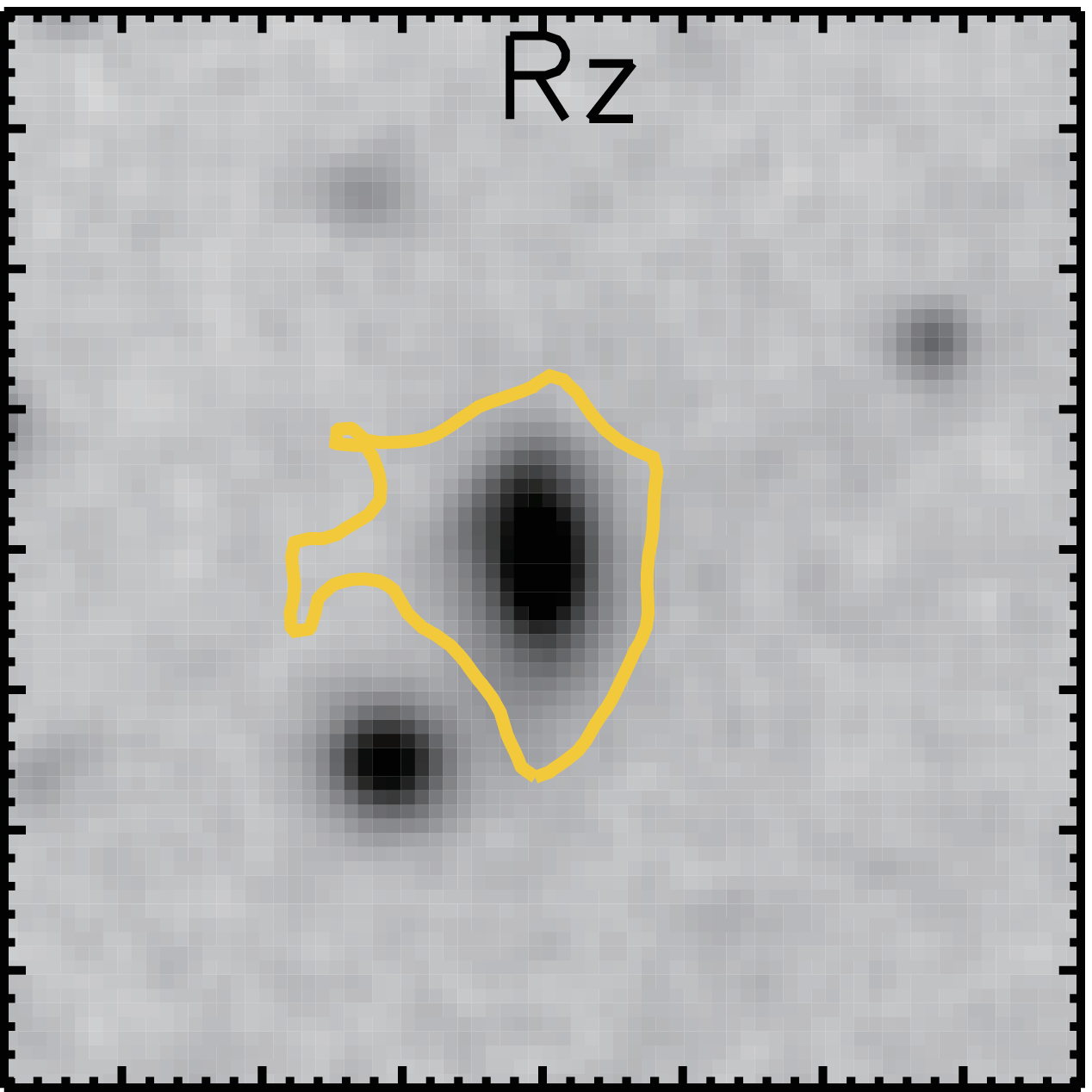} & 
\includegraphics[width=0.15\textwidth]{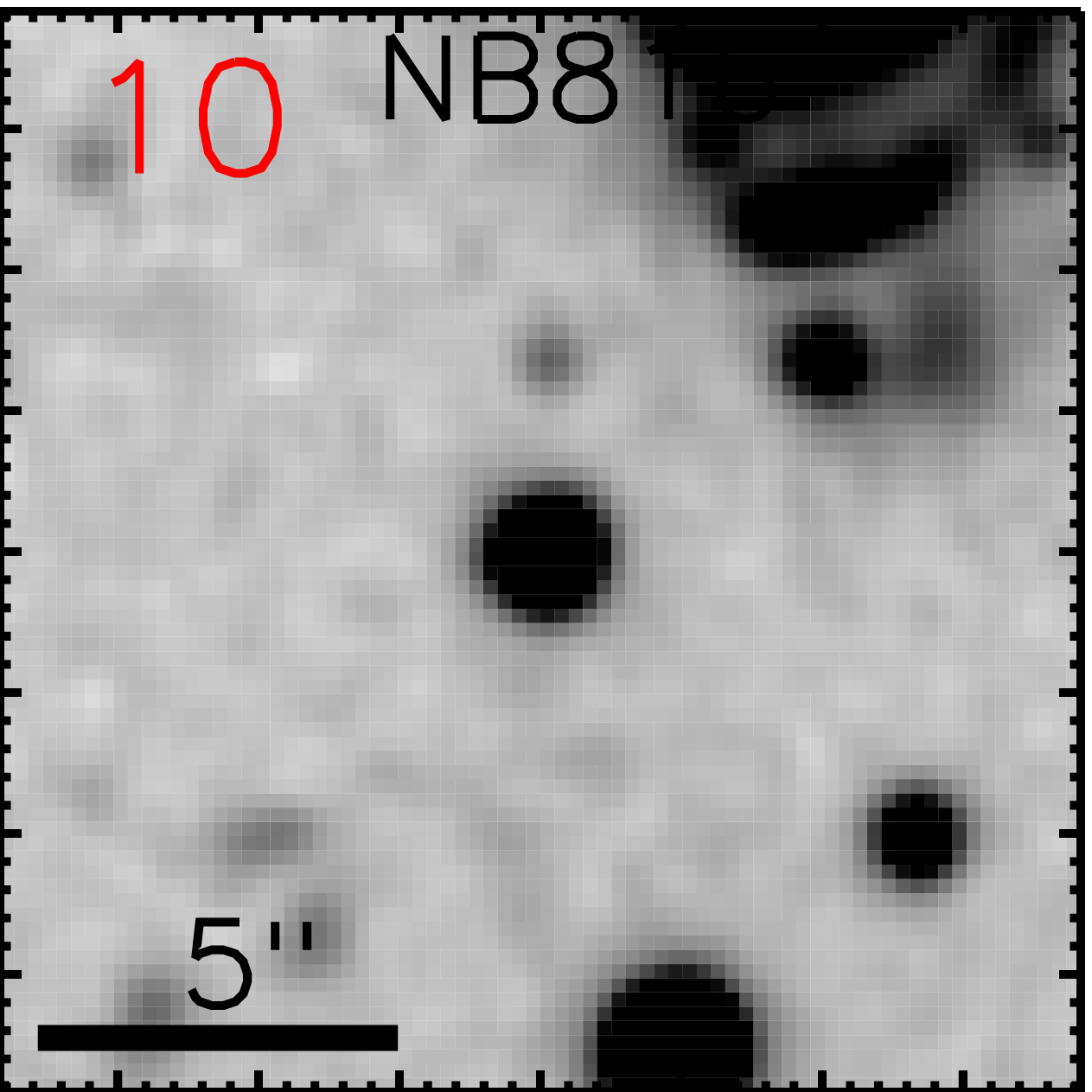} & 
\includegraphics[width=0.15\textwidth]{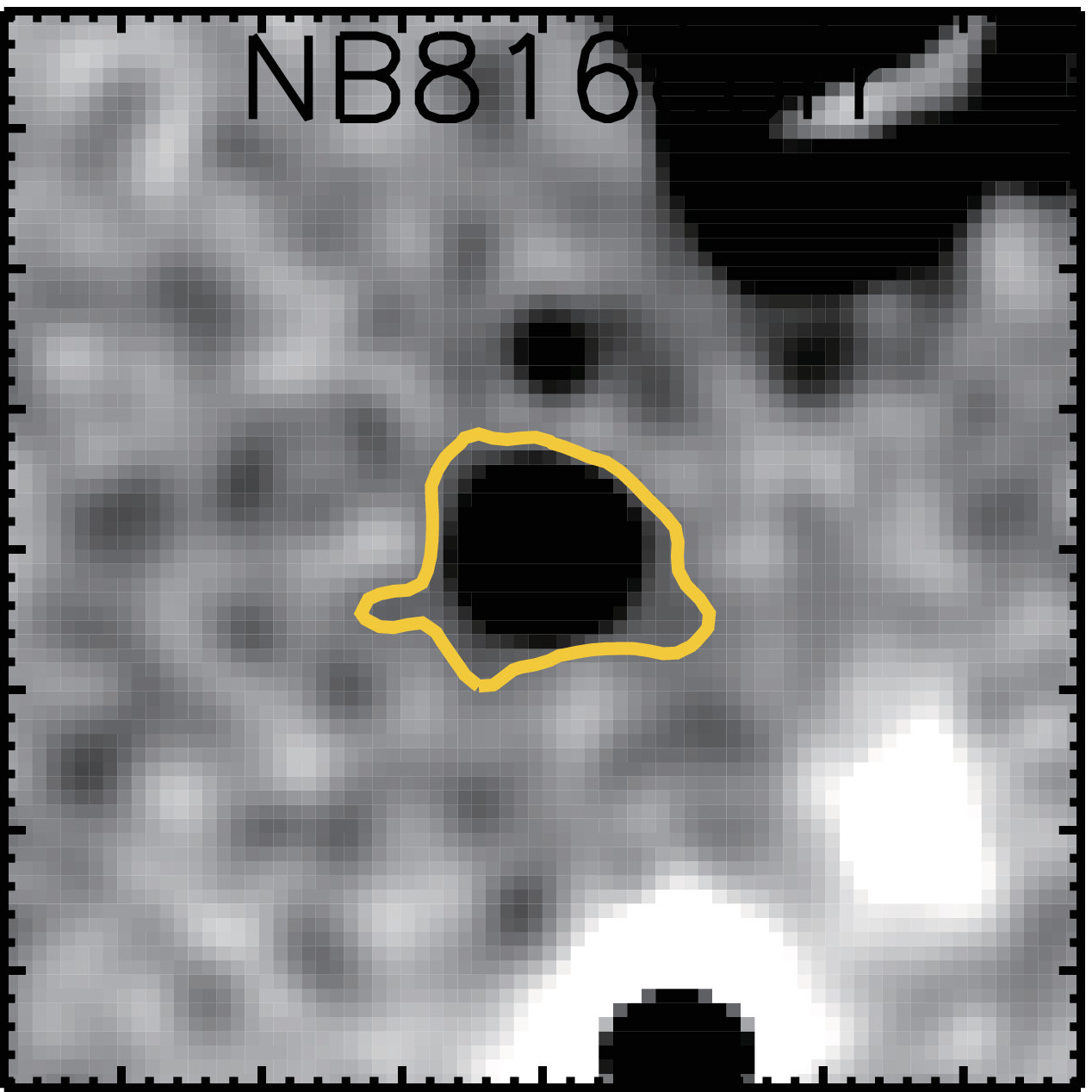} &
\includegraphics[width=0.15\textwidth]{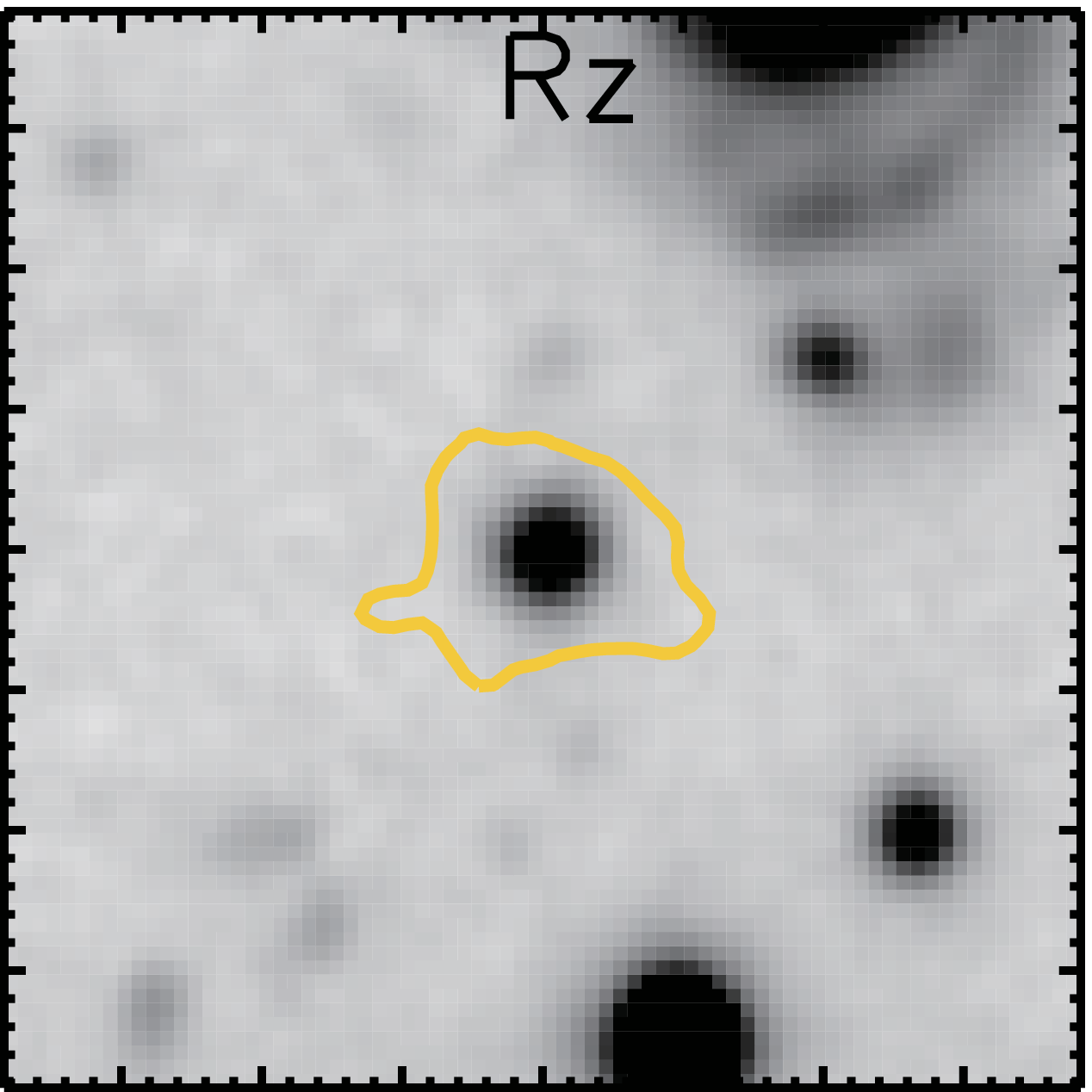} \\

\includegraphics[width=0.15\textwidth]{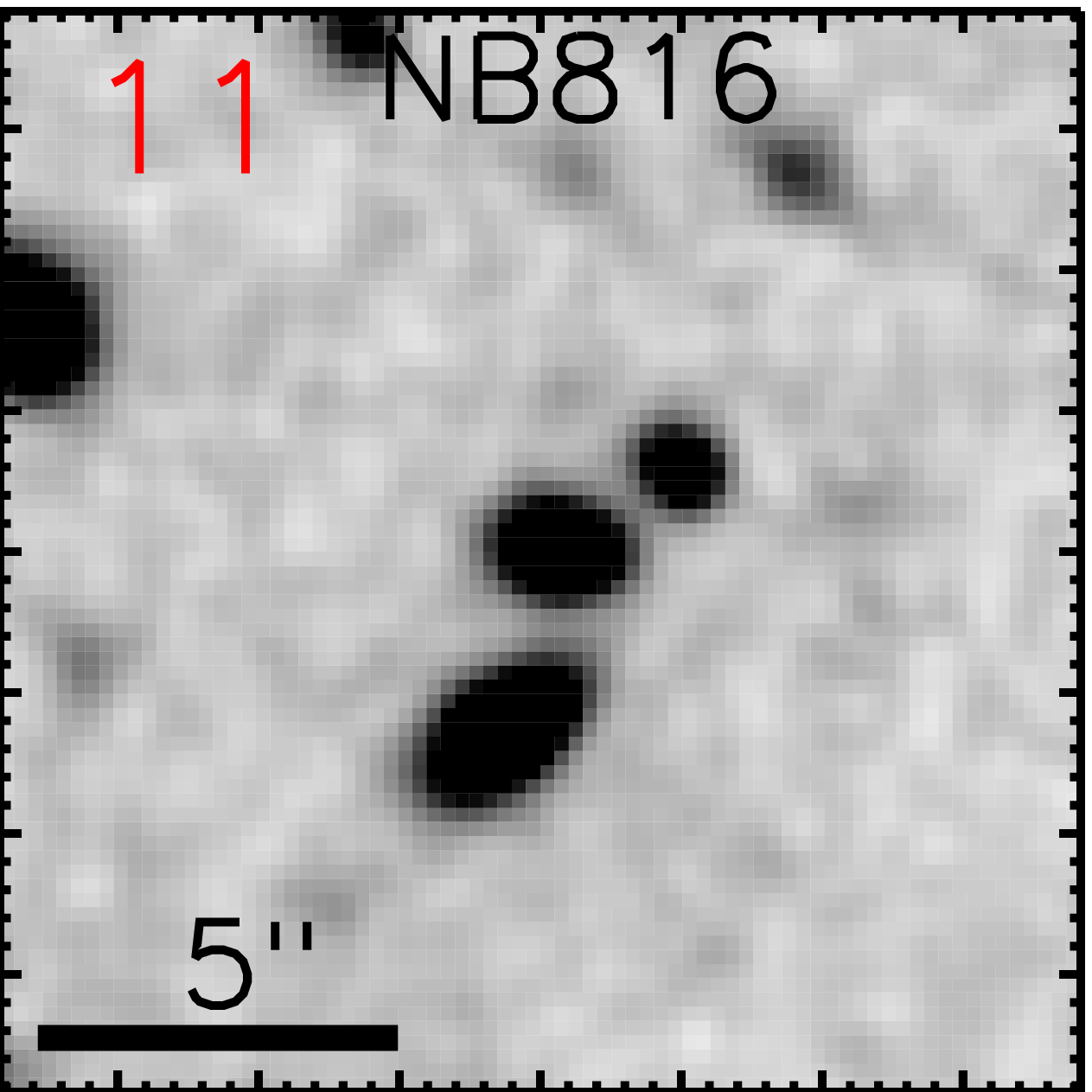} &
\includegraphics[width=0.15\textwidth]{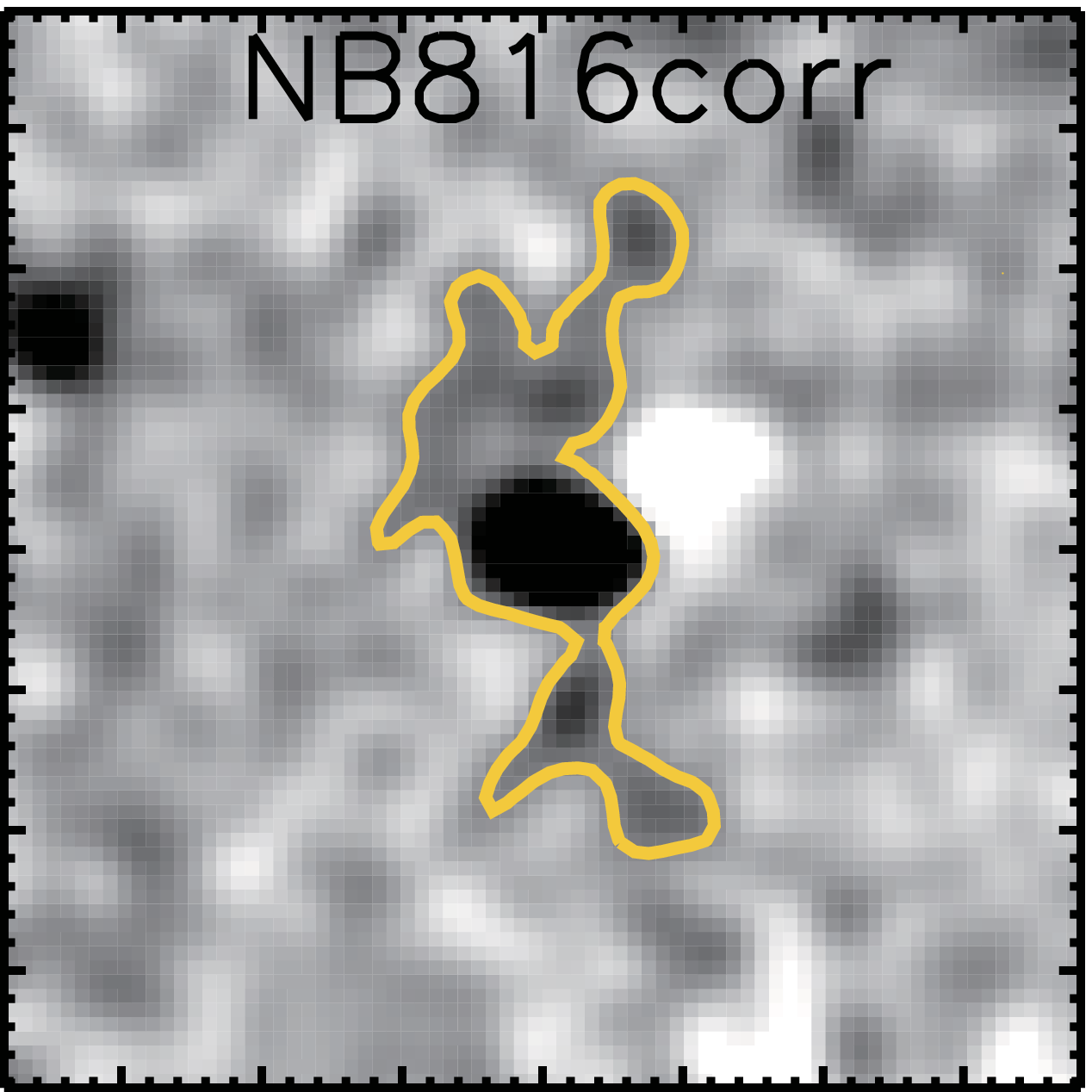} & 
\includegraphics[width=0.15\textwidth]{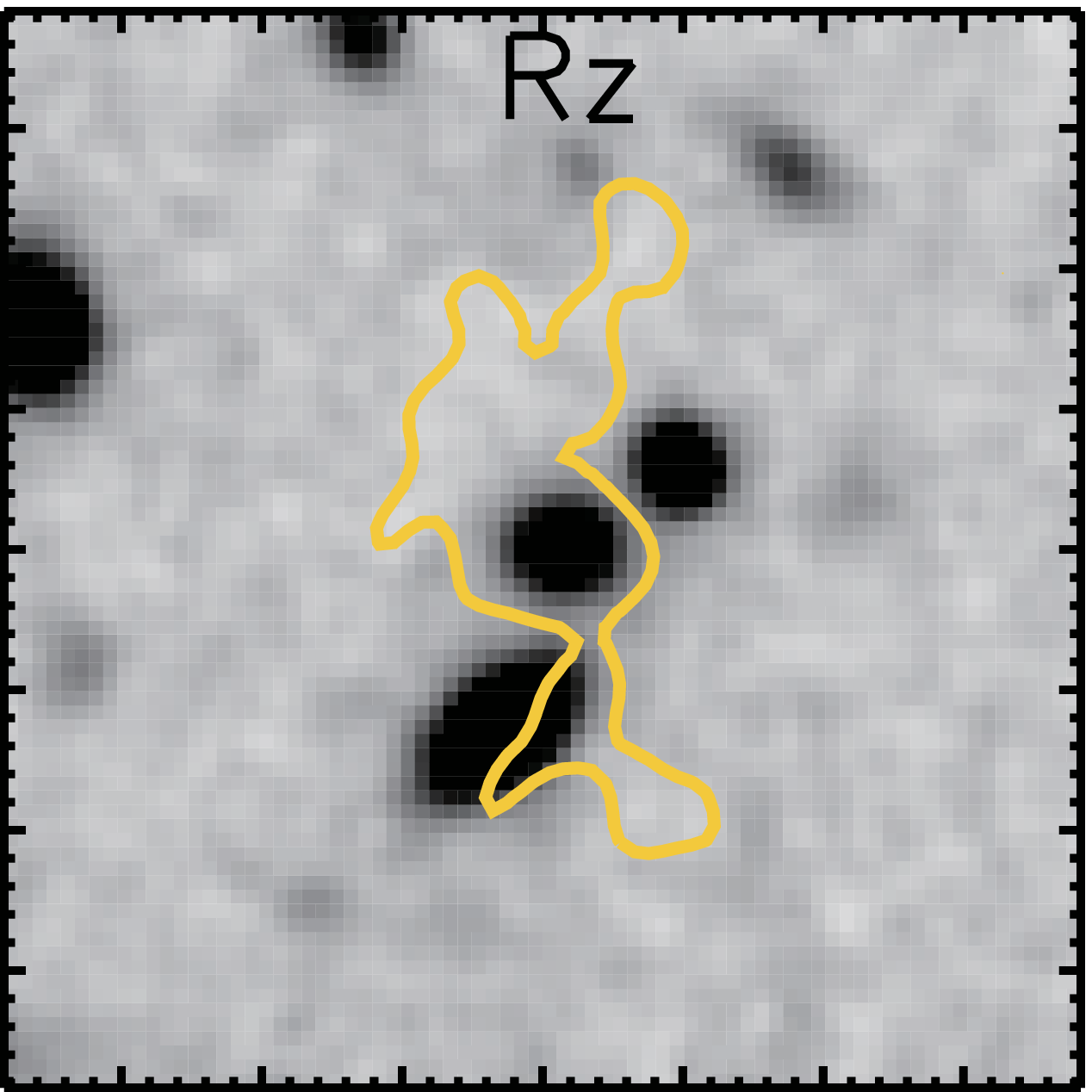} & 
\includegraphics[width=0.15\textwidth]{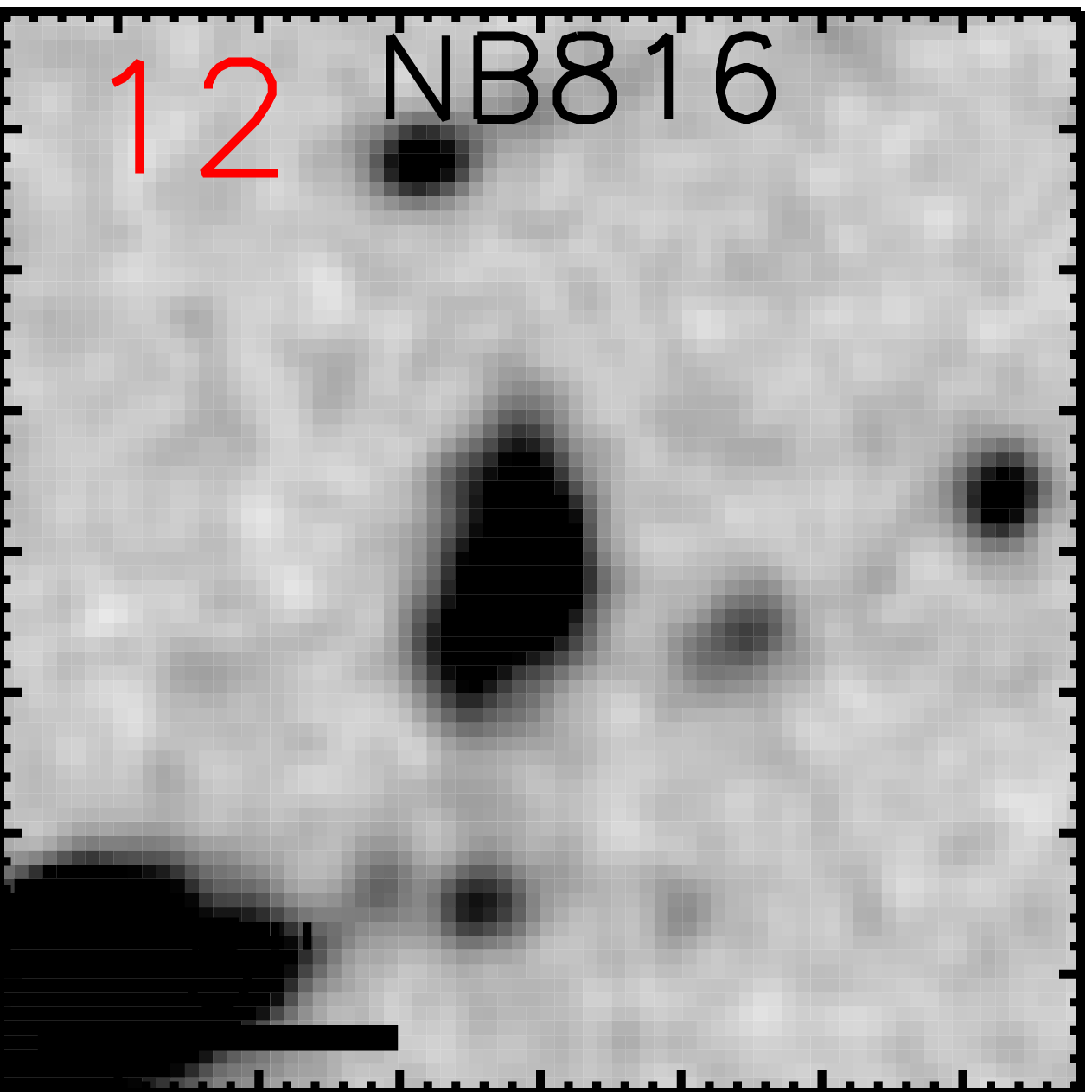} & 
\includegraphics[width=0.15\textwidth]{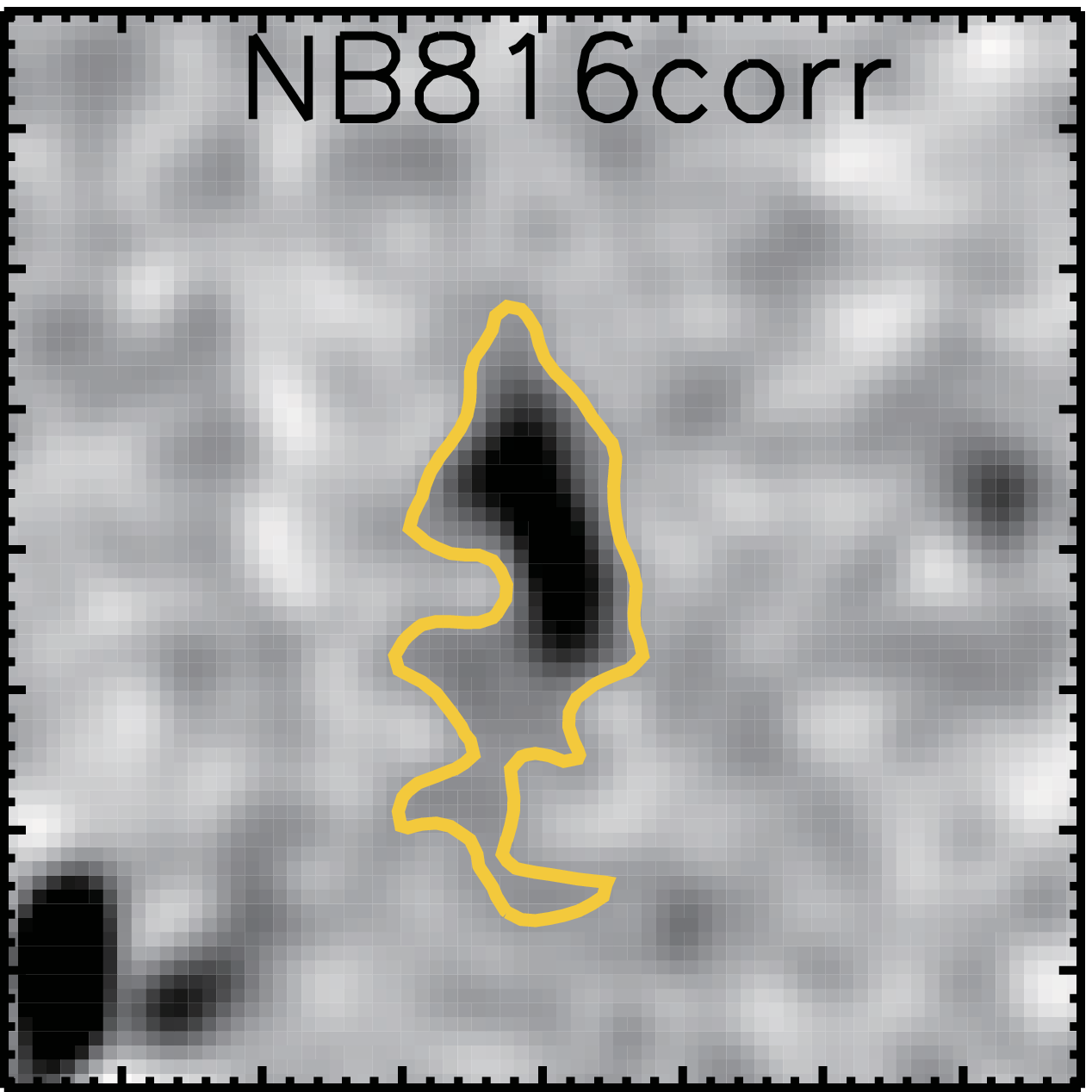} &
\includegraphics[width=0.15\textwidth]{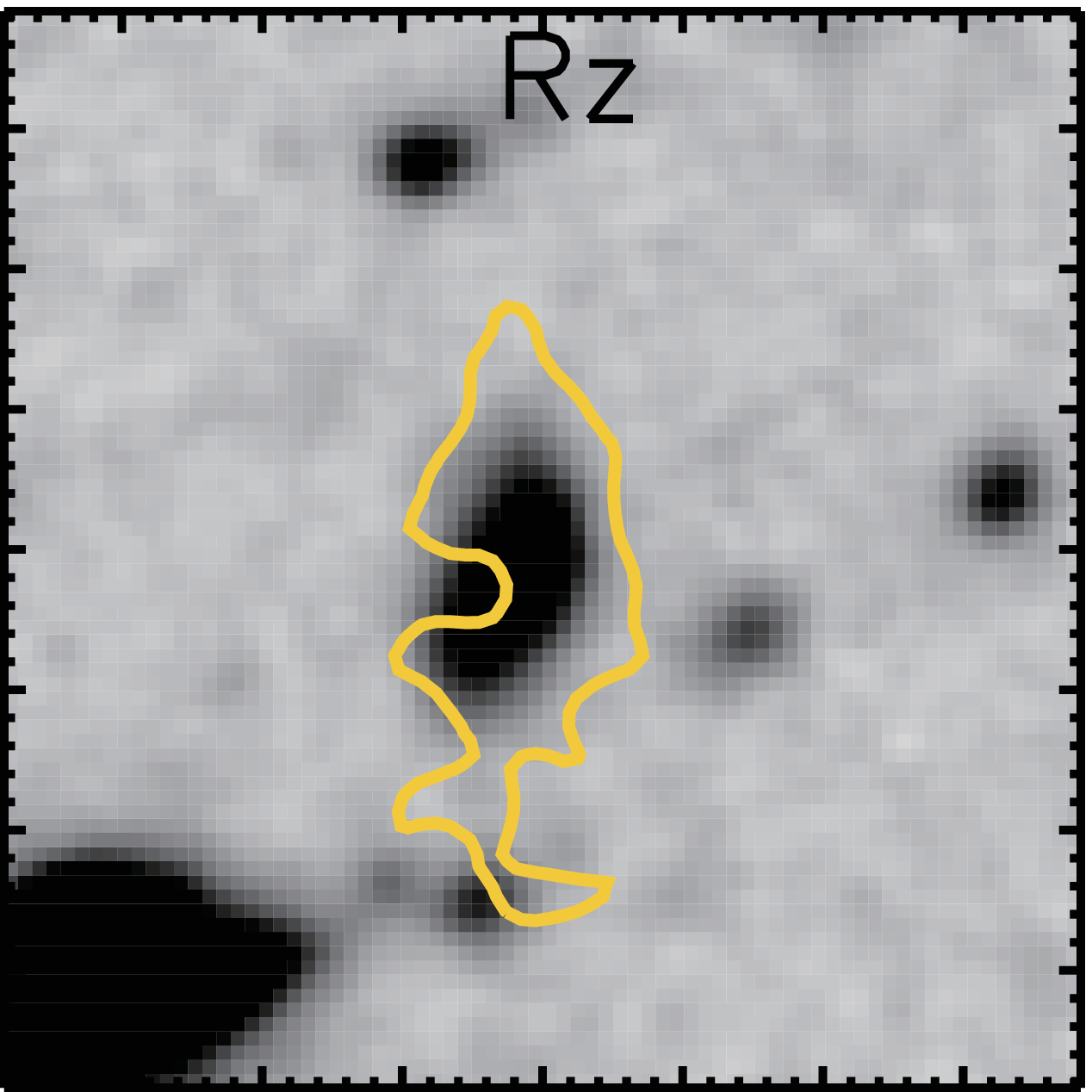} \\
\end{tabular}
\end{center}
\caption{\nb, \nbcorr, and \rz\ images of 12 \oiib s. 
Images are arranged in order of isophotal area from left to right and top to bottom. 
\oiib\ IDs are indicated at the top left corner of each \nb\ image. 
North is up and east is left. 
Each panel is $15\ar\times15\ar$ (corresponding to $124\times124$ kpc at $z=1.19$) 
with the \oiib s at the center. The spatial scale is indicated in the \nb\ images. 
The yellow contours in the \rz\ images show the isophotal area 
above $2\sigma$ arcsec$^{-2}$ in the \nbcorr\ image as explained in Section \ref{select}. 
}
\label{stamp_all}
\end{figure*}

We summarize properties of our 12 \oiib s at $z\sim1.2$ in Table \ref{tab_o2blob_all}. 
\oiib s are called \oiib\ 1 to 12 in order of isophotal area. 
Figure \ref{stamp_all} shows \nb, \nbcorr, and \rz\ images of all 12 \oiib s. 
An isophote of $2\sigma$-level surface brightness is overlaid on the \rz\ images, 
showing asymmetrically extended \oii\ emission with 
relatively compact continuum emission.

\subsection{Photometry and Stellar Properties}\label{photo_all}
We perform photometry of 
our \oiib s with the 11 images of {\it uBVRizJHK} and IRAC 3.6\micron\ and 4.5\micron\ bands. 
Total magnitudes in the Subaru broadbands, {\it BVRiz}, are 
obtained from {\tt MAG\_AUTO} by SExtractor. 
Similarly, we measure {\tt MAG\_AUTO} magnitudes for the {\it u}-band 
image from Megacam on the CFHT and {\it JHK} images 
from DR10 version of the UKIDSS UDS survey.
For each image, we align the broadband images to the \nb\ image 
and crossmatch the \oiib\ positions after visual reconfirmation. 
IRAC and MIPS 24\micron\ imaging data were obtained from the SpUDS survey. 
For the IRAC images, total magnitudes are determined from aperture photometry 
on a 2\farcs4 diameter aperture by applying aperture correction. 
The correction factor is calculated by generating artificial objects with given 
total magnitudes and recovering their aperture photometry. 

We then derive the stellar properties, fitting 
model spectral energy distributions (SEDs) with
the SEDs of 11-band photometry.
The model SEDs are produced with the \cite{bc03} stellar population synthesis code. 
We assume a constant star formation history,
the \cite{salpeter55} initial mass function with standard lower and upper mass cutoffs of $0.1$ \Msun\ 
and $100$ \Msun, respectively, dust attenuation law of \cite{calzetti00}, 
and the solar metallicity. 
We fix the redshift to $z=1.19$, assuming that the \oii\ emission line 
falls in the center of \nb\ filter. 
We set a maximum stellar age of the \oiib s to $\sim5$ Gyr, 
the cosmic age at $z=1.19$.

\begin{figure}
\includegraphics[width=0.48\textwidth]{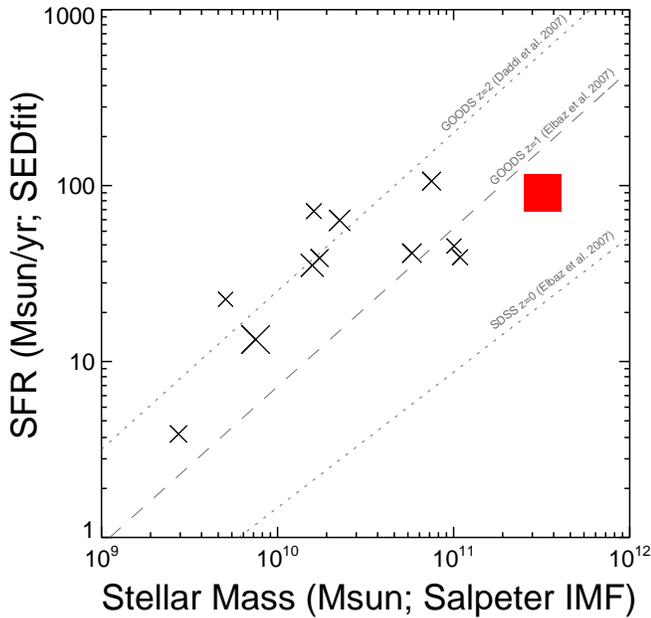}
\caption{Stellar mass--SFR correlation for \oiib s at $z\sim1.2$ in SXDS. 
The red solid square represent \oiib\ 1, the most extended \oiib, while 
black crosses indicate other smaller \oiib s. 
The size of the symbols is proportional to the isophotal area of the \oiib\ each symbol represents. 
The dashed line is the $z=1$ correlation taken from \cite{elbaz07}. 
The dotted lines are the $z=0$ and $z=2$ correlations taken from \cite{elbaz07} and \cite{daddi07}, respectively. 
}
\label{mass-sfr}
\end{figure}

We examine the star-forming nature of the \oiib s with a correlation between their 
stellar masses and SFRs derived from the SED fitting. 
It is known that star-forming galaxies at $z=0$--2 have a tight stellar mass$-$SFR 
correlation referred to as a star formation main sequence ($\sim0.2$ dex scatter). The main
sequence has a similar slope at any redshifts known to date,
but the normalization of the main sequence decreases 
from $z=2$ to $z=0$, which is related to the overall decline 
of cosmic SFR density \citep[e.g.,][]{daddi07, elbaz07}. 
Passive galaxies with high stellar masses are located far below the main sequence and 
they are regarded as progenitors of elliptical galaxies in the local universe. 
Figure \ref{mass-sfr} shows the stellar mass-SFR relation for all \oiib s at $z\sim1.2$. 
In the low mass regime, the \oiib s show a positive correlation between 
stellar mass and SFR. The distribution of \oiib s does not follow 
the main sequence, but presents a plateau at the high-mass end of $\gtrsim10^{10}$ \Msun. 
The \oiib s fall below the main sequence at $>10^{11}$ \Msun. Albeit with 
the small statistics, it would imply that low-mass \oiib s are actively 
star forming, while high-mass \oiib s quench their star formation. 
In Figure \ref{mass-sfr}, \oiib\ 1 is the most massive blob and 
its SFR is below the main sequence, implying that it is quenching 
star-forming activity. Similarly, two other \oiib s, blobs 9 and 12
below the sequence, are probably quenching their star formation.

\section{\oiib\ 1 at $\lowercase{z}\sim1.2$}\label{blob1}

Among 12 \oiib s at $z\sim1.2$, \oii B1 
(${\rm R.A.} = 02^{\rm h}17^{\rm m}08\fs64$, ${\rm decl.} = -04\degr50\arcmin22\farcs88$) 
is the brightest emitter and shows the most extended profile in the \nbcorr\ image 
(red square in Figure \ref{isophotal}) with an isophotal area of $34$ arcsec$^2$. 
The rest-frame equivalent width and \oii\ luminosity of \oiib\ 1 
are estimated from the \nb\ and \rz\ images 
to be $71\pm 10$ \AA\ and $3.2\pm 0.2\times10^{42}$ \ergs, respectively. 
This luminosity is significantly brighter than the typical \oii\ emitters found in the 
literature \citep[e.g.,][]{ly12, drake13}. 

\begin{figure}
\centering
\includegraphics[width=0.5\textwidth]{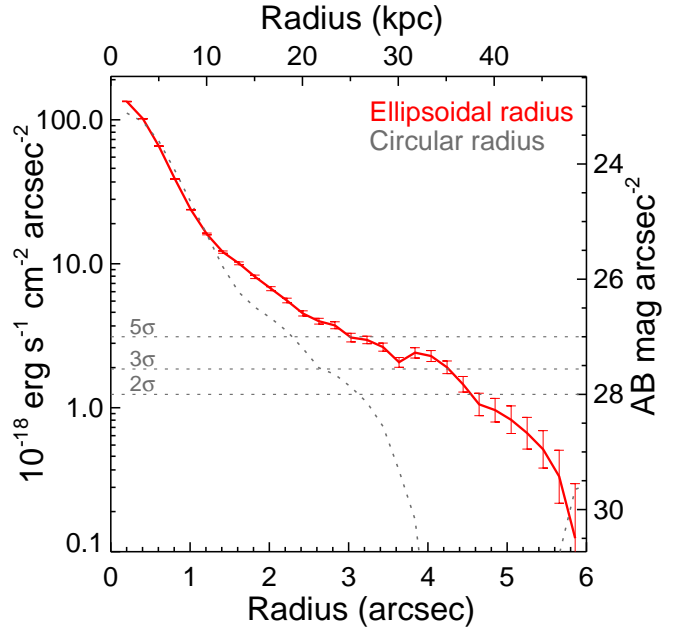}
\caption{
Azimuthally averaged \oii\ surface brightness profiles of \oiib\ 1.
The red and gray lines are the \oii\ profiles 
measured in the \nbcorr\ image with the ellipsoidal radius (semi-major axis)
and circular radius, respectively.
Error bars represent the root-mean-square values
around the average profile. 
Horizontal dotted lines indicate the surface brightness limits
of the $5\sigma$, $3\sigma$, and $2\sigma$ levels,
from top to bottom. 
}
\label{profile_o2blob1}
\end{figure}

Being extended over 9\ar\ or $\sim75$ kpc along the major axis, 
\oiib\ 1 is remarkably larger than the \oiii\ blob at $z=1.6$ reported by \cite{brammer13}. 
However, a direct comparison is difficult due to the different sensitivity limits. 
Figure \ref{profile_o2blob1} presents the azimuthally averaged  \oii\ emission 
surface brightness profiles of \oiib\ 1 measured with the \nbcorr\ image. 
The red line in Figure \ref{profile_o2blob1} is
the surface brightness of ellipsoidal isophote
along the elongated direction,
i.e. east-west direction, 
that is illustrated with the yellow isophote in Figure \ref{stamp_all}. 
The profile of ellipsoidal isophotes 
is extended to the radius of 
$\sim 3\farcs0$, 4\farcs1, and 4\farcs3 along the semi-major 
axis at the $5\sigma$, $3\sigma$, and $2\sigma$ surface brightness levels,
respectively.
This indicates that \oii\ emission of \oiib\ 1 is 
remarkably extended even at the $5\sigma$ surface brightness limit. 
For comparison, we show the surface brightness of circular 
isophotes with the gray dotted line in
Figure \ref{profile_o2blob1}. This surface brightness profile
of circular isophote is truncated at the relatively small 
angular distance, because \oiib\ 1 is highly elongated.

\subsection{Photometric Properties}\label{photo_blob1}
The photometry of \oiib\ 1 in the 11 
bands ($uBVRizJHK$ and IRAC 3.6\micron\ and 4.5\micron) 
is determined and given in Section \ref{photo_all}. 
In addition to the 11-band photometry, 
we use the MIPS 24 \micron\ flux of \oiib\ 1,
$298\pm7 \mu$Jy, 
that is carefully measured by \citet{simpson12}. 
The X-ray photometry of \oiib\ 1 
is obtained from the {\it XMM-Newton} catalog by \cite{ueda08}. 
The flux in 0.3--0.5 keV is calculated by assuming the photon 
index of 1.5, while the index of 2.0 is assumed 
for 0.5--2.0 keV and 2--10 keV \citep{ueda08}. 
Finally, we take the radio 1.4-GHz flux from \cite{simpson12}. 
The radio observation was carried out with the Very Large Array (VLA) of the National Radio 
Astronomy Observatory \citep{simpson06}, and 
the 1.4 GHz radio flux is $216\pm22\mu {\rm Jy}$. 
All of the measurements are listed in Table \ref{tab_sum_photo}. 

\begin{figure}
\centering
\includegraphics[width=0.5\textwidth]{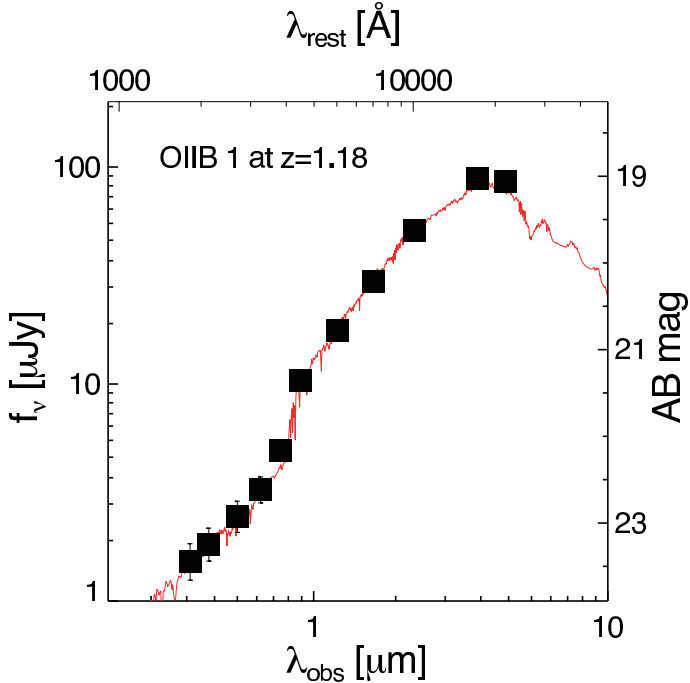}
\caption{SED of \oiib\ 1. 
The black squares show the total fluxes in 11 bands used in SED fitting (Table \ref{tab_sum_photo}). 
The red line represents the best-fitting model described in Section \ref{photo_blob1}. 
}
\label{sedfitting}
\end{figure}

Stellar properties of \oiib\ 1 are obtained in the same manner as those
of the other \oiib s in Section \ref{photo_all},
but with the spectroscopic redshift of $z=1.18$ (see Section \ref{confirm_o2}).
The best-fit SED of \oiib\ 1 has a very old stellar age, 4.7 Gyr,
comparable with the cosmic age at $z=1.18$. 
The stellar mass is $3.2\times10^{11}$ \Msun. 
The SFR is 90 \Msun yr$^{-1}$ and 
the color excess {\it E(B--V)} is 0.3 mag. 
Figure \ref{sedfitting} presents the SED and the best-fit model of \oiib\ 1.

\input{tab_sum_photo}

\subsection{Spectroscopic Properties}\label{spec}

\subsubsection{Observations}\label{spec:obs}

The spectroscopic observation of \oiib\ 1 was carried out with 
the Visible Multi-Object Spectrograph (VIMOS) on the Very Large Telescope (VLT) 
as part of the European Southern Observatory (ESO) program P074.A-033 \citep{simpson12}. 
The observation was conducted in service mode using the MR-orange grating and the 
GG475 order-sorting filter, providing a spectral resolution of $\lambda/\Delta\lambda\simeq 580$ 
(1\ar\ slit) and a spectral dispersion of 2.5 \AA\ pixel$^{-1}$ 
over the wavelength range of 4800--10000 \AA. 
The slit is located at the center of the blob in the north-south direction 
as a default of VIMOS. 
The total exposure time was 5400 s. 
Data reduction is performed using the standard pipeline and described in detail in \cite{simpson12}. 
The wavelength calibration has a root-mean-square (rms) uncertainty of 0.4 pixels or 1.0 \AA. 
The estimated $3\sigma$ limiting flux for detecting an emission line 
is $\sim10^{-17}$ \ergscm\ over 
$5300 {\rm \AA} \lesssim \lambda \lesssim 7700$ \AA\ spectral range. 

\begin{figure}
\centering
\includegraphics[width=0.5\textwidth]{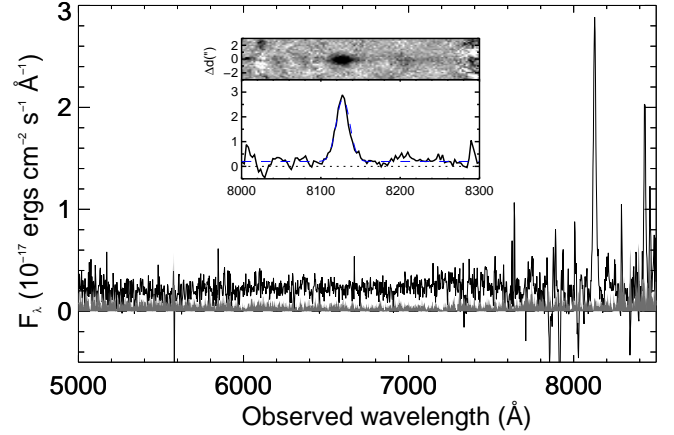}
\caption{VIMOS spectrum of \oiib\ 1 at $z=1.18$ displayed in the observed frame. 
The solid line shows the observed spectrum and the gray shaded region indicates $1\sigma$ sky noise. 
The inset figure illustrates a close-up view of the spectrum around the strong emission line at $8127.9$ \AA\ 
with a two-dimensional spectrum on the top. 
The blue dashed line represents the best-fitting Gaussian profile. 
}
\label{spec_all}
\end{figure}

\subsubsection{Emission-line Identification}\label{confirm_o2}

Figure \ref{spec_all} shows the overall VIMOS spectrum of \oiib\ 1 extracted through 
a 1\ar aperture 
with an enhanced spectrum at the wavelength around a remarkably 
strong emission line in the inset figure. 
A two-dimensional spectrum around the line is also shown in the top panel 
of the inset figure, where the south end of the slit is up. 
The central wavelength of this single line is $8127.9$ \AA. 
We investigate whether this line is \lya, \oii, \oiii, or \ha\ emission. 
We exclude the possibility of \ha\ emission at $z=0.24$, 
as the spectrum shows no detection of \oiii$\lambda5007$ emission  
at $\simeq6209$ \AA\ and \oii$\lambda3727$ emission at $\simeq4621$ \AA. 
It is not \oiii\ at $z=0.62$, either, because of no detection of  
\oiii$\lambda4959$ emission at $\simeq8034$ \AA\ and 
\oii$\lambda3727$ emission at $\simeq 6038$ \AA. 
Furthermore, clear detection of continuum emission blueward of this line 
indicates that this object is not a \lya\ emitter at $z=5.68$. 
Additionally, the derived photometric redshift is quite strongly peaked around $z=1.2$ \citep{drake13}. 
We thus conclude that this line is \oii\ emission at $z=1.1808$.

Because the VIMOS spectral resolution is only $R=580$, 
the \oii$\lambda\lambda3726,3729$ doublet is not resolved. 
We thus simply fit the \oii\ emission line with 
a single symmetric Gaussian profile by using the IDL {\tt MPFITEXPR} routine, 
which is part of the IDL {\tt MPFIT} package 
\citep{mpfit}\footnote{http://www.physics.wisc.edu/~craigm/idl/}. 
We measure the line flux of \oiib\ 1 to be $5.5\pm0.2\times 10^{-16}$ \ergscm, 
corresponding to an \oii\ luminosity of $4.3\pm0.2\times10^{42}$ \ergs. 
We summarize spectroscopic properties of \oiib\ 1 in Table \ref{tab_sum_prop}. 

\input{tab_sum_prop}

\begin{figure}
\centering
\includegraphics[width=0.45\textwidth]{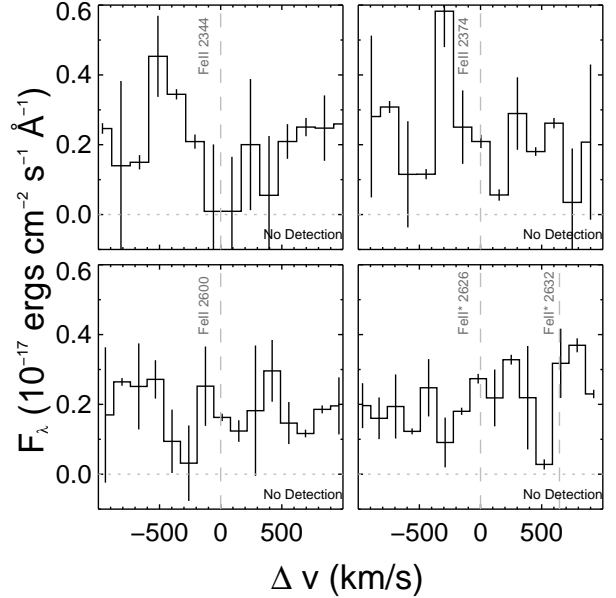}
\caption{Spectra of non-detected \feii\ and \feiis\ lines of \oiib\ 1 as a function of velocity offset 
with respect to the systemic velocity (vertical dashed lines) calculated from the \oii\ emission. 
Error bars indicate $1\sigma$ sky uncertainty. 
}
\label{spec_non-detect}
\end{figure}

\subsubsection{Spectral Properties of Other Lines}\label{other_lines}

\begin{figure}
\centering
\includegraphics[width=0.45\textwidth]{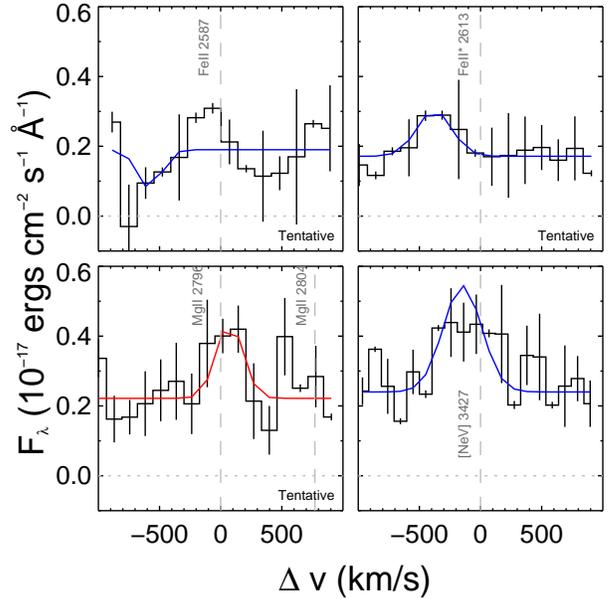}
\caption{Spectra of absorption and emission lines that are tentatively detected of \oiib\ 1 
as a function of velocity offset 
with respect to the systemic velocity calculated from the \oii\ emission (vertical dashed lines). 
Error bars indicate $1\sigma$ sky uncertainty. 
Blue and red dashed lines represent the best-fitting Gaussian profiles of 
blueshifted and redshifted lines, respectively. 
}
\label{spec_detect}
\end{figure}

We also try to detect other emission or absorption lines  
falling into the wavelength range of the VIMOS spectrum: 
\feii\ $2344, 2374, 2587, 2600$, 
\feiis\ $2613, 2626, 2632$, \mgii\ $2796, 2804$, and \nev\ $3427$. 
We fit all lines with a single Gaussian profile using the {\tt MPFIT} package to 
determine the central wavelength, FWHM, and flux. 
Lines with no detection are shown in Figure \ref{spec_non-detect}, 
while those with a tentative signature are shown separately in Figure \ref{spec_detect}. 
Each line is plotted as a function of velocity offset 
with respect to the systemic velocity calculated from the \oii\ emission line. 

The top left panel of Figure \ref{spec_detect} shows a very tentative 
\feii$\lambda2587$ absorption line with only $2.2\sigma$ at $5628.0$ \AA, 
corresponding to a line redshift of $z_{\rm line} = 1.1764$. 
Likewise, we find marginal detection of the fine-structure \feiis$\lambda2613$ 
emission line at the $2.2\sigma$ significant level at 5690.6 \AA\ 
or $z_{\rm line}=1.1772$ (top right panel of Figure \ref{spec_detect}). 
The bottom left panel shows $3\sigma$ marginal detection of \mgii$\lambda2796$ emission 
line at $6097.5$ \AA, whose redshift, $z_{\rm line}=1.1805$, is very close 
to the systemic redshifted derived from the \oii\ emission line. 
The last line, \nev$\lambda3427$, is detected with a high significance level 
of $6\sigma$ at $7467.0$ \AA, or $z_{\rm line}=1.1789$. 
Its flux is derived to be $2.5\pm1.6\times10^{-17}$ \ergscm. 
The properties of these absorption and emission lines are summarized 
in Table \ref{tab_other}.

\section{\oiib \lowercase{s} 4 and 8}\label{other_blobs}

\subsection{Spectroscopic Data}\label{other_obs}

In addition to the VIMOS spectrum of \oiib\ 1, we also have spectroscopic data 
for another two blobs, \oiib s 4 and 8. The observations were carried out 
with the Faint Object Camera and Spectrograph (FOCAS) 
mounted on the Subaru Telescope. 
The spectrum of \oiib\ 4 was obtained with the 150 grism 
and an SY47 order-cut filter 
providing a spectral resolution 
of $\lambda/\Delta\lambda \simeq 250$ (0\farcs8 slit) 
over 4700--9400 \AA. The total exposure time was 3600 s. 
The spectrum of \oiib\ 8 
was taken with the 300 blue grism and the SY47 order-cut filter 
with a spectral resolution of 500 (0\farcs8 slit). 
The total exposure time was 7200 s. 
Data reduction is performed with IRAF following the standard method 
for the optical multi-slit spectroscopy. 
The estimated $3\sigma$ limiting flux density are  
$2\times10^{-18}$ \ergscmA\ and $1\times10^{-18}$ \ergscmA\ 
over $5300 < \lambda < 7700$\AA\ spectral range for 
spectra of \oiib s 4 and 8, respectively. 

\begin{figure*}
\centering
\includegraphics[width=0.7\textwidth]{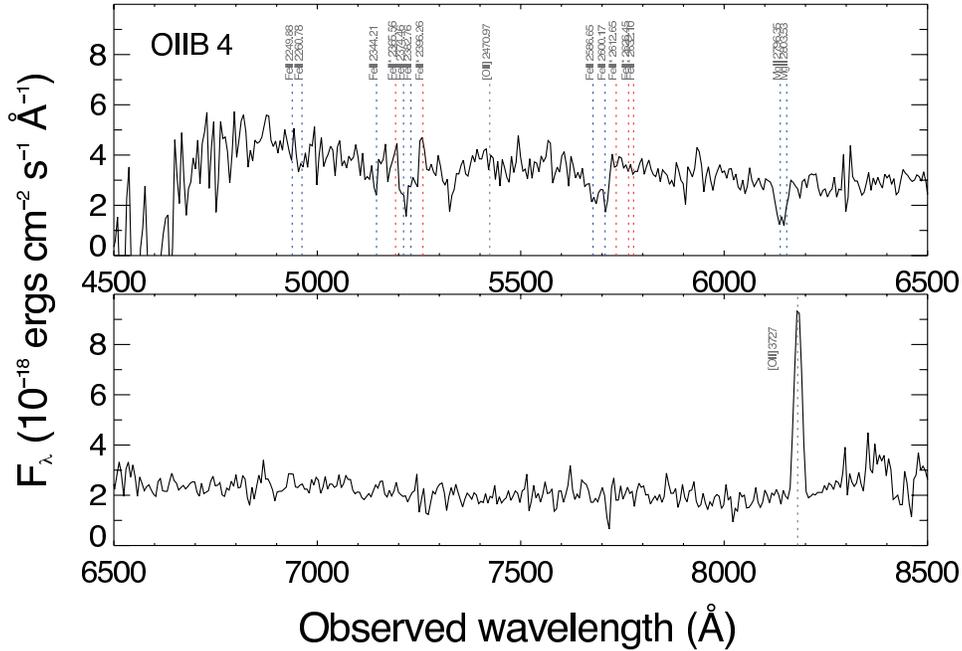}
\caption{
FOCAS spectrum of \oiib\ 4 displayed in the observed frame. 
Red and blue vertical dotted lines indicate the expected emission 
and absorption lines, respectively, 
based on the systemic redshift
determined with the \oii\ emission line (black dotted line). 
}
\label{spec_o2b4}
\end{figure*}

\subsection{Spectroscopic Properties}\label{other_iden}

Figure \ref{spec_o2b4} shows the FOCAS spectrum of \oiib\ 4 
in the observed frame. 
There is a single emission line at 8181.6 \AA. 
Here, we examine if this line is \lya, \oii, \oiii, or \ha\ emission. 
It is not \ha\ emission at $z=0.25$ because of no detection of \oiii$\lambda5007$ 
line at 6242 \AA\ and \oii$\lambda3727$ line at 4646 \AA. 
We also rule out the possibility that this is an \oiii$\lambda5007$ line at $z=0.63$,
because we find no \oiii$\lambda4959$ line at 8103 \AA\ and 
no \oii$\lambda3727$ line at 6090 \AA. Likewise, the line is not \lya\ emission 
at $z=5.73$ as we certainly detect the continuum blueward of the line. 
We thus conclude that the single emission at 8181.6 \AA\ is 
an \oii$\lambda3727$ line emitted at $z=1.195$.
We fit the \oii\ emission line 
with a Gaussian profile by {\tt MPFIT}. 
The best-fit line flux is $1.7\times10^{-16}$ \ergscm, 
corresponding to the \oii\ luminosity of $1.4\times10^{42}$ \ergs. 
The spectrum of \oiib\ 4 shows a possible \mgii$\lambda\lambda2796,2804$ 
doublet absorption in the FOCAS spectrum (Figure \ref{spec_o2b4}). 
The absorption is blueshifted from the systemic redshift determined 
from the \oii\ emission line with an average velocity offset of
$\Delta v\sim -200$ \kms. 

\begin{figure*}
\centering
\includegraphics[width=0.7\textwidth]{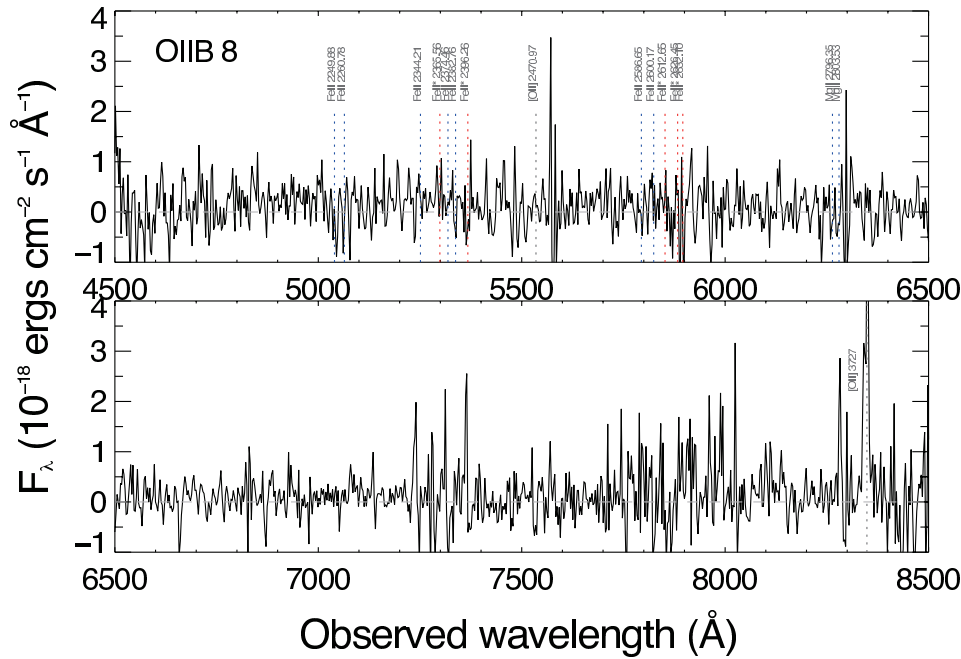}
\caption{
Same as Figure \ref{spec_o2b4}, but for \oiib\ 8.
}
\label{spec_o2b8}
\end{figure*}

The spectrum of \oiib\ 8 is presented in Figure \ref{spec_o2b8}. 
This spectrum is so noisy that we detect no continuum, but
one emission line at 8346.9 \AA.
We rule out the possibility that this is \oiii$\lambda5007$ emission of a source at $z=0.67$,
because we detect neither \oiii$\lambda4959$ line at 8267 \AA\ nor \oii$\lambda3727$ at 
6213 \AA.
Similarly, this line is not \ha\ emission at $z=0.27$,
because of no detection of \oiii$\lambda5007$ 
at 6370 \AA\ and \oii$\lambda3727$ at 4733 \AA.
The line is not the \lya\ emission at $z=5.86$,
due to the clear ($>5\sigma$) detections of continuum
blueward of the line in the {\it uBVR} images.
Using the spectrum, we confirm that the single line 
of \oiib\ 8 at 8346.9 \AA\ is the \oii$\lambda3727$ emission 
at $z=1.24$.
We estimate the \oii\ emission line flux and luminosity to be
$5.6\times10^{-17}$ \ergscm and $4.9\times10^{41}$ \ergs,
respectively, by our Gaussian profile fitting.

\input{tab_other}

\section{Discussion}\label{discuss}

\subsection{AGN Activity}\label{agn}
In this section, we examine the AGN activity of \oiib\ 1 
from several aspects. 
As the first step, we classify \oiib\ 1 as an AGN due to the 
presence of \nev\ emission. 
We have detected \nev\ emission line with a flux of 
$2.5\pm1.6\times10^{-17}$ \ergscm, 
while star-forming regions cannot provide enough energy to produce 
high-ionization lines such as \nev.

The nucleus of \oiib\ 1 is found to be an obscured type-2 AGN. 
The prominent 4000 \AA\ break seen in the observed SED (Figure \ref{sedfitting}) 
indicates that we are not seeing the AGN continuum directly 
and that stellar components dominate the observed SED. 
Moreover, as listed in Table \ref{tab_sum_prop}, the measured equivalent width of \oiib\ 1 
is $71\pm10$ \AA\ for the entire blob. 
It is much higher than the {\it mean} equivalent widths of typical quasars, which is 
$\sim10$ \AA\ \citep{miller92} or more precisely $7.80\pm0.30$ \AA\ for radio-loud quasars 
and $4.77\pm0.06$ \AA\ for radio-quiet samples at $z<1.6$ \citep{kal12}.  
Although quasar samples in \cite{kal12} show a very broad \oii equivalent width distribution 
ranging from EW(\oii) $\sim 0.1$ \AA\ to EW(\oii) $\sim 100$ \AA, 
only a few of their quasars ($<1\%$) show an EW(\oii) as large as $100$ \AA. 
The EW(\oii) of \oiib\ 1 is far above those of the majority of quasars.

\oiib\ 1 can be further classified as a radio-quiet galaxy, 
because of a high mid-infrared to radio flux ratio, 
$q_{24} \equiv \log(S_{24\micron}/S_{1.4 {\rm GHz}})$. 
We obtain $q_{24} = 0.62\pm0.03$ after 
$k$-correcting the 24 \micron\ and 1.4 GHz radio fluxes \citep{simpson12}. 
This $q_{24}$ value is above the $q_{24} < -0.23$ threshold 
used to separate radio-loud galaxies \citep{ibar08}. 
Recently, \cite{liu13} study extended ionized gas nebulae around 11 radio-quiet type-2 quasars 
at $z\sim0.5$ using \oiii$\lambda5007$ emission. 
Observing with an integral field spectrograph, Gemini/GMOS-N IFS, 
they have found that the \oiii\ emission of radio-quiet objects is extended 
and nearly perfectly round rather than clumpy or elongated profiles 
seen in radio-loud galaxies \citep{nesvadba08}. 
\oiib\ 1, however, shows an asymmetric morphology as seen in Figure \ref{stamp_all}, 
unlike radio-quiet quasars at $z\sim0.5$. 
The half-light radius of \oiib\ 1 calculated by SExtractor is $8.3$ kpc. 
It is notably larger than the \oiii\ effective radii of radio-quiet samples 
in \citeauthor{liu13}, 
which range from 2.5 to 5.2 kpc with a median value of 3.2 kpc. 
However, it is noteworthy that the oxygen-line profile studied in this paper 
is the \oii\ emission, while it is \oiii\ emission in \cite{liu13}. 
A comparison of their profiles is not straightforward. 
According to the rest-frame optical spectra of radio-quiet quasars in \citet[][ their Figure 5]{liu13}, 
the \oii\ emission line is significantly weaker than the \oiii, indicating high \oiii/\oii\ line ratios. 
If \oiib\ 1 has the same line ratio, its \oiii\ emission 
would be very strong, being more luminous than the extended \oii\ emission. 
This implies that, similar to the \oii\ emission, \oiii\ of our \oiib\  is possibly more 
extended than those of the radio-quiet quasars at $z\sim0.5$.

\subsection{Outflow Signature}\label{confirm_outflow}

As mentioned in Section \ref{intro}, spatially extended profiles of 
metal emission lines such as \oii\ are likely to originate from outflow 
rather than inflow of pristine gas from metal-poor IGM. 
In this section, we examine whether or not \oiib\ 1 has outflows 
and discuss possible origins of the outflows. 
If interstellar absorption lines are blueshifted against the systemic redshift,
it is strong evidence of outflows.
\nai\ D $\lambda5890, 5896$ doublet is one of the most popular absorption lines 
used to determine the outflow velocity of galaxies at low redshifts \citep[e.g.,][]{martin05,rupke05a,rupke05b}. 
At higher redshifts ($z\gtrsim1.0$), low-ionization interstellar metal lines in the rest-frame UV wavelengths 
such as \feii\, \mgii, or \sii\ lines have been used instead \citep[e.g.,][]{shapley03, martin12, kornei12}, 
as \nai\ lines are redshifted into near-infrared wavelengths and become difficult 
to be observed. 
The \feiis\ emission is a fine structure transition that is connected to 
the \feii\ resonance lines. 
Normally, Fe$^{+}$ ions in the front side of the outflow produce blueshifted absorption and 
those in the far side generate redshifted emission with respect to the systemic velocity, 
thus producing a P-Cygni profile. However, excited Fe$^+$ ions can 
produce fine-structure transitions causing the \feiis\ emission both blueward and 
redward of the systemic velocity \citep{prochaska11, rubin11}. 
Detecting these fine-structure \feiis\ emission lines can be another piece 
of evidence for outflows. 

We have marginally detected both \feii$\lambda2587$ absorption line and 
\feiis$\lambda2613$ emission line with $\sim2.2\sigma$ each. 
The \feii$\lambda2587$ absorption line tentatively shows 
a rest-frame equivalent width (EW$_0$) and a central velocity offset of 
$-2.7\pm1.4$ \AA\ and $-605\pm99$ \kms, respectively. 
The \feiis$\lambda2613$ emission line is blueshifted at $-495\pm138$ \kms\ from 
the systemic velocity. 
It is roughly consistent with the velocity offset of the \feii$\lambda2587$ 
absorption line ($-605\pm99$ \kms) within the $1\sigma$ uncertainties. 
Although the detection of \feii$\lambda2587$ and \feiis$\lambda2613$ lines 
is only marginal at $2.2\sigma$ each, 
the existence of both lines together in our spectrum tentatively supports 
the outflow scenario of \oiib\ 1.

\subsection{Possible Origins of Spatially Extended Emission and Outflows}\label{outflow_origin}

We find one remarkably extended \oiib, \oiib\ 1, at $z=1.18$ 
and its outflow process seems to be supported by the contingent detection of 
blueshifted \feii$\lambda2587$ absorption and \feiis$\lambda2613$ emission lines. 
Several scenarios are plausible to explain the outflow process of galaxies 
including 
star formation activity, photoionization by star-formation/AGN, AGN-driven wind, 
and/or shock excitation.

We constrain the origin of the marginal \feiis$\lambda2613$ emission of \oiib\ 1 
by following the description by \cite{erb12}. 
They use CLOUDY code version 08.00 \citep{cloudy08} 
to construct the photoionization models by adopting the Starburst99 ionization spectrum \citep{starburst99}. 
Constant star formation and a metallicity of $\sim0.6$\Zsun\ are assumed. 
While \cite{erb12} calculate an electron density of 100 cm$^{-3}$ 
from an observed ratio of \oii$\lambda3729/$\oii$\lambda3726$, 
the \oii\ doublet of \oiib\ 1 is not resolved. 
Different values of the electron density lead to difference values in the ionization 
parameter $U$ (the ratio of ionizing photon density to the hydrogen density) 
in the sense that a higher density lowers the ionization parameter. 
The models show that the ratio of \oii\ to \feiis\ lines decrease with 
increasing $U$ \citep[Figure 15 in ][]{erb12}. 
In our case, the \oii$\lambda3727$/\feiis$\lambda2613$ ratio is tentatively $68\pm26$. 
The low \oii/\feiis\ ratio of \oiib\ 1 would imply a very high ionization parameter 
($\log U > -1.0$). 
Although we have not corrected the line ratio for dust extinction, the implied ionization 
parameter would be even higher when taking the dust effect into account, as we expect 
to see more of a decrease in \feiis\ lines relative to the \oii\ line in the presence of dust. 
\cite{lilly03} have found the ionization parameter of galaxies at $0.47 < z < 0.92$ to be 
$-3.2 \lesssim \log U \lesssim -2.5$, which is remarkably lower than the ratio derived in this paper. 
It is thus unlikely that the \feiis\ emission originates from photoionization in \hii\ regions of \oiib\ 1. 
However, as the detection of \feiis\ emission is indeed marginal, the derived \oii/\feiis\ ratio 
is probably just a lower limit and the corresponding ionization parameter is 
subsequently lower than discussed here.

The outflow velocity is an important quantity for determining 
whether outflow is driven by SNe or AGN. 
\cite{heckman00} study outflows in 32 luminous infrared galaxies 
at $z\lesssim0.1$ including galaxies whose luminosities are powered by starburst as well 
as some powered by AGNs. They have found outflow velocities up to 400--600 \kms. 
More recently, \cite{rupke05a} and \cite{rupke13} study outflows from ULIRGs at various redshifts 
and conclude that the central outflow velocities of 
SNe-driven wind are on average $\sim-100$ \kms, 
while those of AGN-powered galaxies are $-200$--$(-1500)$ \kms.  
As described in the previous section, the central velocity offsets of \feii$\lambda2587$ 
absorption line and \feiis$\lambda2613$ emission line for \oiib\ 1 are 
 $-605\pm99$ \kms\ and $-495\pm138$ \kms, respectively. 
These central outflow velocities are generally comparable to those of 
AGN-powered galaxies. 
Thus \oiib\ 1 seems to favor the AGN-driven wind scenario that is thought to be 
one of the major effects of AGN feedback.

Shock excitation is another possible source powering large-scale outflow. 
Radiative shock can be formed at the interface of the fast-moving wind and 
a slow-moving ISM \citep{veilleux05}. 
Many studies found strong evidence of shock excitation dominating the emission 
\citep{veilleux02, veilleux03, lipari04, monreal10, sharp10, rich11}. 
Shock can be distinguished from the others by examining various line ratios 
(e.g., \nii$\lambda6583$/\ha\ or \sii$\lambda\lambda6716, 6731$/\ha), which 
are not available in this study. 
Further spectroscopic observations especially with an integral field unit spectrograph 
covering those lines are necessary to determine whether 
the outflow of \oiib\ 1 is powered by shock heating.

\subsection{Can Gas Escape Out of the \oiib\ 1?}\label{escape_velo}

We calculate the escape velocity inferred from the virial theorem using the simple equation 
of $V_{\rm esc} \simeq \sqrt{2GM_{\rm halo}/R}$, 
where $M$ is the halo mass and $R$ is the radius of the galaxy. 
The halo mass of \oiib\ 1 is obtained from the derived stellar mass in Section \ref{photo_blob1} by 
applying the relationship between the stellar and halo masses given by \cite{leauthaud12}. 
For the stellar mass of $3.2\times10^{11}$ \Msun, the halo mass ranges $1.0\times10^{13}$ \Msun. 
We define the radius of \oiib\ 1 of $R_{\rm blob} = 17.6$ kpc as twice the Petrosian radius $r(\eta=0.2)$ 
measured in the \rz\ continuum image. 
The Petrosian radius is first introduced by \cite{petrosian76} 
as the radius at which the surface brightness 
is $\eta$ times the average surface brightness within this isophote. 
It is found that the magnitude within twice the Petrosian radius with $\eta=0.2$ 
is approximately equal to the total magnitude for objects with exponential and 
Gaussian profiles and is $\sim90\%$ of the total 
magnitude in the case of objects with $r^{1/4}$ profiles \citep{bershady00}. 

The estimated escape velocity is $2208$ \kms. This is larger than the velocity of 
the \feii$\lambda2587$ absorption and \feiis$\lambda2613$ emission lines ($495-605$ \kms). 
If the detection of \feii\ and \feiis\ lines is real, it indicates that 
the majority of gas cannot escape out of the gravitational potential well 
of \oiib\ 1 and cannot chemically enrich the IGM.

\subsection{How Rare are the \oiib s?}\label{numden}

\cite{drake13} calculate the comoving volume of their \oii\ emitter survey, 
corresponding to the FWHM of \nb\ times the survey area, 
to be $1.9\times10^5$ Mpc$^3$. 
The number density of giant \oiib s like \oiib\ 1 is thus 
$5.3\times10^{-6}$ Mpc$^{-3}$, 
whereas it is $6.3\times10^{-5}$ Mpc$^{-3}$ for all \oiib s (including \oiib\ 1) 
selected by the same criterion as in \cite{matsuda04}. 
The fraction of $z\sim1.2$ \oii\ emitters 
from \citeauthor{drake13} that are blobs is 0.1\% and 1.2\% for 
giant \oiib\  and small blobs, respectively.

\cite{faber07} study the evolution of the galaxy luminosity function up to $z\sim1$ and 
propose a mixed scenario of star formation quenching, 
moving star-forming galaxies to the red sequence where they further merge with 
other small galaxies. 
They define blue and red galaxies according to the location on 
the $U-B$ versus $M_B$ diagram 
and find the mean density of blue galaxies at $1<z<1.2$ to be $2.08\times 10^{-3}$ 
Mpc$^{-3}$ down to $M_B\simeq-20.0$ mag in DEEP2 survey \citep{davis03}. 
To compare with this result, 
we estimate the rest-frame $U-B$ color from the observed $R-z$ color. 
The relationship between the $U-B$ and $R-z$ colors is determined using models with 
constant star formation history from \cite{bc03}. 
All but one \oiib s at $z\sim1.2$ is located in the blue cloud region 
of the color-magnitude diagram. 
Down to the same UV limit, we calculate the number density of \oiib s 
at $z\sim1.2$ to be $6.3\times10^{-5}$ Mpc$^{-3}$. 
It implies that $\sim3\%$ of blue galaxies at $1.0\leq z \leq 1.2$ 
are quenching star formation by the same process as for our \oiib s.

As \oiib\ 1 at $z=1.18$ is classified as an obscured type-2 AGN (Section \ref{agn}), 
it is important to compare the estimated number density with those of AGNs at similar redshifts. 
The number density of AGNs at $z\sim1.0$ with hard X-ray luminosities of 
$L_{\rm 2-8keV} = 10^{44}-10^{45}$ 
\ergs\ is $\sim10^{-5}$ Mpc$^{-3}$ \citep{barger05, barger_cowie05}. 
An overall fraction of AGNs displaying outflows is approximately 60\% \citep[e.g., ][]{ganguly08}, 
albeit some variety depending on AGN properties. 
Consequently, the number density of AGNs with outflows at $z\sim1.0$ 
is roughly $5.4\times10^{-6}$ Mpc$^{-3}$, 
which is in good agreement with the number density of giant blobs like \oiib\ 1. 
It is likely that the outflows from \oiib\ 1 can be explained solely by AGN. 
However, the number density of all 12 \oiib s is larger than those of AGNs with outflows, 
implying that not all \oiib s are powered by AGNs. 
Additional observations are desirable to understand 
the nature of the entire population of \oiib s.

Since there is no systematic study of \oiib s up to date, 
it is difficult to state whether \oiib\ 1 is a rare object or not. 
As a baseline of discussion, we compare the above number density to 
those corresponding to extended LABs at various redshifts. 
Studies of LABs at $z\sim2.3$ \citep{yang09} and at $z\sim3$ \citep{matsuda04, saito06} 
have determined their number densities of $3\times10^{-6}$ Mpc$^{-3}$, 
$3\times10^{-4}$ Mpc$^{-3}$, and $1\times10^{-5}$ Mpc$^{-3}$, respectively. 
At higher redshift, \cite{ouchi09} discovered an extended \lya~emitter at $z\sim6.6$ and calculate the 
number density of LAB to be $1.2\times10^{-6}$ Mpc$^{-3}$. 
The number density of \oiib\ 1-type objects is at least an order of magnitude smaller than 
those of LABs at $z\sim3$ by \cite{matsuda04} and \cite{saito06}, 
but comparable to the number densities of 
LABs at $z\sim2.3$ by \cite{yang09} and at $z\sim6.6$ by \cite{ouchi09}. 
On the other hand, the number density of LABs at $z\sim3$ is comparable to our study if 
it is compared with the number density of the entire \oiib\  samples. 
Given that LAB surveys have shown that constraining the number density of 
the extended sources is challenging because of the intrinsic clustering or 
the selection criteria, we stress that the larger survey for \oiib s is 
required to further constrain their clustering and number density.

\subsection{Diffuse \oii\ Halos}\label{o2_halo}

\cite{steidel11} and \cite{matsuda12} have identified
the diffuse, extended \lya\ emission down to 
a very faint surface brightness limit 
in the composite images of 
UV-selected galaxies and/or LAEs at $z\sim3$. 
The diffuse \lya\ emission is thought to be a generic property 
of star-forming galaxies at high redshifts. 
\cite{steidel11} argue that \lya\ emission of
diffuse halos are produced by the resonance scattering
of neutral \hi\ gas in the circumgalactic medium (CGM).

\begin{figure}
\centering
\includegraphics[width=0.45\textwidth]{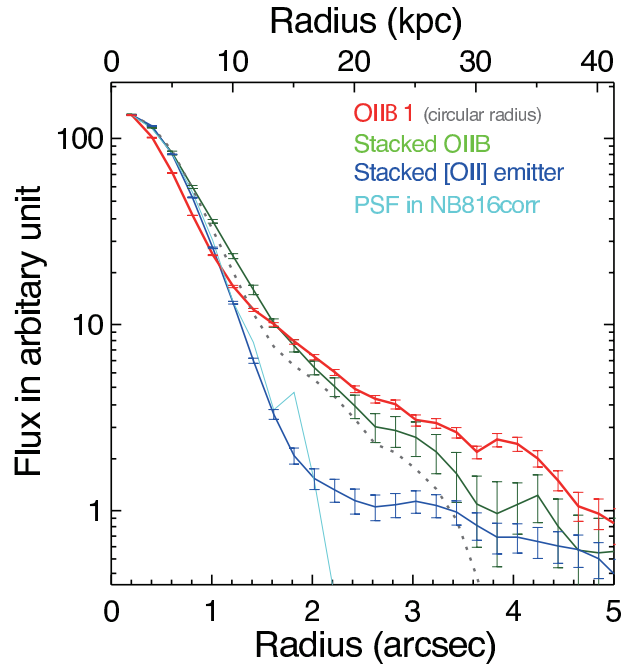}
\caption{
Surface brightness profiles of \oii\ emission.
The green solid line denotes the stacked \oii\ profile of all 12 \oiib s including 
\oiib\ 1, whereas the blue line is the stacked profile of all $z=1.2$ \oii\ emitters of 
\citet{drake13}. 
The red solid and gray dotted lines represent the \oiib\ 1
surface profiles of ellipsoidal and circular isophotes,
which are the same as those presented in Figure \ref{profile_o2blob1}. 
These surface brightness profiles are normalized at 
the innermost radius bin. 
Errors of the stacked profiles are estimated by
Monte-Carlo simulations that produce
a large number of sky image composites
in the same manner as the images for the stacked profiles.
The surface brightness errors of \oiib\ 1 are the same as
those of Figure \ref{profile_o2blob1}, which are
rms values of surface brightness at the given radius bin.
As a reference, the PSF profile is also shown.
}
\label{profile_stack}
\end{figure}

In this section, we investigate whether extended \oii\ emission
is universally found in any \oii\ emitters, using
\oii\ emission image (\nbcorr) composites.
Figure \ref{profile_stack} presents the stacked \oii\ profiles
for all \oiib s (the green line) and \oii\ emitters (the blue line). 
We compare these surface brightness profiles with
the one of \oiib\ 1.
%
%
Similar to \oiib\ 1, the stacked image of the \oiib s shows 
an extended component at the radius larger than 1\ar over the PSF. 
On the other hand, the stacked profile of the all \oii\ emitters of \citet{drake13} 
is similar to that of point sources (stars). 
We conclude that the extended nature of \oii\ emission is not a common feature 
of star-forming galaxies. Unlike \lya\ emission, 
\oii\ emission is not resonantly scattered by \hi\ gas in the CGM. 
Thus, the \oii\ halos can probably be found only in 
high-$z$ galaxies with strong outflows.

\section{Summary}\label{summary}

We present the first systematic search for galaxies with spatially extended metal-line 
\oii$\lambda\lambda3726, 3729$ emission at $z\sim1.2$, which we call ``\oiib s". 
We have discovered a giant blob with a spectroscopic redshift of $z=1.18$, 
\oiib\ 1.
We have also discovered 12 small \oiib s. 
The \oiib\ 1 extends over 75 kpc above 28 mag arcsec$^{-2}$ along its major axis. 
The rest-frame equivalent width (EW$_0$) of the entire blob calculated from 
the \nb\ and \rz\ images 
is $237^{+37}_{-33}$ \AA\ and the \oii\ luminosity is $1.5\pm0.2\times10^{43}$ \ergs. 
The major results of our study are summarized as follows. 

\begin{enumerate}

\item 
A strong emission line has been detected at $8127.9$ \AA\ in the VIMOS spectrum. 
We conclude that it is the \oii\ emission at $z=1.18$, because we do not find 
other strong emission lines at the bluer wavelengths which are expected 
if the line is \ha\ at $z=0.24$ or \oiii\ at $z=0.62$, 
because we detect continuum emission blueward of the line 
which is not expected if the line is \lya\ at $z=5.68$, 
and because the probability distribution of its photometric redshifts 
is peaked around $z=1.2$.

\item
We classify \oiib\ 1 as a radio-quiet obscured type-2 AGN according to the following 
properties. 
(1) \mgii$\lambda2796$ and \nev$\lambda3427$ emission lines are marginally detected in 
the VIMOS spectrum. 
(2) \oii\ line width of $537\pm4$ \kms\ is larger than those observed for 
normal star-forming galaxies but smaller than those for typical broad-line AGNs. 
(3) A much higher \oii\ equivalent width than typical AGNs and a prominent 4000 \AA\ break indicate 
that the AGN continuum is largely obscured. 
(4) The ratio of mid-infrared to radio fluxes ($q_{24}=0.624$) is consistent with 
those for radio-quiet galaxies.

\item 
The outflow signature of \oiib\ 1 is supported by the marginal existence of 
both \feii$\lambda2587$ absorption and \feiis$\lambda2613$ emission lines. 
Despite marginal detections, both lines show the consistent velocity offsets from the systemic 
velocity ($-605\pm99$ \kms\ and $-495\pm138$ \kms\ for \feii$\lambda2587$ 
and \feiis$\lambda2613$ lines, respectively).

\item 
Investigating the physical origins of the spatially extended \oii\ emission and outflow, 
we have found that 
\oiib\ 1 appears to favor the AGN-driven outflow scenario, 
though shock excitation and photoionization by AGN cannot be ruled out. 
Photoionization by star formation is unlikely, because 
the ionization parameter inferred from the \oii/\feiis\ ratio of $\log U > -1.0$ 
is higher than 
the average values found in star-forming galaxies at similar redshifts. 
The AGN-driven wind is, in contrast, supported by 
the central outflow velocities derived from the  
\feii$\lambda2587$ absorption and \feiis$\lambda2613$ emission lines, 
which are larger than those found in normal galaxies without an AGN, 
though the detection is tentative.

\item 
At $z\sim1.2$, the number density of giant \oiib s like \oiib\ 1 is 
$5.3\times10^{-6}$ Mpc$^{-3}$, 
while it is $6.3\times10^{-5}$ Mpc$^{-3}$ for the entire \oiib\  sample. 
We estimate that approximately 3\% of blue galaxies at $1.0\leq z \leq 1.2$ may be 
experiencing star formation quenching process similar to the \oiib s. 
Comparing the number densities of the \oiib s with those of AGNs with outflow features 
at similar redshifts, we find that \oiib\ 1 seems to be explained by AGN, but it is 
unlikely for the other \oiib s.

\end{enumerate}

We thank the anonymous referee for careful reading and valuable comments that 
improved clarity of the paper. 
We are also grateful to Masao Hayashi, Shota Kisaka, Tomoki Saito, and Wiphu Rujopakarn 
for their useful comments and discussions. 
This work was supported by KAKENHI (23244025) Grant-in-Aid for Scientific Research (A) 
through Japan Society for the Promotion of Science (JSPS). S.Y. acknowledges support 
from the JSPS through JSPS postdoctoral fellowship for foreign researchers. 
C.S. and A.B.D. acknowledge support from the UK Science and Technology Facilities Council. 

\bibliographystyle{apj}
\bibliography{o2blob_2013}

\end{document}

%% file: tab_o2blob_all.tex
\begin{deluxetable*}{l c cccc}
\tabletypesize{\footnotesize}
\tablecaption{Properties of 12 OIIBs at $z\sim1.2$\label{tab_o2blob_all}}
\tablewidth{0pt}
\tablehead{
\colhead{} & 
\colhead{\nbcorr\tablenotemark{a}} &
\colhead{{\it f}(\oii)\tablenotemark{b}} &
\colhead{{\it L}(\oii)\tablenotemark{c}} &
\colhead{Area\tablenotemark{d}} &
\colhead{$\langle\ {\rm SB}\ \rangle$\tablenotemark{e}} \\
\colhead{ID} & 
\colhead{(mag)} & 
\colhead{(\ergscm)} &
\colhead{(\ergs)} &
\colhead{(arcsec$^{-2}$)} & 
\colhead{(mag arcsec$^{-2}$)}  
}
\startdata
OIIB1   & 22.01 & 3.1(--16) & 2.5(+42) & 34 & 25.8 \\
OIIB2   & 23.31 & 9.3(--17) & 7.5(+41) & 26 & 26.9 \\
OIIB3   & 22.78 & 1.5(--16) & 1.2(+42) & 21 & 26.1 \\
OIIB4   & 23.03 & 1.2(--16) & 9.7(+41) & 19 & 26.2 \\
OIIB5   & 23.72 & 6.4(--17) & 5.1(+41) & 18 & 26.8 \\
OIIB6   & 23.18 & 1.1(--16) & 8.4(+41) & 17 & 26.3 \\
OIIB7   & 23.15 & 1.1(--16) & 8.7(+41) & 16 & 26.2 \\
OIIB8   & 24.23 & 4.0(--17) & 3.2(+41) & 16 & 27.2 \\
OIIB9   & 23.00 & 1.2(--16) & 1.0(+42) & 14 & 25.9 \\
OIIB10 & 22.89 & 1.4(--16) & 1.1(+42) & 14 & 25.8 \\
OIIB11 & 23.74 & 6.3(--17) & 5.0(+41) & 13 & 26.6 \\
OIIB12 & 23.97 & 5.1(--17) & 4.1(+41) & 13 & 26.8 
\enddata
\tablenotetext{a}{Isophotal magnitude in \nbcorr\ images.}
\tablenotetext{b}{\oii\ emission line flux.}
\tablenotetext{c}{\oii\ luminosity calculated by fixing redshifts at $z=1.19$.}
\tablenotetext{d}{Isophotal area measured on the smoothed \nbcorr\ image.}
\tablenotetext{e}{Average surface brightness.}
\end{deluxetable*}

%% file: tab_sum_photo.tex
\begin{deluxetable}{lc}
\tabletypesize{\footnotesize}
\tablecaption{Photometry of OIIB1\label{tab_sum_photo}}
\tablewidth{0pt}
\tablehead{
\multicolumn{1}{l}{Band} & 
\multicolumn{1}{c}{Total Magnitude/Flux}
}
\startdata
{\it f}(0.3--0.5 keV)\tablenotemark{a} & $2.8\pm2.5\times10^{-16}$ \\
{\it f}(0.5--2.0 keV; soft)\tablenotemark{a} & $1.1\pm0.4\times10^{-15}$ \\
{\it f}(2--10 keV; hard)\tablenotemark{a} & $1.7\pm0.6\times10^{-14}$ \\ 
{\it u} & $23.45\pm0.21$ \\
{\it B} & $23.25\pm0.19$ \\
{\it V} & $22.93\pm0.18$ \\
{\it R} & $22.62\pm0.15$ \\
\nb     & {\bf $21.06\pm0.07$ }\\
{\it i}   & $22.16\pm0.12$ \\
{\it z}  & $21.36\pm0.09$ \\
{\it J}  & $20.78\pm0.06$ \\
{\it H} & $20.22\pm0.05$ \\
{\it K} & $19.63\pm0.04$ \\
{\it m}(3.6\micron) & $19.03\pm0.05$ \\
{\it m}(4.5\micron) & $19.07\pm0.05$ \\
{\it f}(24\micron)    & $298\pm7\mu$Jy \\
{\it f}(1.4GHz)        & $216\pm22\mu$Jy 
\enddata
\tablenotetext{a}{The X-ray fluxes are taken from \cite{ueda08} in 
units of \ergscm.}
\end{deluxetable}

%% file: tab_sum_prop.tex
\begin{deluxetable}{lc}
\tabletypesize{\footnotesize}
\tablecaption{Properties of OIIB1\label{tab_sum_prop}}
\tablewidth{0pt}
\tablehead{
\multicolumn{1}{l}{Quantity} & 
\multicolumn{1}{c}{Value}
}
\startdata
Redshift ($z$) & $1.1804\pm0.0003$\tablenotemark{a}\\
Isophotal area\tablenotemark{b} in {\it NB816}$_{corr}$ & 34 arcsec$^2$ \\
Major axis\tablenotemark{c} in \nbcorr & 9\ar or 75 kpc\\
Half-light radius\tablenotemark{d} in {\it NB816}$_{corr}$ & 1\farcs0 or 8.3 kpc \\
{\it f}(\oii)\tablenotemark{e} (VIMOS) & $5.5\pm0.2\times10^{-16}$ \ergscm \\ 
{\it L}(\oii)\tablenotemark{e} (VIMOS) & $4.3\pm0.2\times10^{42}$ \ergs\\
{\it f}(\oii) (\nb\ and \rz\ images) & $4.1\pm0.3\times10^{-16}$ \ergscm \\ 
{\it L}(\oii) (\nb\ and \rz\ images)   & $3.2\pm0.2\times10^{42}$ \ergs\\
$M_B$\tablenotemark{f} & $-22.34\pm0.09$ mag\\
\oii\ EW$_0$\tablenotemark{g} (VIMOS) & $126\pm14$ \AA \\
\oii\ EW$_0$ (from \nb\ and \rz) & $71\pm10$ \AA \\
Stellar mass & $3.2\times10^{11}$ \Msun\\
Stellar age & $4.7$ Gyr\\
Color excess ({\it E(B-V)}) & 0.3 mag \\
SFR from UV\tablenotemark{h} & $52\pm4$ \Msun\ yr$^{-1}$\\
SFR from SED fit & $90$ \Msun\ yr$^{-1}$\\
SFR from \oii\tablenotemark{h} &  $45\pm 13$ \Msun\ yr$^{-1}$ \\ 
Number density & $5.3\times10^{-6}$ Mpc$^{-3}$\\
OIIB1 fraction & 0.1\%
\enddata
\tablenotetext{a}{The spectroscopic redshift is obtained from cross-correlation between 
the target spectrum and a template spectrum as described in \cite{simpson12}. }
\tablenotetext{b}{The isophotal area is determined above $2\sigma$ arcsec$^{-2}$
(28 mag arcsec$^{-2}$) in the {\it NB816}$_{corr}$. }
\tablenotetext{c}{Major axis is determined in the smoothed \nbcorr\ image by using 
the isophotal area shown in the SEGMENTATION image obtained from SExtractor. }
\tablenotetext{d}{Half-light radius is a radius containing 50\% of the total light 
from the object obtained by SExtractor.}
\tablenotetext{e}{The VIMOS spectrum was obtained using 1\ar\ slit, which  
apparently do not cover the entire blob. 
}
\tablenotetext{f}{The rest-frame $B$-band magnitude is calculated from 
the $z$-band photometry covering the rest-frame 3900--4500 \AA.
It is close to bandpass of the $B$ band; therefore, we do not apply $k$-correction 
to the magnitude. }
\tablenotetext{g}{The rest-frame \oii\ equivalent width (EW$_0$) is calculated from 
the VIMOS spectrum covering 1\ar\ at the center of the \oiib.}
\tablenotetext{h}{The SFRs are derived, respectively, from the UV continuum and \oii\ luminosity 
by applying the relation from \cite{kennicutt98} without any dust extinction. }
\end{deluxetable}

%% file: tab_other.tex
\begin{deluxetable*}{c c c c c c}
\tabletypesize{\scriptsize}
\tablecaption{Line Properties of OIIB1\label{tab_other}}
\tablewidth{0pt}
\tablehead{
\colhead{Line} & 
\colhead{$\lambda_{\rm rest}$ (\AA)} &
\colhead{$\lambda_{\rm obs}$ (\AA)\tablenotemark{a}} &
\colhead{$z_{\rm line}$} &
\colhead{Flux\tablenotemark{a}} \\
\colhead{(1)} &
\colhead{(2)} &
\colhead{(3)} &
\colhead{(4)} &
\colhead{(5)} &
}
\startdata
\feii\tablenotemark{b} & 2587 & $5628.0$ & $1.1764$ & Absorption? \\
\feiis\tablenotemark{b} & 2613 & $5690.6$ & $1.1772$ & Emission?  \\
\mgii\tablenotemark{b} & 2796 & $6097.5$ & $1.1805$ & Emission? \\ \\
\nev  & 3427 & $7467.0\pm4.1$ & $1.1789\pm0.0012$ & $2.5\pm1.6\times10^{-17}$\\ 
\oii    & 3727 & $8127.9\pm8.6$ & $1.1808\pm0.0023$ & $5.5\pm0.2\times10^{-16}$
\enddata
\tablecomments{
(1)--(2) Line names and their rest-frame wavelengths. 
(3) Observed wavelengths of the lines obtained from best-fitting 
Gaussian profiles by {\tt MPFIT}. 
(4) Redshifts. (5) Observed fluxes in the units of \ergscm. 
}
\tablenotetext{a}{An error on the observed wavelengths is a standard deviation obtained from fitting a line with a single 
Gaussian profile. }
\tablenotetext{b}{These lines are just marginally detected. Discussion using these lines should be done with caution. }
\end{deluxetable*}